\documentclass[usenatbib]{emulateapj}

\pdfoutput=1

\def\Msun{M_\odot}
\def\Lsun{L_\odot}

\def\Mstar{M_\star}
\def\Lstar{L_\star}
\def\Rstar{R_\star}

\def\Mdot{\dot{M}}
\def\etareim{\eta_{\rm Reim}}

\def\Teff{T_{eff}}
\def\Menv{M_{env}}

\shorttitle{A Parameter Space Study of EHB Stars}
\shortauthors{Yaron et al.}

\begin{document}


\title{An Extensive Grid of Models Producing Extreme Horizontal Branch Stars}


\author{
Ofer~Yaron\altaffilmark{1},
Dina~Prialnik\altaffilmark{2},
Attay~Kovetz\altaffilmark{2,3} and
Michael~M.~Shara\altaffilmark{4}
}

\altaffiltext{1}{Department of Particle Physics and Astrophysics, Weizmann Institute of Science, Rehovot 76100, Israel}
\altaffiltext{2}{Department of Geophysics and Planetary Sciences, Sackler Faculty of Exact Sciences, Tel Aviv University, Ramat Aviv 69978, Israel}
\altaffiltext{3}{School of Physics and Astronomy, Sackler Faculty of Exact Sciences, Tel Aviv University, Ramat Aviv 69978, Israel}
\altaffiltext{4}{Department of Astrophysics, American Museum of Natural History, Central Park West and 79th street, New York, NY 10024-5192}
\email{ofer.yaron@weizmann.ac.il, dina@planet.tau.ac.il, attay@etoile.tau.ac.il, mshara@amnh.org}

\begin{abstract}

Horizontal branch (HB) morphology is a complex multiple-parameter problem. 
Besides the metallicity, two other leading parameters 
have been identified to be the mass loss rate (MLR) and the initial helium abundance of the HB progenitors.
Using the {\it STAREV} stellar evolution code,
we produce a wide array of Extreme Horizontal Branch (EHB) stars and also examine their post-HB evolution.
EHB stars are produced in our calculations by the so called `delayed (late) hot core flash' scenario.
The MLR is increased on the red giant branch (RGB) to the extent that, prior to reaching core flash conditions, 
only a very thin hydrogen-rich envelope remains and helium 
ignition takes place at hotter positions on the Hertzsprung-Russell diagram (HRD). 
We perform an extensive, self-consistent parameter study, covering populations I and II ($Z=0.0001-0.03$),
for both normal initial helium abundances and He-enriched models (up to $Y=0.40$).
For completeness of the study and in order to point to  complete trends, we chose not to cut out several combinations (or results of) that may extend beyond realistic limits.
We present results and complete evolutionary tracks for the covered parameter space, showing in particular that: 
a)~Increased He abundance {\it on its own} -- without having a significant-enough MLR on the RGB --  {\it does not} lead to the production of EHB stars; however, 
b)~The bluest (hottest) HB positions do result from the {\it combined} effect of He-enhancement and increased MLR; 
c)~The general trend is that the effective temperature on the HB increases with decreasing metallicity, but there is an 
indication for a halt, or even a reversal of this trend, as Z further decreases below $10^{-3}$; 
d)~EHB stars can serve as major contributors to the UV flux emanating from their host system. 
Thus, the present comprehensive study both complements and lends support to previous, more restricted studies of the HB phase,
and adds results for unexplored regions of the parameter space.

\end{abstract}



\section{Introduction} \label{intro}

Globular clusters (GCs) are recognized as the oldest star clusters in the Milky Way, with ages
in excess of $10$ Gyr. The concentration of blue (and presumably young) main-sequence-like
stars, also known as Blue Straggler Stars (BSS), in the cores of GCs was first discovered by \cite{1991Natur.352..297P}. 
These objects were initially thought to be due to stellar collisions and mergers,
but additional processes have been proposed and investigated in recent years, especially 
mass-transfer in interacting binaries (\cite{2009Natur.457..288K, 2013MNRAS.428..897L, 2014ApJ...783L...8G}). 
It is now clear that most BSS form via mass transfer in close binaries.

The presence of extremely blue stars in GCs
was first noted almost a century earlier by \cite{1900ApJ....12..176B} and by \cite{1915mtwilson}.
Shapley also pointed out that the stars in M3 became bluer as they became fainter, in contrast to field dwarfs 
(i.e. main sequence (MS) stars). Using Shapley's data, \cite{1927tenBrug..book} 
plotted the first colour-magnitude diagrams (CMDs). He confirmed Shapley's
finding of red giant branches (RGB) that became bluer as they became fainter.
But he also discovered a horizontal branch of stars that separated from the cluster RGBs and that
extended to very blue colours. It is these stars, also observed by \cite{1939ApJ....90..387G} 
and by \cite{1955AJ.....60..317A}, that we refer to as the extreme (or extended) horizontal
branch (EHB) stars. \cite{1955ApJS....2....1H} realized that EHB stars must be core helium-burning, post red giant stars. 
An excellent review of EHB stars was given by \cite{2001PASP..113.1162M};
and a most extensive work reviewing both the theoretical and observational aspects of HB morphology
was carried out by \cite{2009Ap&SS.tmp...18C}.
Recently, a detailed theoretical and observational analysis aimed at setting the framework for the determination
of HB temperature distributions has been described in \cite{2015ApJ...800...52L}.

The pioneering observations by \cite{1955AJ.....60..317A} showed that the morphologies of
horizontal branches in GC CMDs are not smooth, continuous or uniform; and that
they vary much more than any other feature of cluster CMDs. In Arp's words, 
"The resulting colour-magnitude diagrams show that the form or population of the blue
horizontal sequence differ most from cluster to cluster".

Over the next fifteen years, three explanations for EHB stars emerged. 
\cite{1960ApJ...131..598S} were the first to point out a correlation between
HB morphology and metallicity: metal-poor clusters have bluer horizontal
branches. \cite{1966ApJ...144..978F} demonstrated that a high helium-shell abundance 
($Y = 0.35$) could also account for the observed EHB morphologies. The modern interpretation
of the horizontal branch emerged with the seminal paper of \cite{1970ApJ...161..587I}. 
They found that a helium abundance of $Y \sim 0.29 \pm 0.05$ best described cluster stars if 
$Z = 10^{-3} - 10^{-4}$. For these compositions, model tracks during core helium burning were all found to be 
short compared with the colour width of observed horizontal branches. Thus they concluded 
that there might be a spread in stellar mass along the horizontal branch, of the order of $0.1-0.2\ \Msun$.

The message of a variable envelope mass along the EHB, and the likely evolutionary phase
for this to happen were pointed out by \cite{1973ApJ...184..815R}, who noted that a
HB "...made up of stars with the same core mass and slightly varying
total mass, produces theoretical CMDs very similar to those observed... A mass loss
of perhaps $0.2\ \Msun$, with a random dispersion of several hundredths of a solar mass is
required somewhere along the giant branch". This conclusion is unchanged even when the
effects of semi-convection are included in stellar evolution calculations (\cite{1987ApJS...65...95S}). 
The most extreme (i.e. bluest) of EHB stars are likely to have very low-mass 
envelopes ($\lesssim 0.02\ \Msun$) (\cite{1972A&A....20..357C}). Such objects can evolve directly
towards the WD sequence rather than follow through and ascend the asymptotic giant branch (AGB).

To clarify, in accordance with the current canonical definition, 
we refer to EHB stars as those having black-body temperatures in excess of approx. $20,000$ K,
which are distinguished by their lack of an active hydrogen-burning shell due to their small envelope masses.
The term blue HB (BHB) is commonly used to refer to stars that reside on the HRD between the RR-Lyrae instability strip 
and up to $T_{eff}\approx20,000$ K.

\cite{1979ApJ...228...95C} and \cite{1982A&AS...50..247D} discovered that the
energy distributions of elliptical galaxies present significant ultraviolet (UV) excesses over
those expected from old stellar populations. These authors also noted that blue
HB stars' energy distributions were very similar to those of elliptical galaxies. 
Understanding EHB stars thus became important for the study of galaxies at all redshifts.
Indeed, both observational and theoretical investigations of the last two decades have shown that low-mass, small-envelope, 
evolved helium-burning stars in the EHB and subsequent evolutionary phases
may be major contributors to the ultraviolet upturn phenomenon (`UVX') as identified in old stellar populations,
mainly early-type --elliptical-- galaxies (\cite{1990ApJ...364...35G, 2000ApJ...532..308B, 2008ApJ...682..319B}).

The unexpected discovery of EHB stars in the cores of metal-rich bulge GCs
(\cite{1997ApJ...484L..25R}) "...demonstrates directly for the first time that a major
population of hot HB stars can exist in old, metal-rich systems".
\cite{1997ApJ...484L..25R} also noted that "the high central densities in NGC 6388 and 6441
suggest that the existence of the blue HB (BHB) tails might be caused by stellar interactions in
the dense cores of these clusters, which we calculate to have two of the highest
collision rates among GCs in the Galaxy. Tidal collisions might act in
various ways to enhance loss of envelope mass and therefore populate the blue side of the
HB."
The possibility that planets help strip some of the atmospheres of red giants has been noted by \cite{2011MNRAS.414.1788B}.  
The initial velocities of rotation of the progenitors of red giant stars might also cause different rates of mass loss on the red giant branch, 
as noted in \cite{2010A&A...519A..25G, 2013A&A...551L...4G, 2015MNRAS.446.1469D}.
Recently, \cite{2015ApJ...806..178S} modeled sdB stars using the MESA code (\cite{2011ApJS..192....3P, 2013ApJS..208....4P}),
with emphasis on a more detailed treatment of properties that affect the internal structure (like the extent of the convective core) 
for comparison with measured asteroseismology data.

In summary, it seems almost certain that there is more than one effect at work in
producing EHB stars in globular clusters, and that different effects may be active to
different degrees in different cluster environments. Thus, an extensive, systematic parameter study is indicated,
to probe the limits of properties that may be obtained by the combined effect of several factors. This is the
aim of the present paper, which may be viewed as an extension of 
the `delayed core He flash' scenario 
proposed by \cite{1993ApJ...407..649C} and thoroughly investigated by e.g., 
\cite{1993ApJ...419..596D, 1995ApJ...442..105D, 1996ApJ...466..359D, 2001ApJ...562..368B}, among others.
The added value of the current investigation is the \emph{fast} and \emph{continuous} calculation of evolutionary
tracks through all phases, using the code developed by \cite{2009MNRAS.395.1857K}, 
hereafter Paper 1, which yields both self-consistent results, 
and enables the calculation of the largest and most complete set of EHB evolutionary tracks obtained to date. 
For example, we do not need to stop during, say, core He flash, and recommence calculations with an assumed zero-age HB (ZAHB) configuration, 
consisting of an assumed (however reasonable) core mass. 
All calculations and evolutionary tracks presented here were obtained by continuous evolution from a pre-MS configuration, with no skips or shortcuts.

We stress that within the scope of this comprehensive parameter study we did not have a specific GC in mind,
nor did we try to simulate specific GC configurations.
We did not intend to include in this parameter-space study only those exact combinations of parameters (mass, composition) 
that produce specific measured GC ages.
{\it For completeness of the study we chose not to cut out some combinations (or results of) that extend beyond realistic limits 
(e.g. helium abundance below the conventional primordial value, or obtained ages at HB surpassing the age of the Universe), 
and this is mainly in order to show and point to the complete trends.} 

In the following section we briefly mention some of the features of the evolution code, referring to Paper 1 for a more detailed description. 
In \S\ref{calc} we present the results of numerical calculations
for the more common metallicity values representing populations I \& II ($Z=0.001$ to $0.02$),
followed in \S\ref{extend} by results obtained for an expanded metallicity range to extreme high and low values, 
stretching the parameter space to its limits.
A brief comparison with observationally-deduced atmospheric parameters of 
`hot subdwarfs' -- blue HB stars and their UV-bright progeny -- is illustrated in \S\ref{comp}.
A discussion and some concluding remarks are presented in \S\ref{conc}.


\section{The Evolution Code} \label{code}

As mentioned, over the last two decades sophisticated stellar evolution codes have been used for 
investigating HB stars and identifying the factors that lead to the 
hot HB stars that are likely connected to the UVX phenomenon (see extensive review by \cite{1999ARA&A..37..603O}) 
and to the blue HB tails and `blue hook' stars observed in CMDs of GCs.

The majority of the codes are not suited for a very extensive parameter study, and indeed, model calculations have been limited to 
some part of the parameter space. Since evolution codes differ in many details, assumptions and parameters, comparing or combining
the results of different studies into one picture is difficult. Moreover,
many (not all) of the evolution codes used for these studies were not capable of following all evolutionary stages in a continuous manner;
in particular, they do not continue through the core He flash, but rather stop at the tip of the RGB, or
wherever the runaway ignition of helium takes place,
and resume with a ZAHB model, with an \emph{assumed} value for the mass of the H-depleted core.
There are usually also difficulties in following through the post-HB stages and down the cooling WD track.

This prompted us to use our evolution code,
presented in detail in Paper 1, which has the capability of running continuously through all evolutionary phases---notwithstanding 
their widely varying timescales and behaviour---without any need for interruption, skips, or intervention.
For the low to intermediate-mass stars, this means following the evolution from a pre-MS initial model to a cooling WD.
The code is fast and is found to be stable for a wide range of masses and metallicities of all stellar populations.
It is therefore ideally suited for an extensive, self-consistent parameter study that requires the
calculation of hundreds of evolution tracks.

The evolution code is based on a fully implicit, non-Lagrangian adaptive-grid numerical scheme 
that---following \cite{1971MNRAS.151..351E}, \cite{1972MNRAS.156..361E}---simultaneously solves for structure, mesh and composition.
The equation of state is derived from the free energy, a sum of ionic, radiative and electronic (positronic) contributions,
together with corrections for pressure ionization (\cite{1995MNRAS.274..964P}), Coulomb interactions and quantum effects.
Ionization equilibria for H and He are taken into account, as well as creation of H$_2$ and electron-positron pairs.
Opacities are calculated as a harmonic sum of radiative and conductive opacities, with the radiative ones based on
\cite{1996ApJ...464..943I} OPAL Rosseland mean opacity tables, and on \cite{2005ApJ...623..585F}
for the low temperatures.  Conductive opacities are taken from the \cite{2007ApJ...661.1094C} tables.
For a specified value of Z, a set of 49 opacity tables is created, in which interpolations are performed
(cubic Hermite splines). 

Nuclear reaction rates are taken from \cite{1988ADNDT..40..283C}, and the reactions network used 
follows the changes in abundances of eight active isotopes:
$^1$H, $^4$He, $^{12}$C, $^{14}$N, $^{16}$O, $^{20}$Ne, $^{24}$Mg, and $^{28}$Si.
Other isotopes (like e.g. $^{56}$Fe) are regarded as inert; they contribute to the equation of state, but their abundances do not change.
Convection is treated according to the mixing-length recipe (\cite{1978stat.book.....M}), with a mixing-length parameter calibrated by a solar model.
We should stress here that such a code is not meant for a detailed analysis of isotopic abundances, which require a more extended
reactions network and a more detailed treatment of mixing at the boundaries of convective zones (e.g., convective overshoot, or thermohaline mixing)
than considered here. For such an analysis, see, e.g., \cite{2003ApJ...582L..43C}, or \cite{2008A&A...491..253M}.

The MLR is a more problematic issue. A unique, reliable law---one that would be valid for 
any stellar configuration and evolutionary stage---is still missing 
(see e.g. the discussion of RGB mass-loss by \cite{2009Ap&SS.tmp...18C}, and references therein). 
Instead, mass-loss prescriptions are used, the majority being mostly empirical in nature;
a few try to address more specific physical mechanisms that should affect the manner and efficiency
by which material is carried away from the star (be it pulsations, molecule/dust-driven winds etc...).
Some mass-loss laws are meant to be valid for particular evolutionary phases (e.g. AGB), 
and some have been constructed for particular stellar configurations (such as specific spectral types).
Most mass-loss prescriptions are functions of the basic stellar parameters $\Mstar, \Lstar, \Rstar$, and are variations of the original
formula proposed by \cite{1975MSRSL...8..369R}, which was derived from observations of RGB stars:
$\Mdot_{\rm Reim}=-4\times 10^{-13}\etareim \frac{\Lstar \Rstar}{\Mstar}$,
with its $\etareim$ coefficient as a free parameter, usually taken to lie between $0.3$ and $3.0$.

Despite differences between the various mass-loss laws---variations of the original Reimers's formula---a
common aim is to obtain an enhanced wind during advanced stages high up along the AGB.
In the present calculations we use Reimers's MLR throughout RGB evolution, adopting varying $\etareim$ values, 
as will be addressed in detail later on.
For the post-HB phases we apply the \cite{1995A&A...297..727B} $\Mdot_{\rm B1}$ MLR formula
(as in Paper 1).
It should be noted that since our purpose is to test the occurrence of hot HB stars due to enhanced mass-loss on the RGB, 
the question as to which mass-loss prescription should be used for the post-RGB phases is less important.
However, since we are also interested in identifying the post-HB evolution,
whether normal post-AGB or the `post-early AGB' (P-EAGB)/AGB-$\rm{manqu\grave{e}}$---progeny of hotter HB stars 
(see e.g. \cite{2000ApJ...532..308B} for a description of the possible post-HB evolution classes)---post-HB mass loss is
still relevant.


\section{Evolutionary Calculations} \label{calc}

We perform calculations of complete stellar evolutionary tracks ending with cooling WDs,
for various combinations of: initial metallicity -- $Z$, initial He mass fraction -- $Y$, and initial mass -- $M$.
For each such $(Z,Y,M)$ combination we adopt a sequence of increasing values for the mass-loss efficiency 
parameter $\etareim$, so as to cover a \emph{complete} `spectrum' of evolutionary patterns.

We first focus on the more common Pop. I and II metallicity values, spanning the range $Z=0.001$ to $0.02$;
metallicities covering the extended range $Z=0.0001$ to $0.03$ will be addressed in \S\ref{extend}. 
The initial He mass fractions---$Y$---are either taken to be suitable to the adopted $Z$ 
(see e.g. the relation for $Y(X,Z)$ as proposed by \cite{1998MNRAS.298..525P}):
Y=$0.24$ for Z of $0.001$ or Y=$0.28$ for Pop. I (around solar metallicity);
or, for the He-enhanced models, increased up to $Y=0.40$ for both populations.
Initial masses are taken to range from $0.80$ to $1.20\ \Msun$, so as to include
typical MS-turnoff (MSTO) masses of galactic GCs, that are normally around $0.80$ to $0.90\ \Msun$ 
(e.g. the estimated $M_{TO}\backsimeq 0.85\ \Msun$ for NGC 2808, \cite{2001ApJ...562..368B}).
Although the aim is to test mainly those combinations for which the age at ZAHB
is of order of $10\pm 5$ Gyr, for completeness, we also consider combinations that yield ages at ZAHB deviating from this range.
To clarify, we did not aim to compute sequences that give specific GC ages, which can be done by adjustment of the ZAMS mass 
per given composition (Z,Y values), but to present the complete set of results that have been obtained for the chosen parameter-space combinations. 

The mixing length parameter---$\alpha$---is taken to be $2.5$ for the Pop. I models,
according to the solar calibration model, as presented in Paper 1; $2.0$ for the Pop. II models, 
and $2.2$ for an intermediate $Z=0.005$, as will be mentioned in \S\ref{extend}.

We note that for each $(Z,Y,M)$ combination, different sequences of $\etareim$ values
are required to account for all the different evolutionary behaviours.
Lowest and highest $\etareim$ values, as well as the span ($\eta_{R,max}-\eta_{R,min}$) vary from one combination to another.
In this context, we would like to point out that there is no meaning in comparing results of different studies for the same $\etareim$, since
results depend on the input physics adopted and also, to some extent, on numerical parameters. Rather, the general spread
and trends should be compared as functions of the free parameters - $Z,Y$ and $M$. 

In the description and discussion of the results, we relate to the following four evolutionary patterns (following the categorization 
suggested by \cite{1996ApJ...466..359D}): 
\begin{enumerate}
\item `Tip Flasher' - A `classic' core He flash taking place at the tip of the RGB; 
obtained for the lower end of $\etareim$ values, leading to `normal' (red) HB positions.
\item `Post-Tip Flasher' - core flash develops only after turning from the tip of RGB (`RGB peel-off') towards higher $\Teff$; leading to bluer HB positions.
\item `WD Flasher' - core flash occurs while already descending along the WD cooling curve; obtained for high $\etareim$ values; leading to the hottest HB positions.
\item `He-WD' - core does not reach the required conditions for ignition of He and we are left with a He WD; obtained for yet higher $\etareim$ values.
\end{enumerate}

For each series of runs for a given $(Z,Y,M)$ combination and increasing values of $\etareim$, 
we choose four representative ones that cover the four evolutionary patterns mentioned above.
These are also the cases indicated in the last column ("Flash") of the tables and
displayed (for each combination) in the HRD figures and accompanying tables.
(Only for two specific combinations, we show results---in Tables~1 and 3 and the corresponding figure, see \S\ref{comp}---for 
a more `densely sampled' sequence of $\etareim$ values.)

Within each sequence of the four cases, the first two lower $\etareim$ values---corresponding to `Tip Flasher' and `Post-Tip Flasher'---
should not be regarded as lower or upper-limiting values, but just as representative ones.
However, the $\etareim$ value corresponding to `WD Flasher' is the maximal one resulting in 
that outcome, thus an upper limit. 
This is the value for which the bluest HB position was obtained for the corresponding $(Z,Y,M)$ combination.
The $\etareim$ value corresponding to `He-WD' is the minimal one for which a He WD was obtained;
any higher $\etareim$ value would also lead to a He WD, that is, prevent ignition of He.

Figures.~\ref{fig:hrds-001}--\ref{fig:hrds-02},\ref{fig:hrds-001He}--\ref{fig:hrds-02He} show sequences of four tracks for each
$(Z,Y,M)$ combination. We note that both the MS and HB branches have a width on the HRD, marked as thick red and blue bands, respectively,
on all the displayed evolutionary tracks. 
The beginning and end-points of the HB---ZAHB and TAHB (Terminal-Age HB), respectively---have to be 
defined with some care if one wishes to analyze the HB morphology, in comparison with other studies.
Still, we must keep in mind that there is some degree of arbitrariness in any such definition.
We set the ZAHB when thermal equilibrium is achieved ($L=L_{nuc}$) after He ignition.
We determine the TAHB by requiring that the central He mass fraction---$Y_c$---be less than $10^{-6}$ and a turnoff in the HR-diagram has begun. 

Because the HB has a significant width (in both $\log L$ and $\log \Teff$), we list in the tables the resulting properties 
for both the beginning and end points of the HB, as have been identified by the above criteria. Properties are also
listed for an intermediate point on the HB, where they may be taken to be representative of the HB phase. 
It should be noted that the treatment of the TAHB following the above criterion is theoretically-based 
and is less meaningful observationally-wise.
The duration of HB stars around the TAHB is short, the evolution speeds up considerably towards the end of the HB,
so observationally, very few stars will actually be observed around this phase. The HB widths we present 
(luminosities, temperatures) are thus likely larger than what is effectively observed. 
 
\begin{figure}
\begin{center}
\includegraphics[scale=.4]{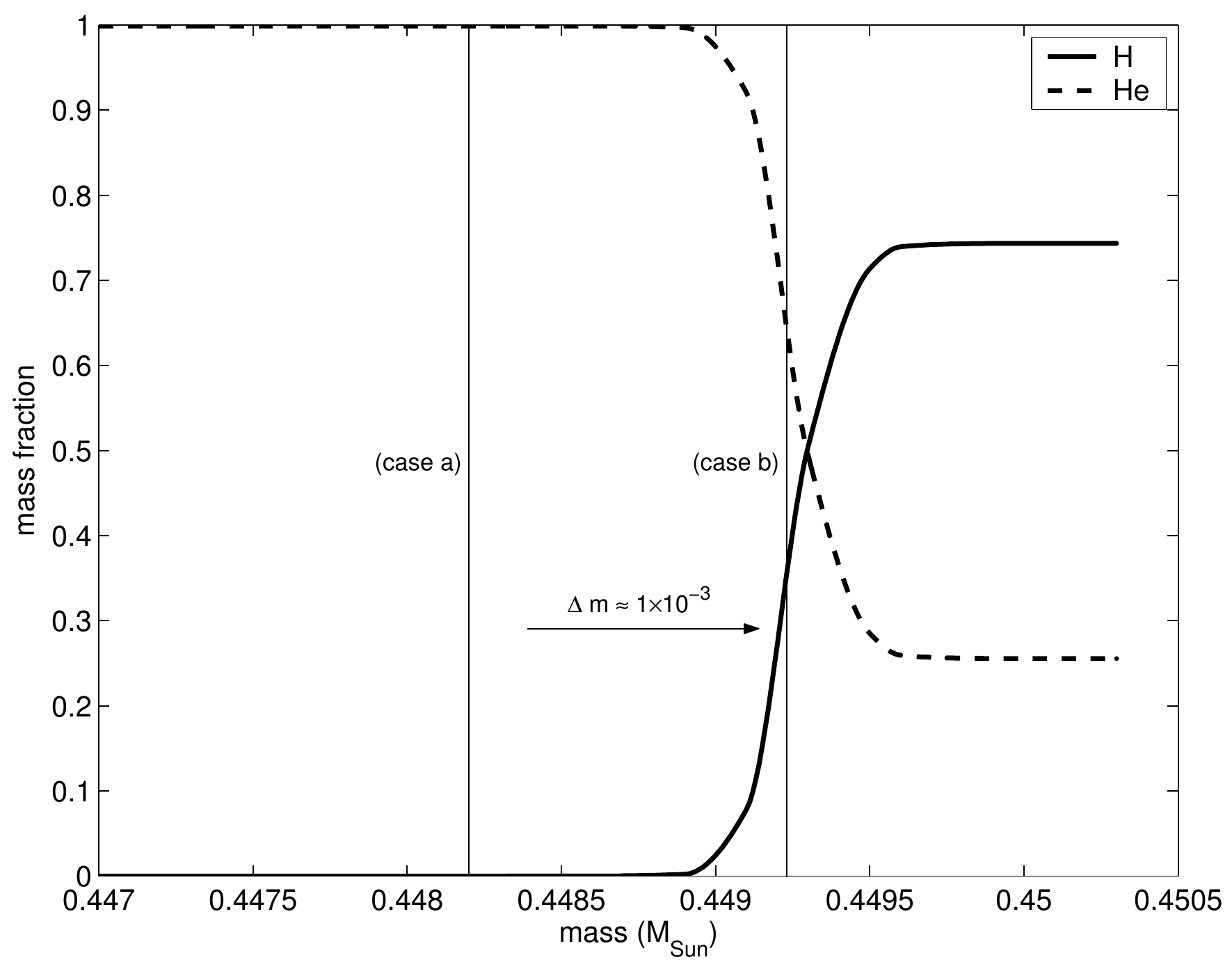}
\end{center}
\caption{H and He mass fraction profiles at the boundary between the helium (H-depleted) core and the H-rich envelope
at time of core He flash, for model combination $(Z,Y,M,\etareim)=(0.001,0.24,0.90,1.80)$.
The boundary as set by the code --- marked {\it case a} (see text, \S\ref{calc}) --- is denoted by the vertical solid line to the left; 
a different definition --- marked {\it case b} --- results in the vertical line to the right.
}
\label{fig:coremass}
\end{figure}

We define the outer boundary of the core as the outermost point where $X\,<\,10^{-6}$ ({\it case a} in Figure~\ref{fig:coremass}).
Alternatively, one may mark the core boundary at the point where the hydrogen content has dropped to half its maximum (initial) value
({\it case b} in that figure). Clearly, the core mass will be larger by the latter definition, and the envelope mass, correspondingly smaller.
In most cases, the difference between the two definitions is small, of the order of a 1\%; 
however, when the envelope mass becomes very small, of order
$10^{-3}\Msun$ or less, different definitions may yield results that diverge by a factor of $2$, although they refer to the same model.
In our case, since the definition of core mass tends to minimize it, envelope masses may sometimes appear larger when compared
to those of other studies. 

The green circle on each of the evolutionary tracks marks the position where the core He flash has just begun,
with $L_{nuc}\,>\,10^5\ \Lsun$.
The peak nuclear energy generation rate $L_{nuc,max}$ attained during the thermonuclear runaway (TNR) for all calculations 
is found to be in the range $1-7\times 10^9\ \Lsun$, with $L_{nuc,max}$ decreasing with increasing stellar mass for a given $Z$,
and also decreasing with increasing MLR for each $(Z,M)$ combination.
The time elapsed from the core flash until the ZAHB is of the order of a few $10^6$~yr for all parameter combinations and MLRs.

We divide our sets of calculations into three groups and address each group separately.
First, we describe sequences of increasing MLR ($\etareim$ values) for various $(Z,Y,M)$ combinations, 
adopting `normal' $Y$ values for the corresponding $Z$ (\S\ref{mlr}).
We then examine in \S\ref{vary} consequences of varying $Y$ on HB morphology 
by testing sequences of increasing $Y$ for various $(Z,M,\etareim)$ combinations, 
$\etareim$ being fixed for each $(Z,M)$, so that the realistic $Y(Z)$ leads to a `normal' `Tip Flasher'.
Finally, we present in \S\ref{yenrich} calculations of increasing MLR for He-enriched initial models, 
adopting $Y=0.32$, 0.36 and 0.40 for the various $(Z,Y,M)$ combinations.

\subsection{Increasing MLR for Pop. I and Pop. II compositions} \label{mlr}

\begin{figure}
\begin{center}
{\fbox{\includegraphics[scale=.45]{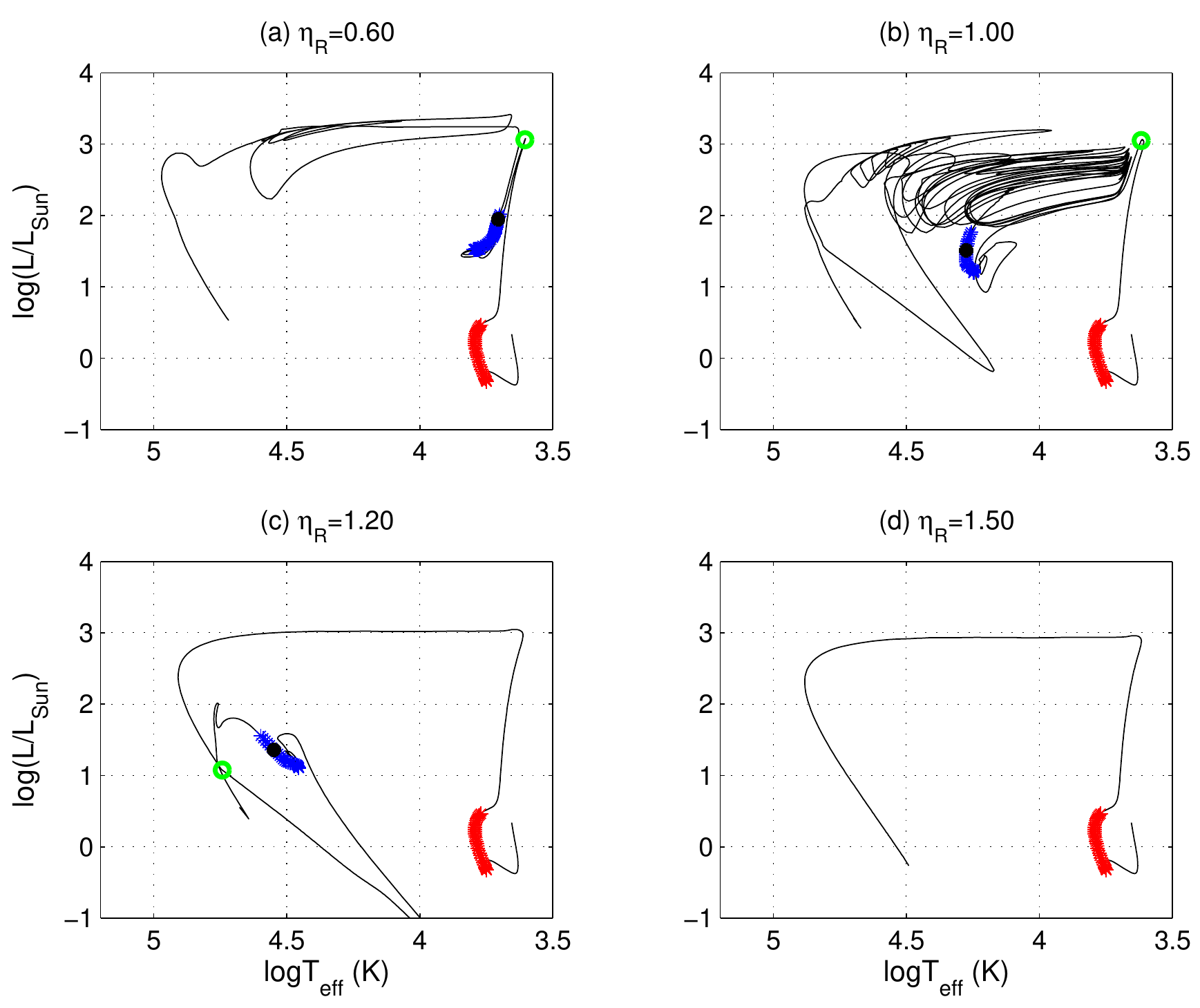}}
\fbox{ \includegraphics[scale=.45]{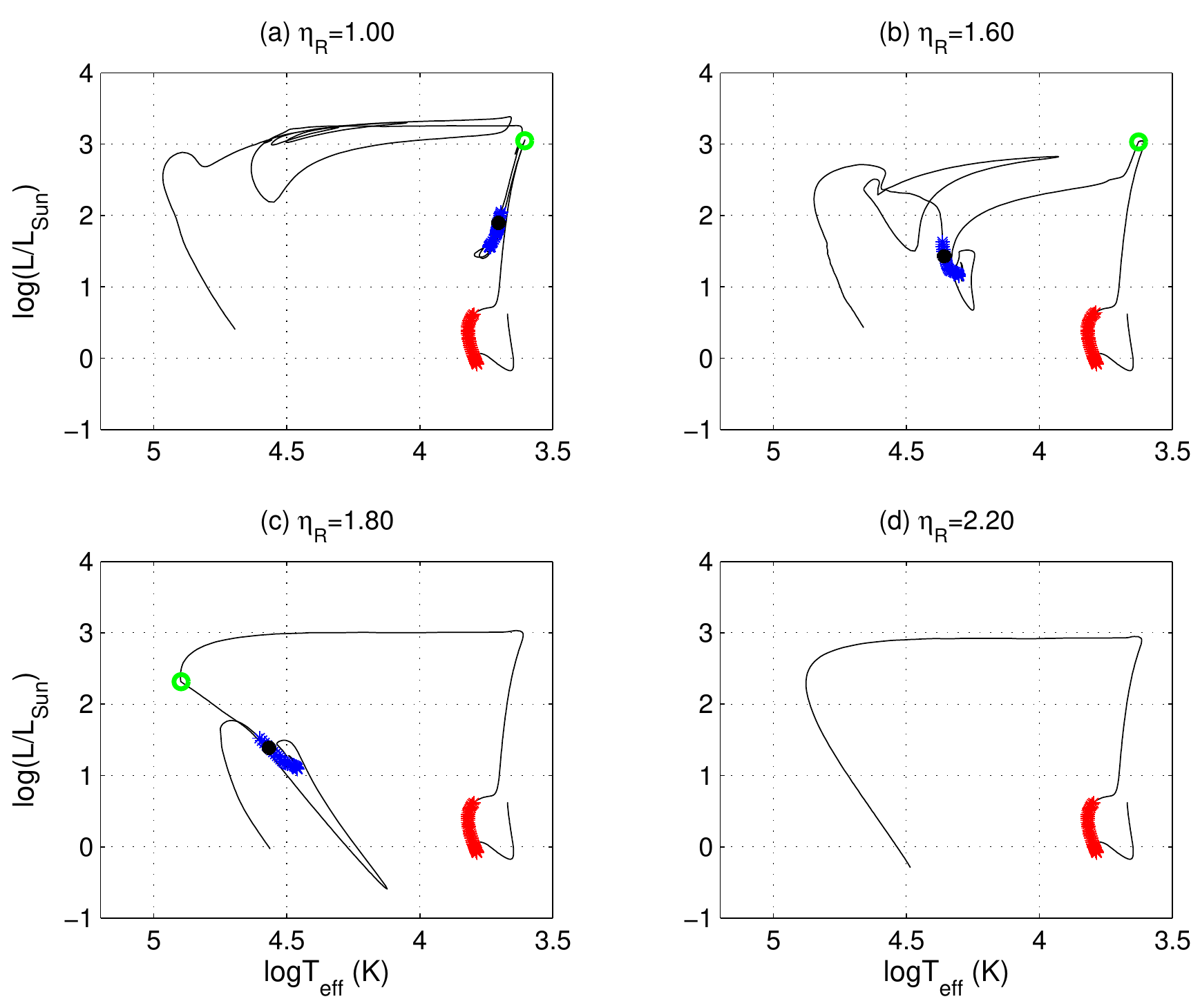}}}
{\fbox{\includegraphics[scale=.45]{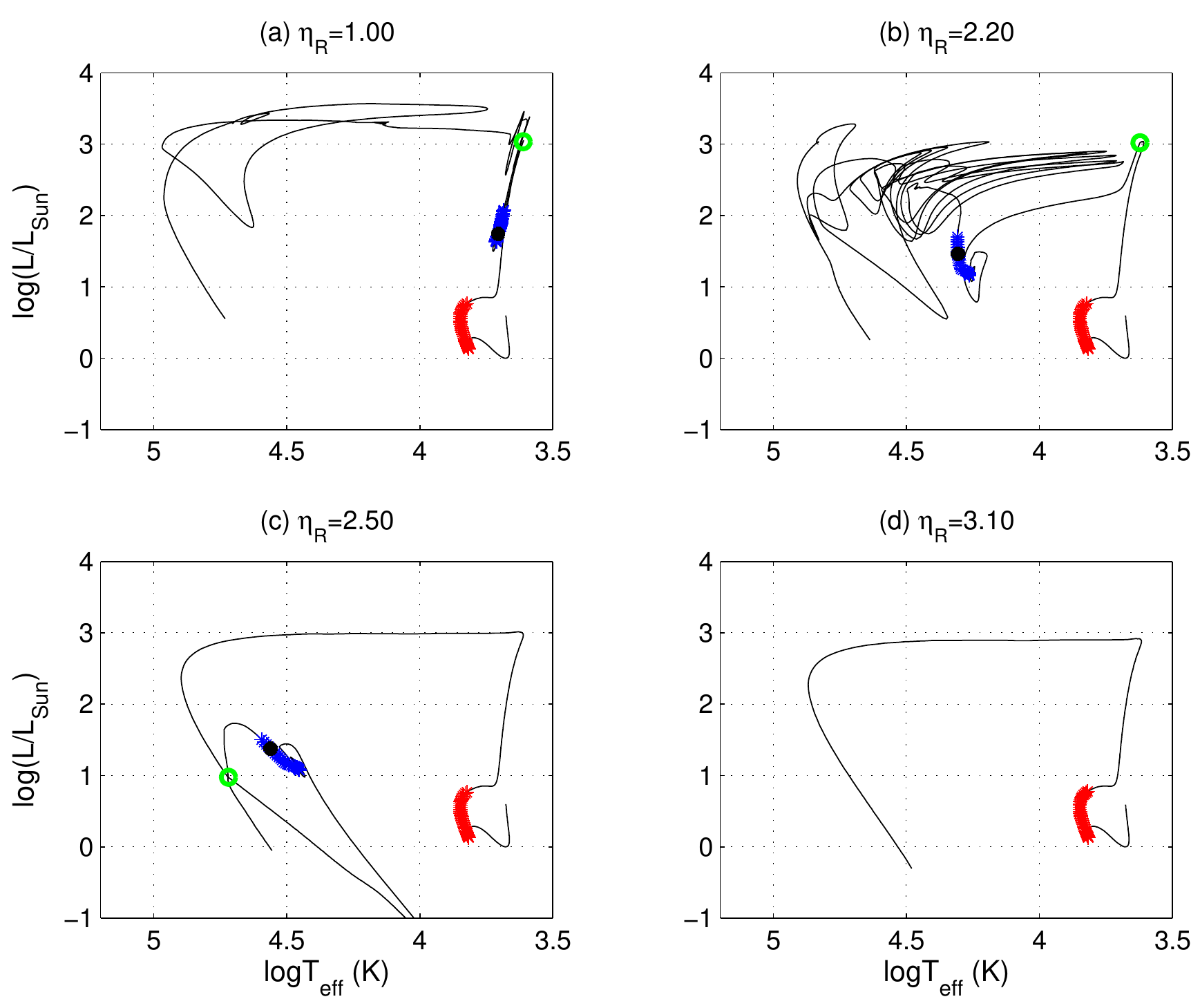}}
\fbox{ \includegraphics[scale=.45]{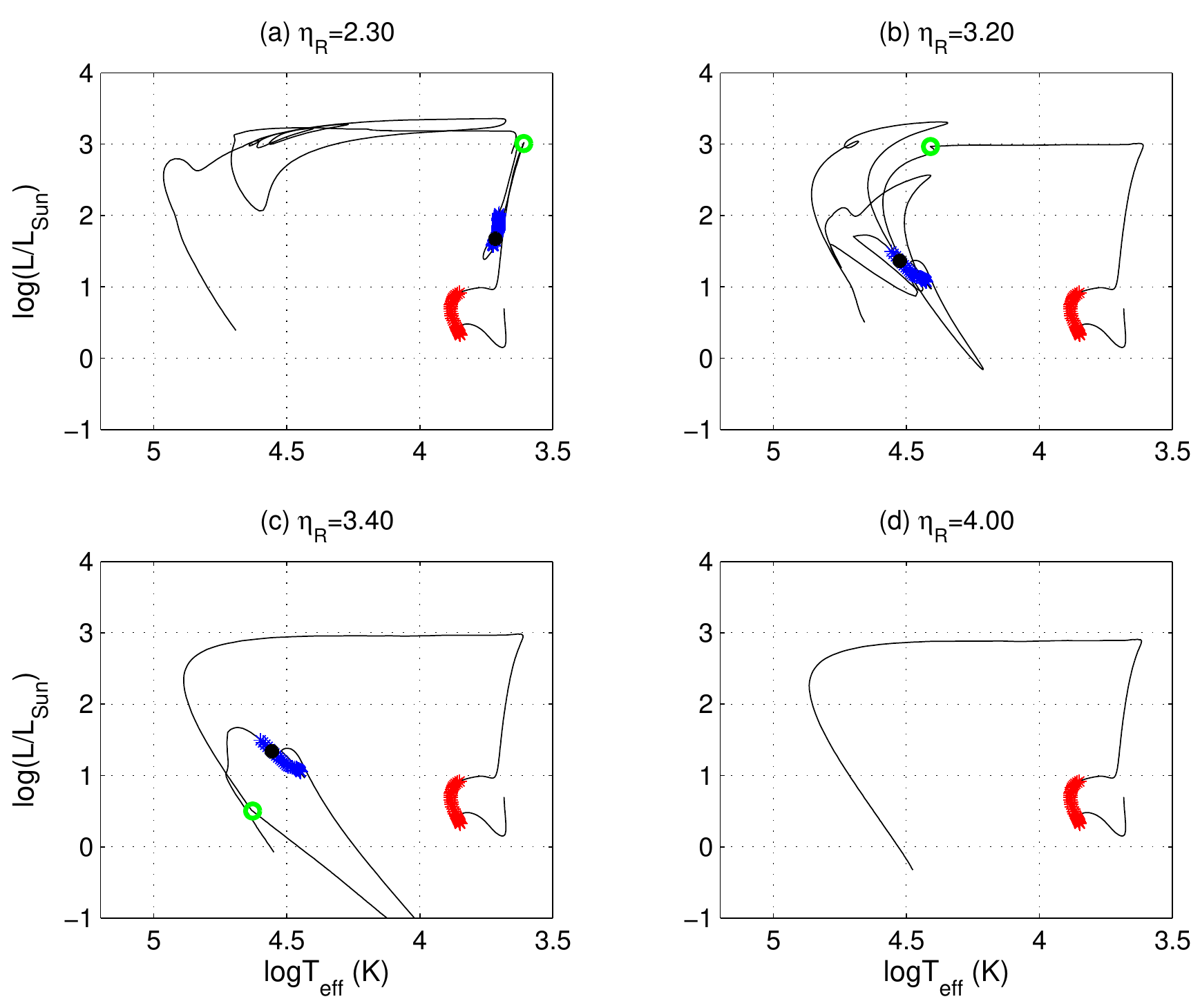}}}
\end{center}
\caption{Pop. II $(Z,Y)=(0.001,0.24)$ -- Complete evolutionary tracks on HRDs for sequences of increasing MLR ($\etareim$ values, shown in title of each HRD).
The four large panels correspond to masses of 0.8 (top left), 0.9 (top right), 1.0 (bottom left) and 1.1 $\Msun$ (bottom right).
Asterisks forming thick red and blue bands correspond to MS and HB phases, respectively;
the darker asterisk within the blue band marks `Mid'-HB - position of maximal timestep duration within HB (see text);
green circle marks the position of onset of core He flash.
Ages at ZAHB are 15.7, 10.0, 6.8 and 4.8 Gyr for the four masses, respectively.
See table~1 for more details related to the displayed sequences.
}
\label{fig:hrds-001}
\end{figure}

\begin{figure}
\begin{center}
{\fbox{\includegraphics[scale=.45]{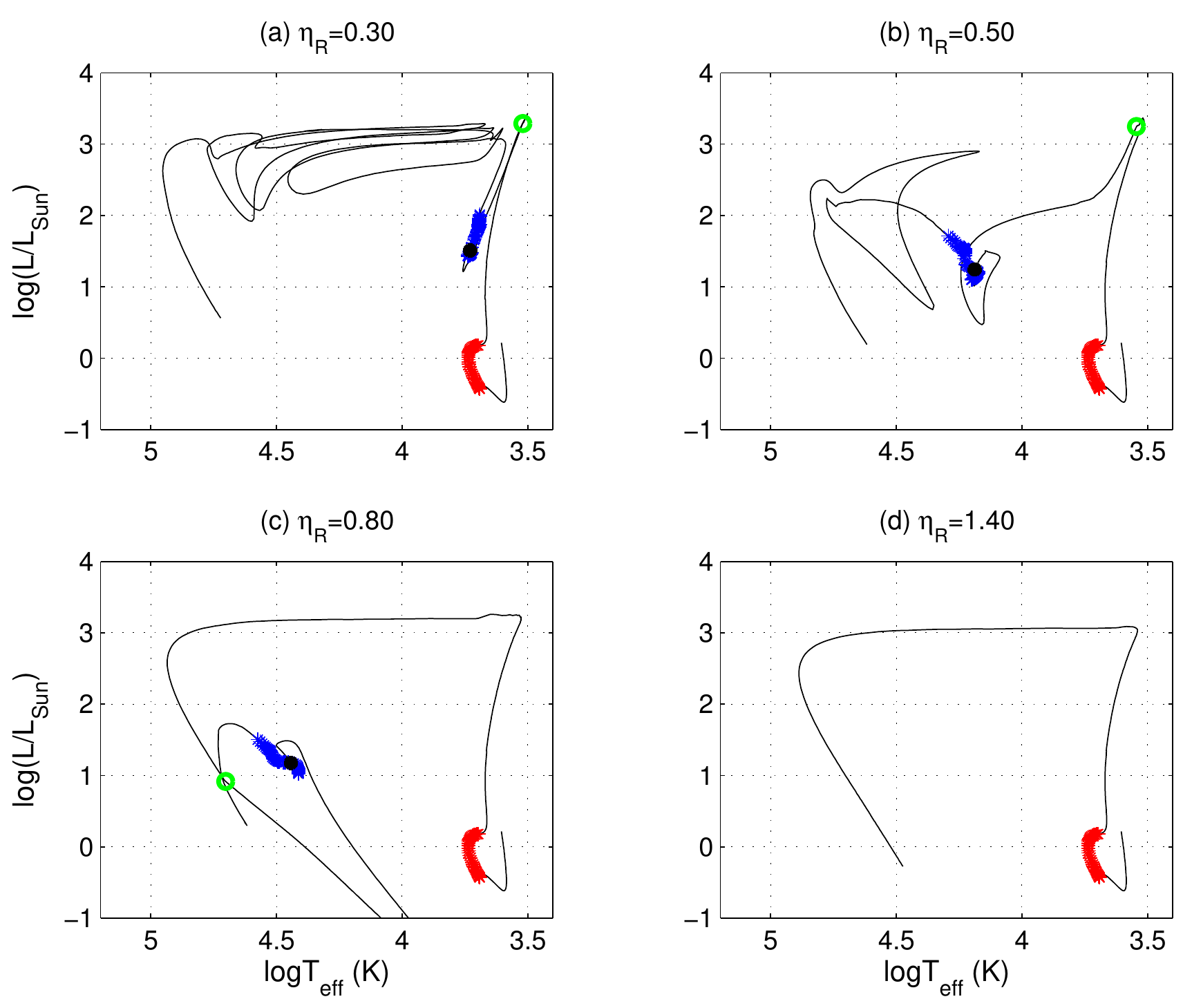}}
\fbox{ \includegraphics[scale=.45]{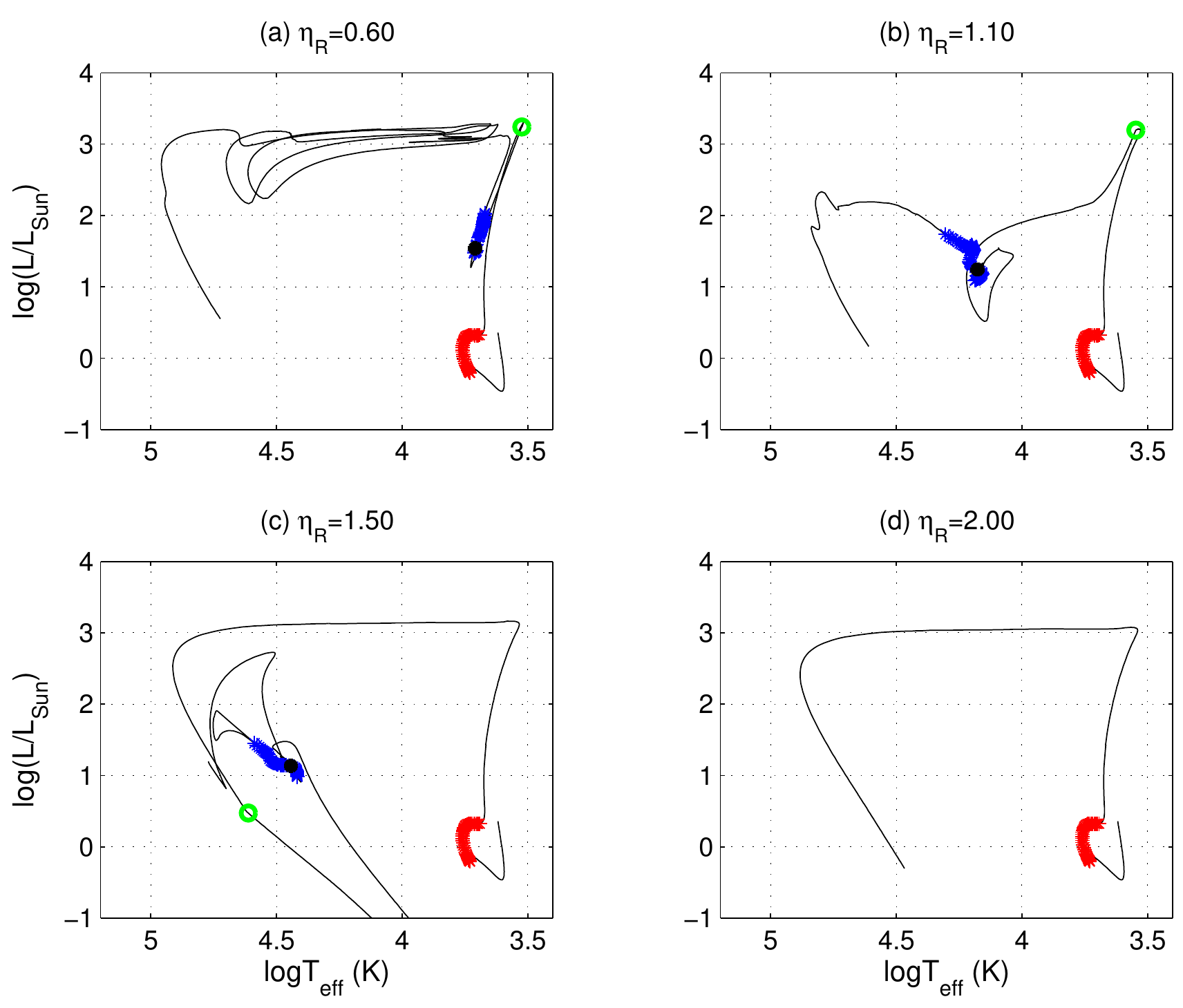}}}
{\fbox{\includegraphics[scale=.45]{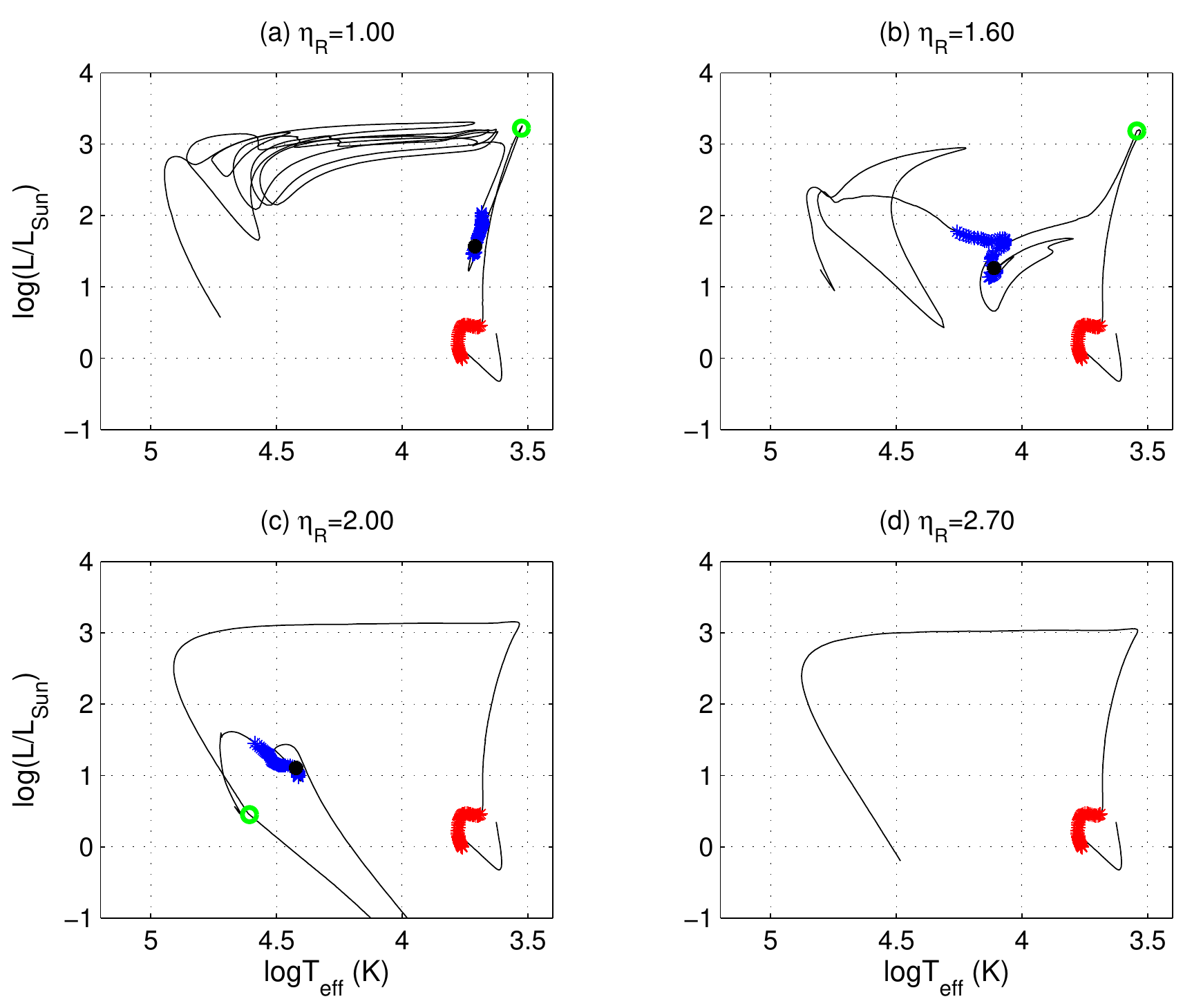}}
\fbox{ \includegraphics[scale=.45]{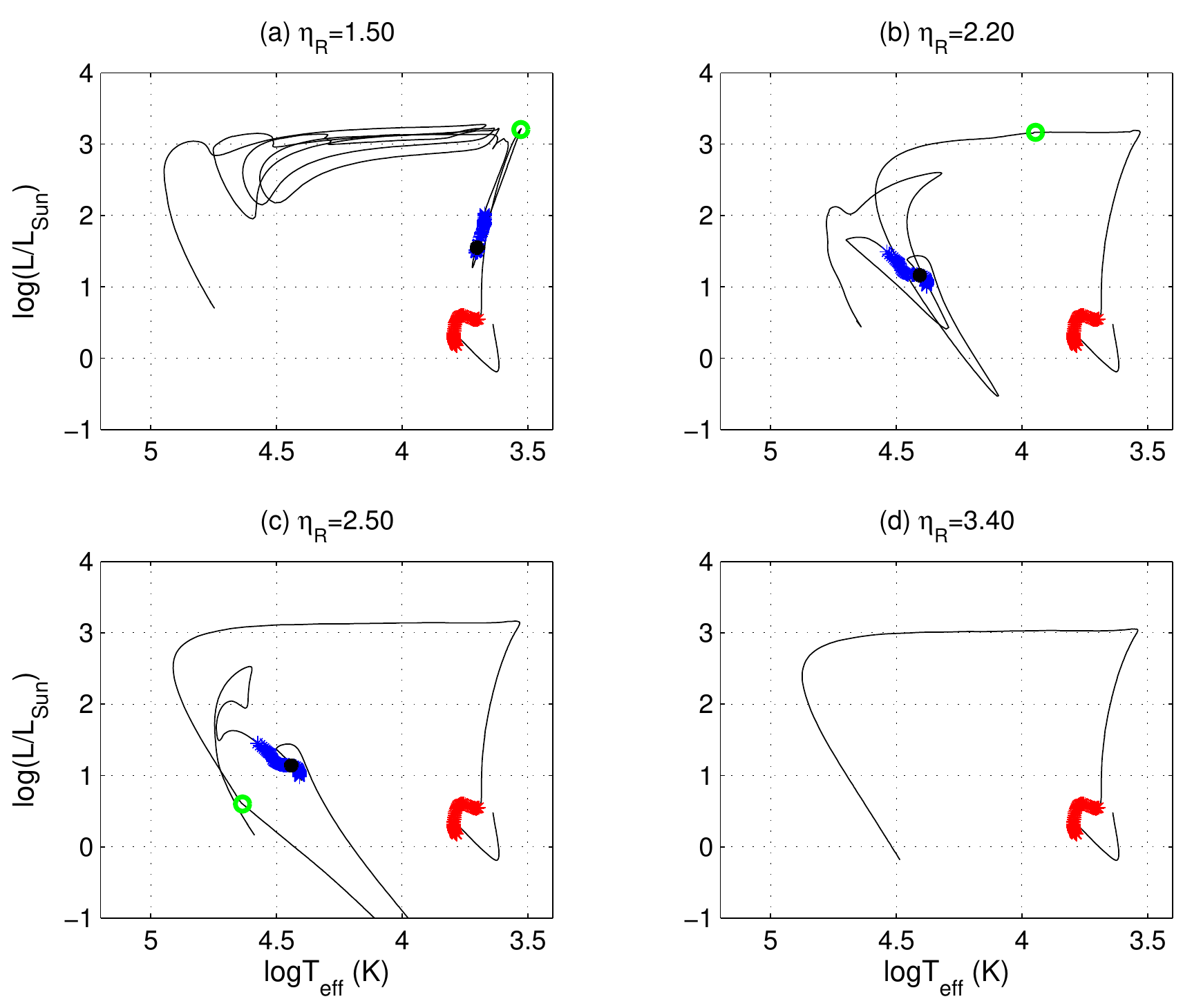}}}
\end{center}
\caption{Pop. I $(Z,Y)=(0.02,\ 0.28)$ -- Complete evolutionary tracks on HRDs for increasing MLR ($\etareim$ values, shown in title of each HRD).
The four large panels correspond to masses of 0.9 (top left), 1.0 (top right), 1.1 (bottom left) and 1.2 $\Msun$ (bottom right).
(See figure~\ref{fig:hrds-001} for explanation of the plotted symbols.)
Ages at ZAHB are 19.8, 13.1, 9.0 and 6.5 Gyr for the four masses, respectively.
See table~1 for more details related to the displayed sequences.
}
\label{fig:hrds-02}
\end{figure}

\begin{figure}
\begin{center}
{\fbox{\includegraphics[scale=.45]{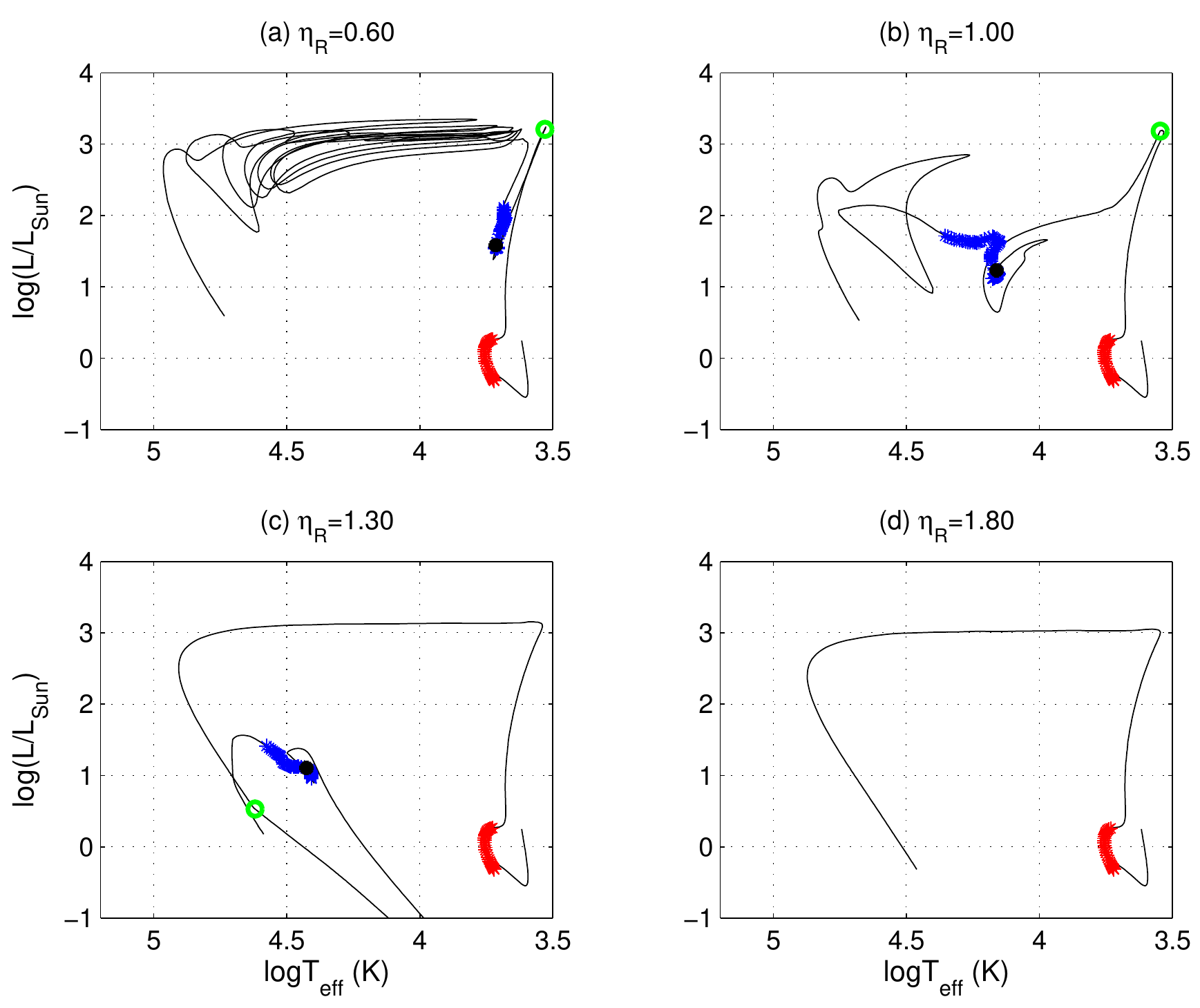}}
\fbox{ \includegraphics[scale=.45]{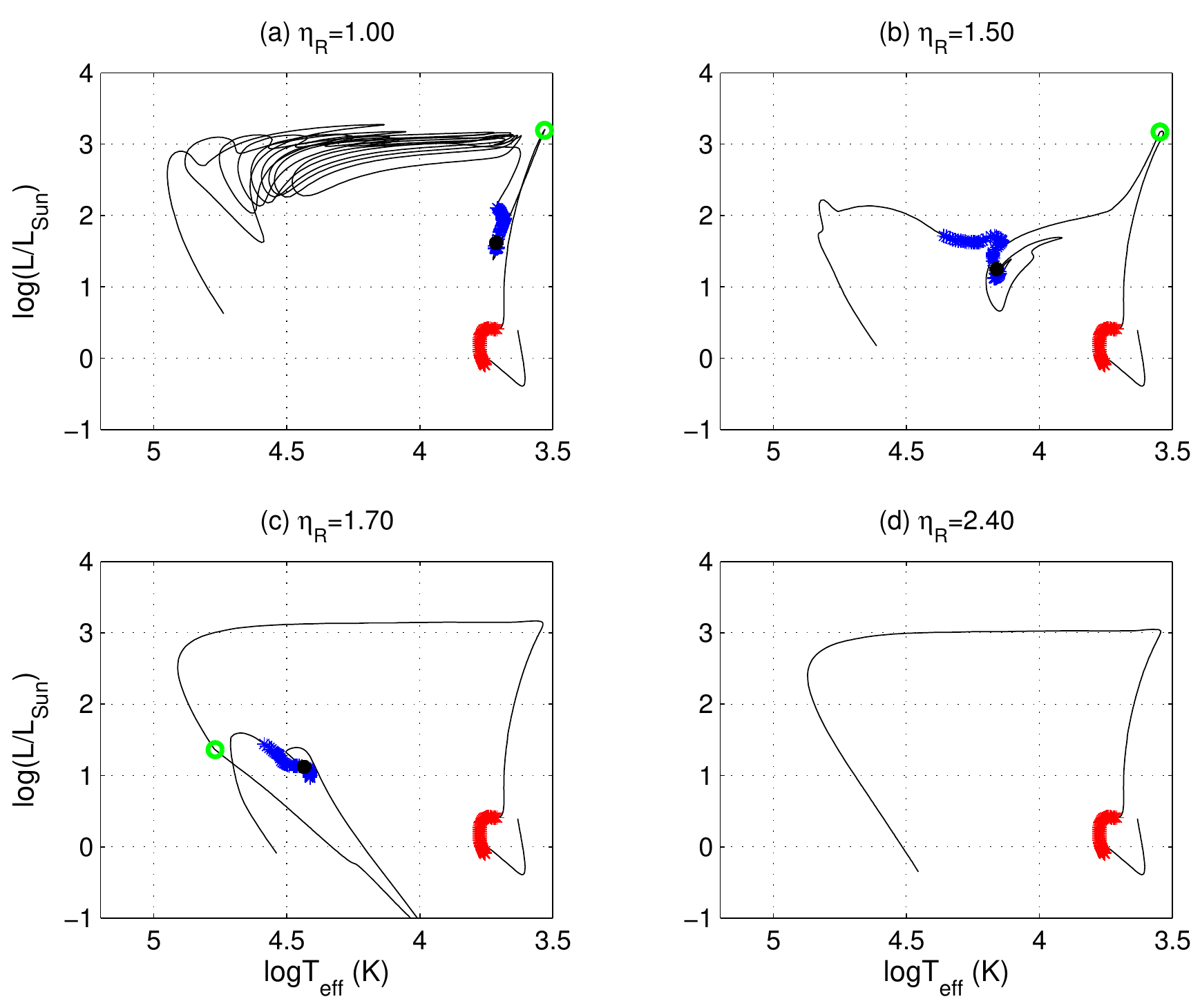}}}
{\fbox{\includegraphics[scale=.45]{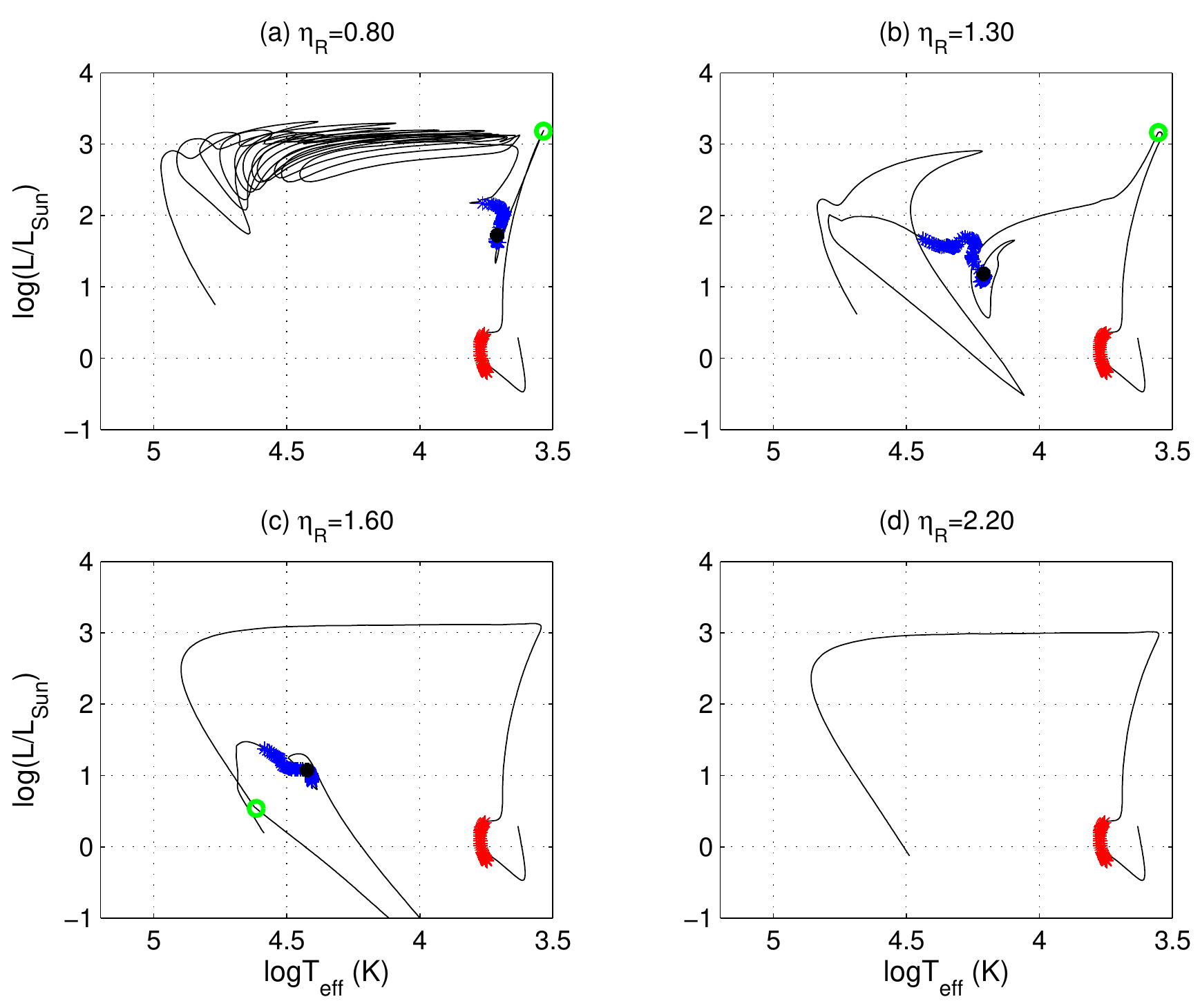}}
\fbox{ \includegraphics[scale=.45]{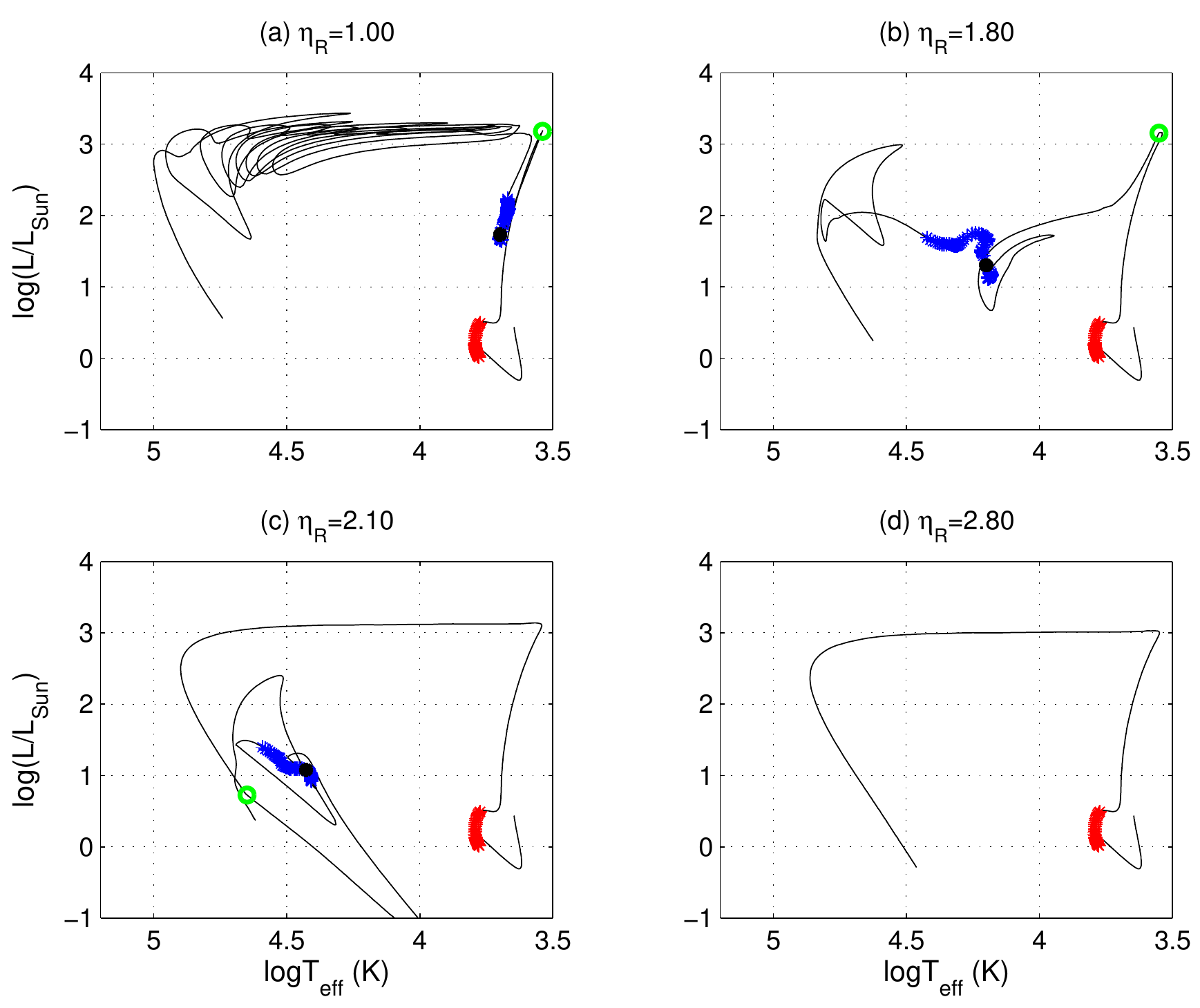}}}
{\fbox{\includegraphics[scale=.45]{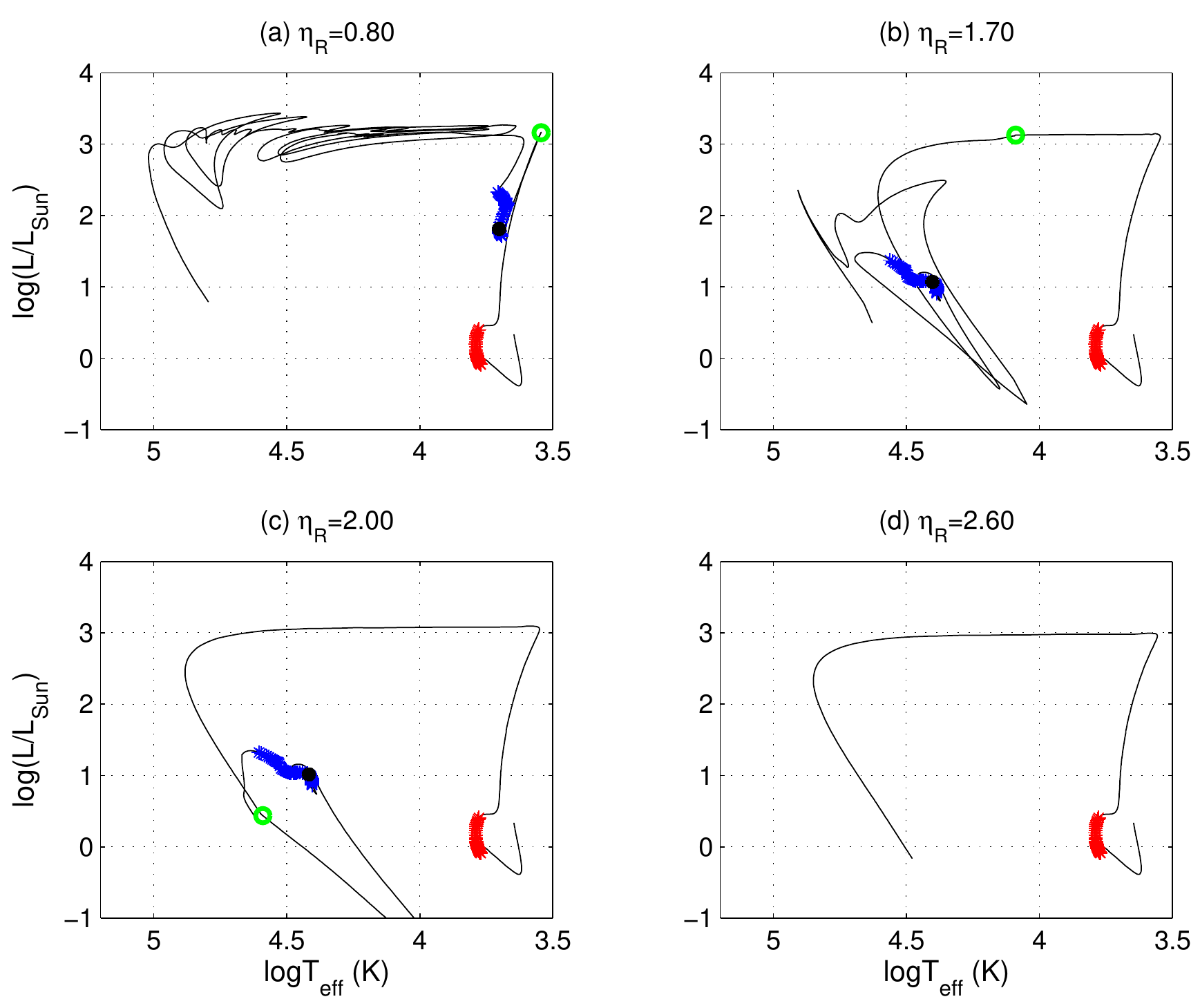}}
\fbox{ \includegraphics[scale=.45]{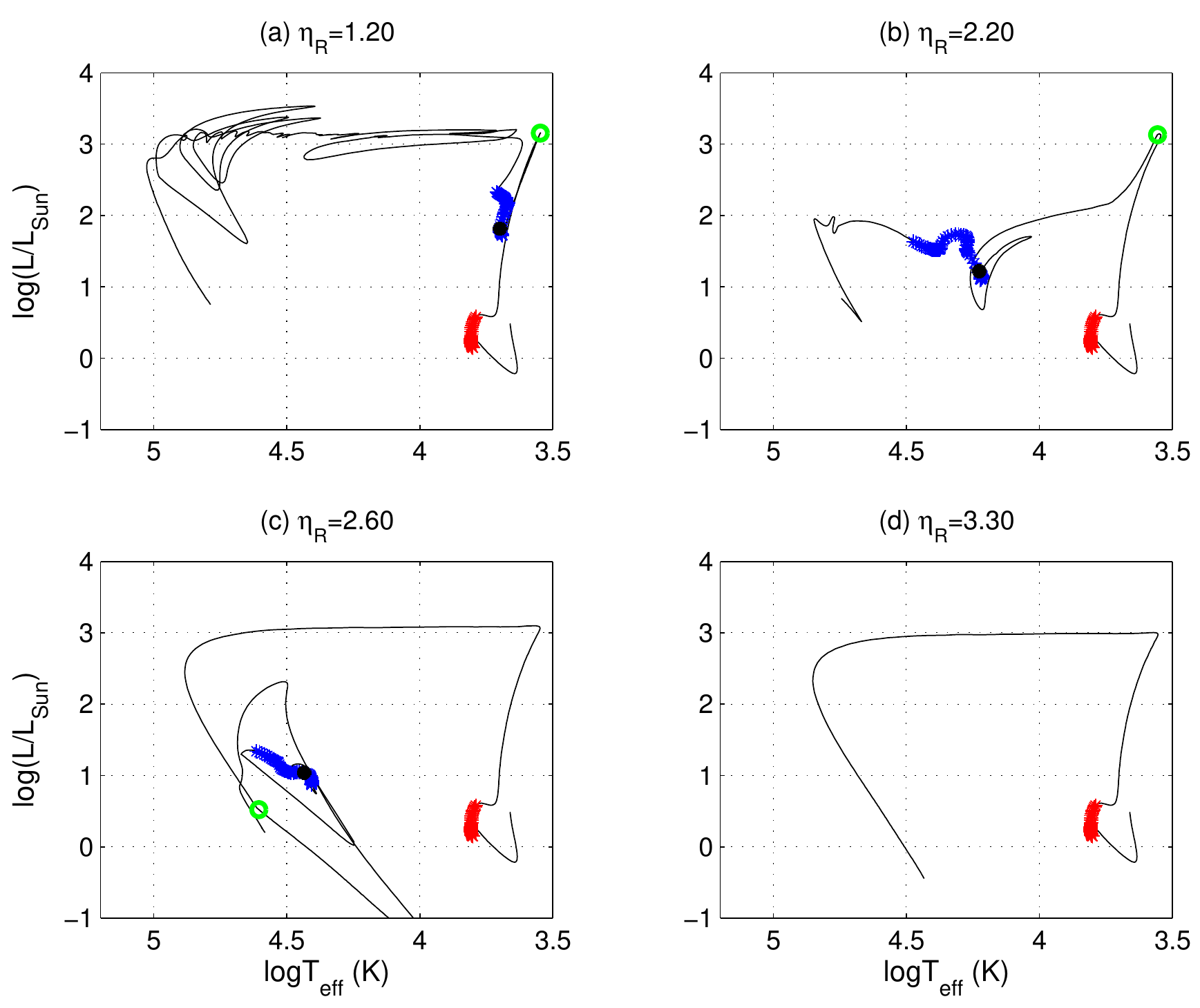}}}
\end{center}
\caption{Pop. I -- $Z=0.02$ -- He-enriched -- Complete evolutionary tracks on HRDs for increasing MLR ($\etareim$ values, shown in title of each HRD).
Left and right columns are for masses of $0.9$ and $1.0\ \Msun$, respectively;
within each column, the three panels relate to increasing initial He abundance---$Y$---of $0.32, 0.36$ and $0.40$, top down.
(See figure~\ref{fig:hrds-001} for explanation of the plotted symbols.)
For the $0.9\ \Msun$ models, ages at ZAHB are $14.6, 10.8$ and $7.9$ Gyr for $Y=0.32, 0.36$ and $0.40$, respectively.
For the $1.0\ \Msun$ models, ages at ZAHB are $9.7, 7.2$ and $5.3$ Gyr for the three increasing $Y$'s.
See table~3 for more details related to the displayed sequences.
}
\label{fig:hrds-02He}
\end{figure}

Table~1 and the accompanying Figures~\ref{fig:hrds-001} and 3 
display results for sequences of increasing $\etareim$ for the following combinations:
$(Z,Y)$ of $(0.001, 0.24)$ and $(0.01, 0.28)$ and initial masses: 0.80, 0.90, 1.00, 1.10, 1.20~$\Msun$;
and $(Z,Y)$ of $(0.02, 0.28)$ and initial masses: 0.90, 1.00, 1.10, 1.20~$\Msun$ 
(the age at HB for a $(Z,M)=(0.02,0.80)$ model is over $20$~Gyr).
As a more detailed example, we show in Table~1 results for $(Z,Y,M)=(0.001, 0.24, 0.90)$ for $1.0\le\etareim\le2.2$, at intervals of 
$0.1$, where asterisks indicate models for which the evolutionary tracks are shown in Figure~\ref{fig:hrds-001}.

The monotonic trends are clearly apparent in the table, for the increasing MLRs within each combination, and among different parameter combinations.
With increasing MLR, core masses as well as envelope masses at the onset of the core He flash decrease.
Within each increasing $\etareim$ sequence, $\Teff$ increases while $L$ decreases in all cases.

On the evolutionary tracks of the models in the HRD, we note a few interesting features.
For example, during the post-HB evolution
we identify the two commonly recognized behaviours:   
P-EAGB (leaving the AGB before the TP phase) and `AGB-$\rm{manqu\grave{e}}$' 
(staying at high temperatures from TAHB throughout the remainder post-HB evolution).
In some cases of low $\etareim$ values, for which envelope masses during HB are still high enough,
multiple 'last' shell flashes are apparent during post-HB evolution.
Like the well-known shell flashes that occur on the AGB, they are caused by the double-shell burning instability. However, the HRD flashes are characterized by horizontal oscillations (large variations in $\Teff$, whereas the AGB flashes appear as nearly vertical oscillations (meaning almost constant $\Teff$). This is due to the significant difference in envelope mass: for the large convective envelopes of AGB stars, the $\Teff(L)$ variation is constrained by the Hayashi forbidden zone, while no such constraint is imposed on the small envelopes of stars undergoing last flashes. 
Eventually, when the envelope mass has been completely dissipated, whether after a single flash, multiple ones or none, the star settles on the WD cooling curve.

\begin{figure}
\centering
\scalebox{0.33}
{\includegraphics{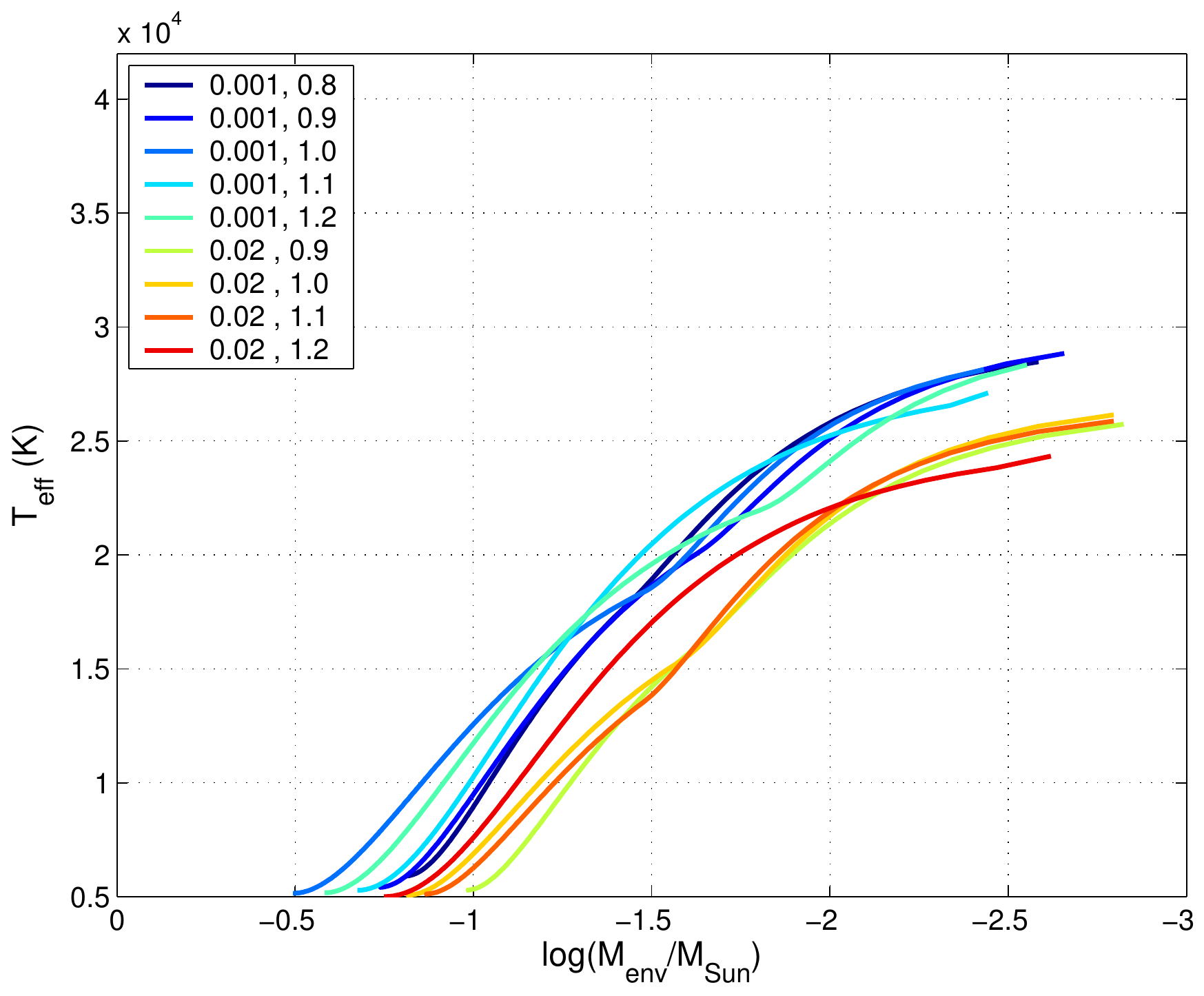}
 \includegraphics{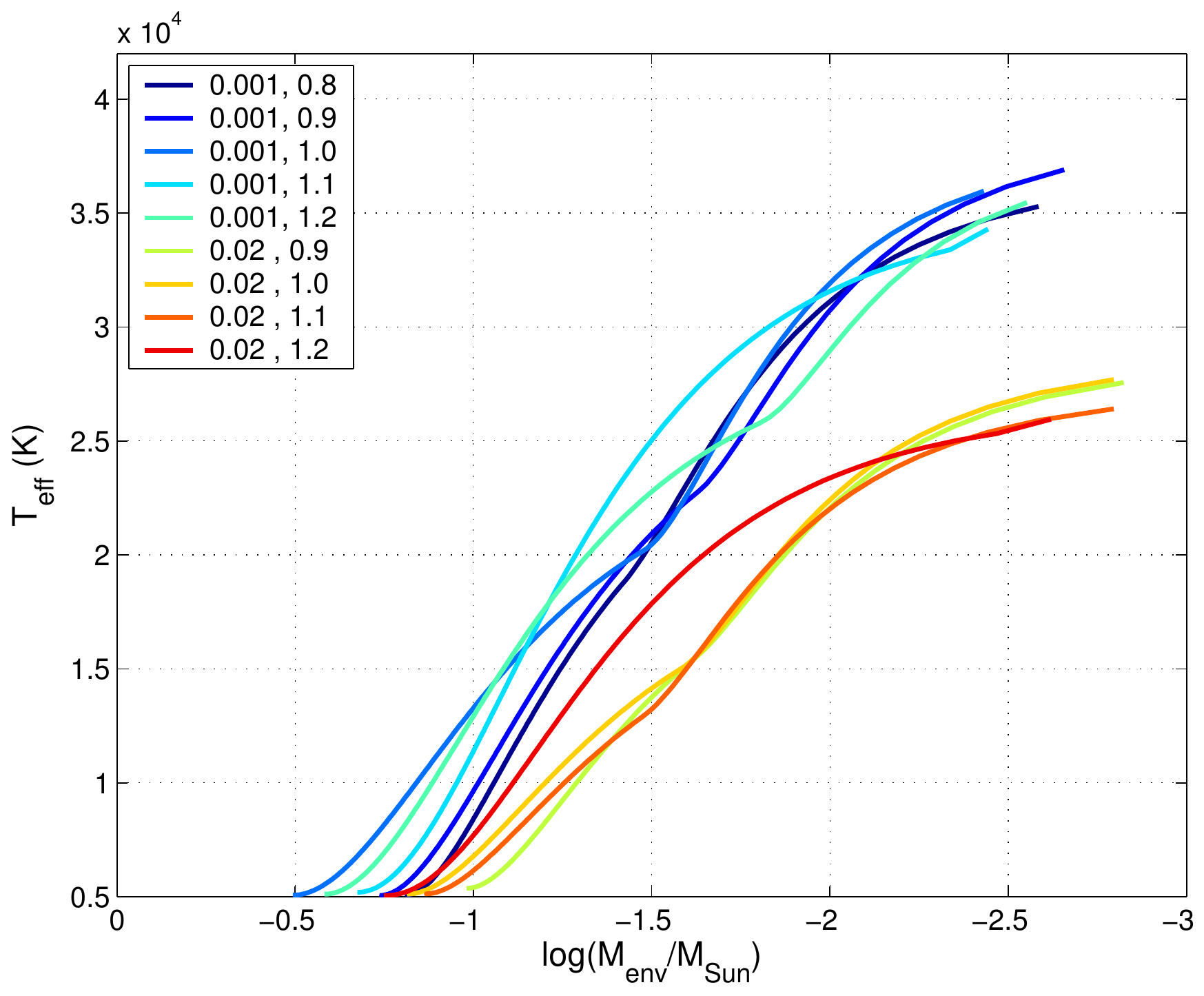}
 \includegraphics{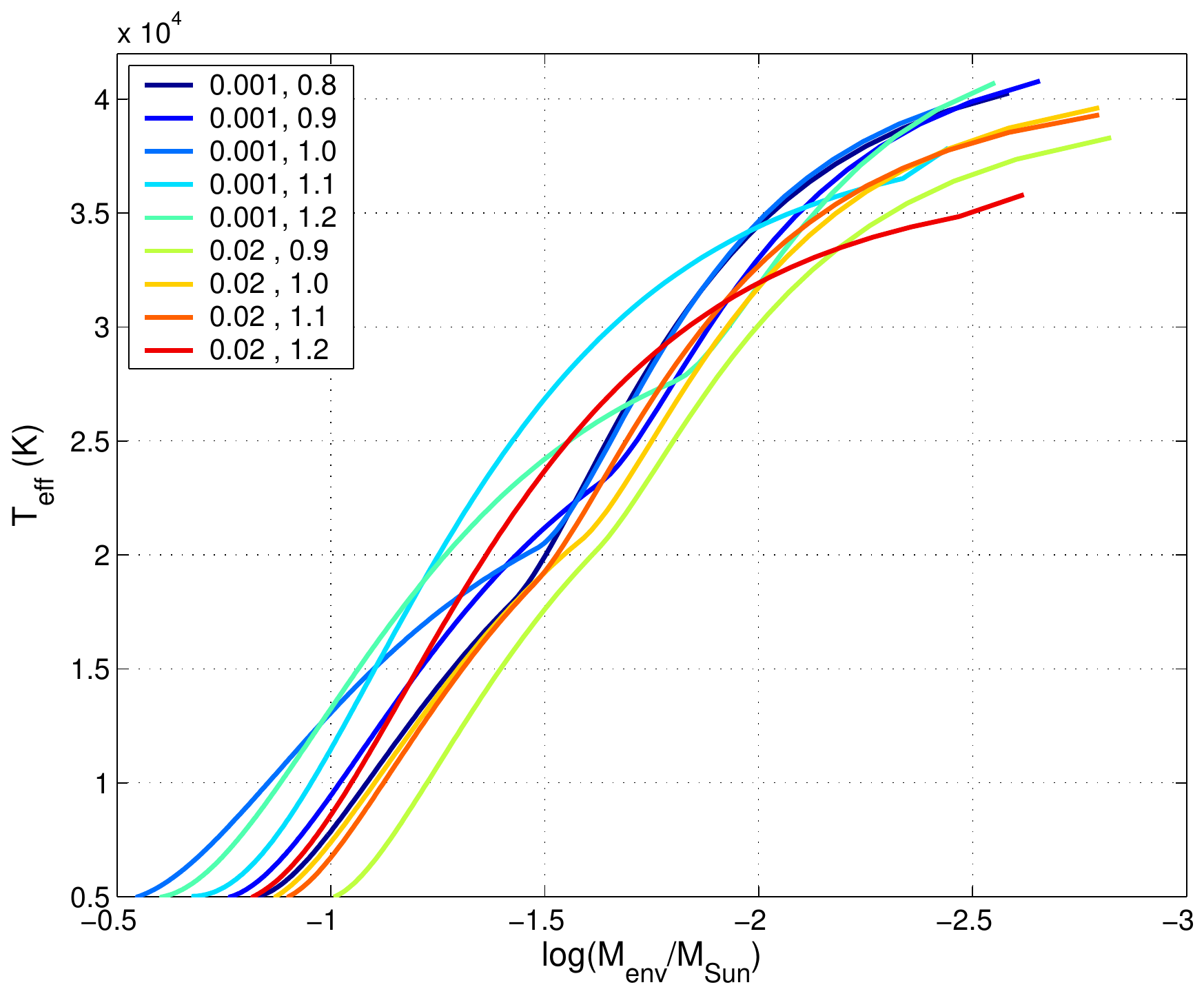}}
\caption{Effective temperature at ZAHB (left), `Mid'-HB (middle; see text) and TAHB (right)
versus the envelope mass -- decreasing left to right.
Legends display $(Z,M)$ values; $Y=0.24, 0.28$ for $Z=0.001, 0.02$, respectively.
Decreasing envelope masses correspond to increasing mass-loss on RGB.
Particularly evident at the mid-HB phase, the function $\Teff(\Menv)$ is steeper for Pop. II, for which higher effective temperatures are attained.
}
\label{fig:menv_teff}
\end{figure}

Figure~\ref{fig:menv_teff} shows the effective temperature as function of the envelope mass at the onset of the core He flash, $\Menv$, for $Z$ values of $0.001$ and $0.02$, at given phases of the HB: ZAHB, `Mid-HB' and TAHB. 
While there is an obvious monotonic anti-correlation between $\etareim$ and $\Menv$, the latter is more appropriate for defining the HB population.
We note the difference between Pop. I ($Z=0.02$) and Pop. II ($Z=0.001$) stars, which is particularly evident at the mid-HB phase, 
where the function $\Teff(\Menv)$ is steeper for Pop. II and higher effective temperatures are attained.
This difference become less conspicuous towards the TAHB, where the highest  $\Teff$ are attained: - above $35,000$~K for $\Menv \lesssim 3\times 10^{-3}\ \Msun$.

\subsection{Varying He-abundance for fixed metallicity} \label{vary}

Various GCs show evidence for a large variation in primordial helium abundance, possibly a result of self-enrichment in helium 
from the material ejected by massive AGB stars of the preceding generation/s (e.g. \cite{2005A&A...435..987C}).
We now examine the consequences of varying He abundances for \emph{fixed} combinations of ($Z,M,\etareim$).
For each $(Z,M)$ combination we choose an $\etareim$ value such that the core He flash takes place at the tip of the RGB for the `normal' $Y(Z)$ (see \S\ref{calc}). 

Table~2 shows the results for sequences of increasing initial He-abundance for the various ($Z,M,\etareim$) combinations,
with $Z$ assuming the values: $0.001, 0.018$ (solar model) and $0.02$.
The lowest $Y$ for each sequence is given by $0.20+0.04n$, where $0\le n \le 2$ is chosen so that ages at ZAHB are less than $20$ Gyr;
the highest $Y$ value terminating each sequence is 0.40.
Generally, we find that for each $(Z,M,\etareim)$ combination, $\Teff$ decreases and $L$ increases with increasing $Y$, at all HB phases. Since a higher He abundance ($Y$) also leads to a higher $\Menv$ 
(as the model is evolving more quickly on the RGB and thus experiences less mass-loss), 
this result is consistent with the results of the previous section. 

\begin{figure}
\begin{center}
\includegraphics[scale=.6]{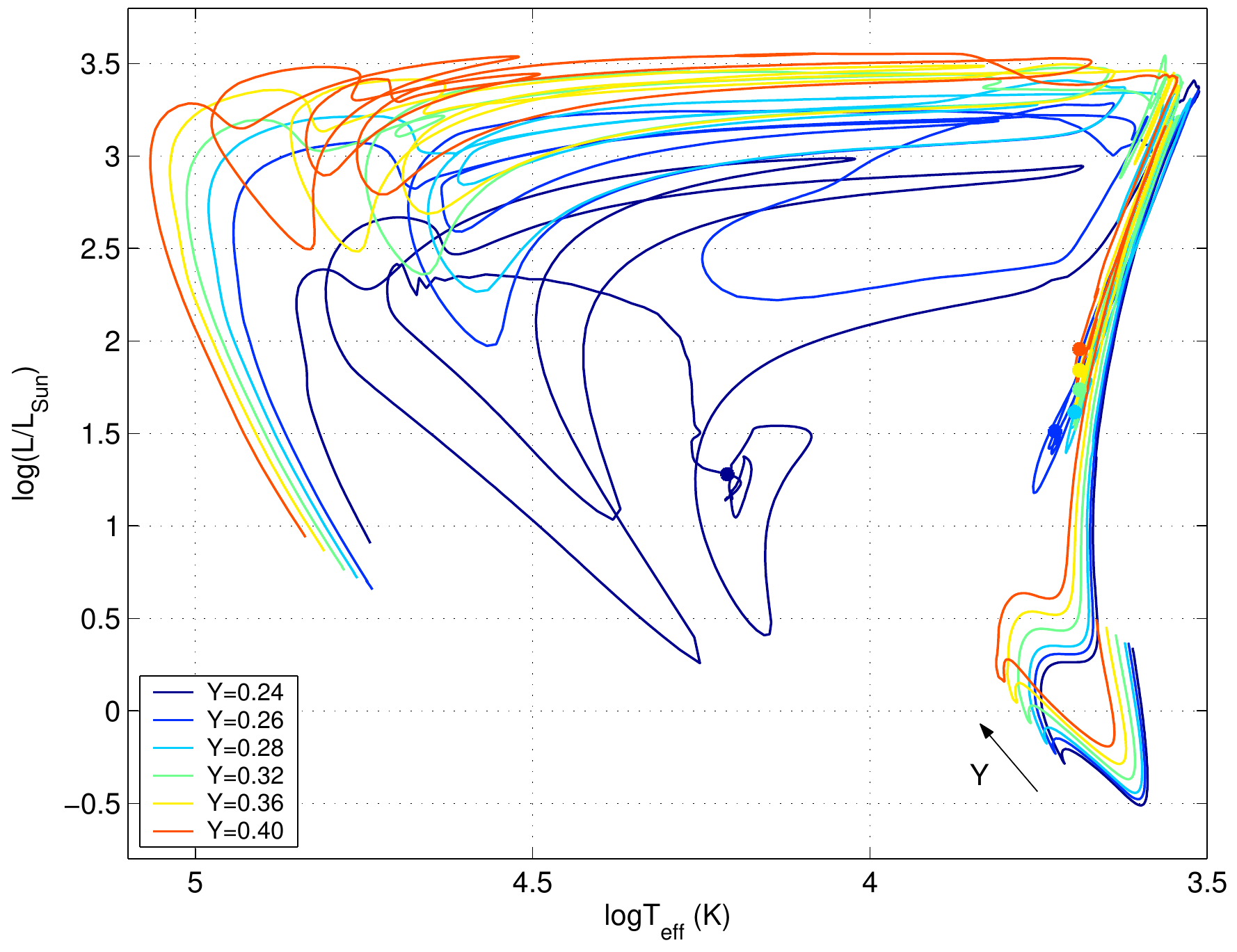}
\end{center}
\caption{Varying initial He abundance -- Complete tracks on HRD for \emph{fixed} $(Z,M,\etareim)=(0.018,1.0\ \Msun,0.6)$ 
with increasing initial $Y$ in the range $0.24$ to $0.40$.
$L$ and $\Teff$ of MS increase with increasing $Y$. Only for the smallest value of $Y$ do we obtain a significantly bluer HB, 
due to the expelling of the envelope by end of RGB and the development of a delayed core He flash; not a direct influence of the He abundance.
If not accounting for the decrease in the MS turnoff mass (per given age) for increasing He abundance,
an increase in the initial He abundance on its own does not lead to bluer HB positions.
}
\label{fig:vary_sol}
\end{figure}

Figure~\ref{fig:vary_sol} shows complete evolutionary tracks for the set of six $Y$ values with solar metallicity
(corresponding to the middle section of Table~2), where increasing $Y$ does lead to higher $\Teff$ and $L$ during \emph{MS} evolution.
As mentioned, observational evidence for a spread in MS structures and the existence of multiple stellar populations in GCs 
has been attributed in recent years mainly to processes of He self-enrichment (`{\it primordial contamination}') and
to the probable existence of multiple star formation epochs within a single GC 
(see e.g. \cite{2005ApJ...631..868D, 2005ApJ...621L..57L, 2007ApJ...661L..53P}, relating $\omega$ Cen and NGC 2808).

However, for the low values of $\etareim$ considered here, an increase in initial He abundance does not lead to bluer HB stars, in apparent contradiction to the presumed effect observed in GCs, of bluer HB positions for increasing initial He-abundance (e.g. \cite{2005ApJ...621L..57L}, \S~7.1 in \cite{2009Ap&SS.tmp...18C} and references therein). Only for the lowest value of $Y=0.24$ do we obtain a significantly bluer HB, due to the peeling-off of the H-rich envelope by the 
end of RGB and the development of a delayed core He flash (with $\Menv\approx 0.02\ \Msun$). Since there is no a priori reason for adopting a low value of $\etareim$, we shall now consider the entire possible range of values.

\subsection{Increasing MLR for He-enriched compositions} \label{yenrich}

\begin{figure}
\begin{center}
{\fbox{\includegraphics[scale=.45]{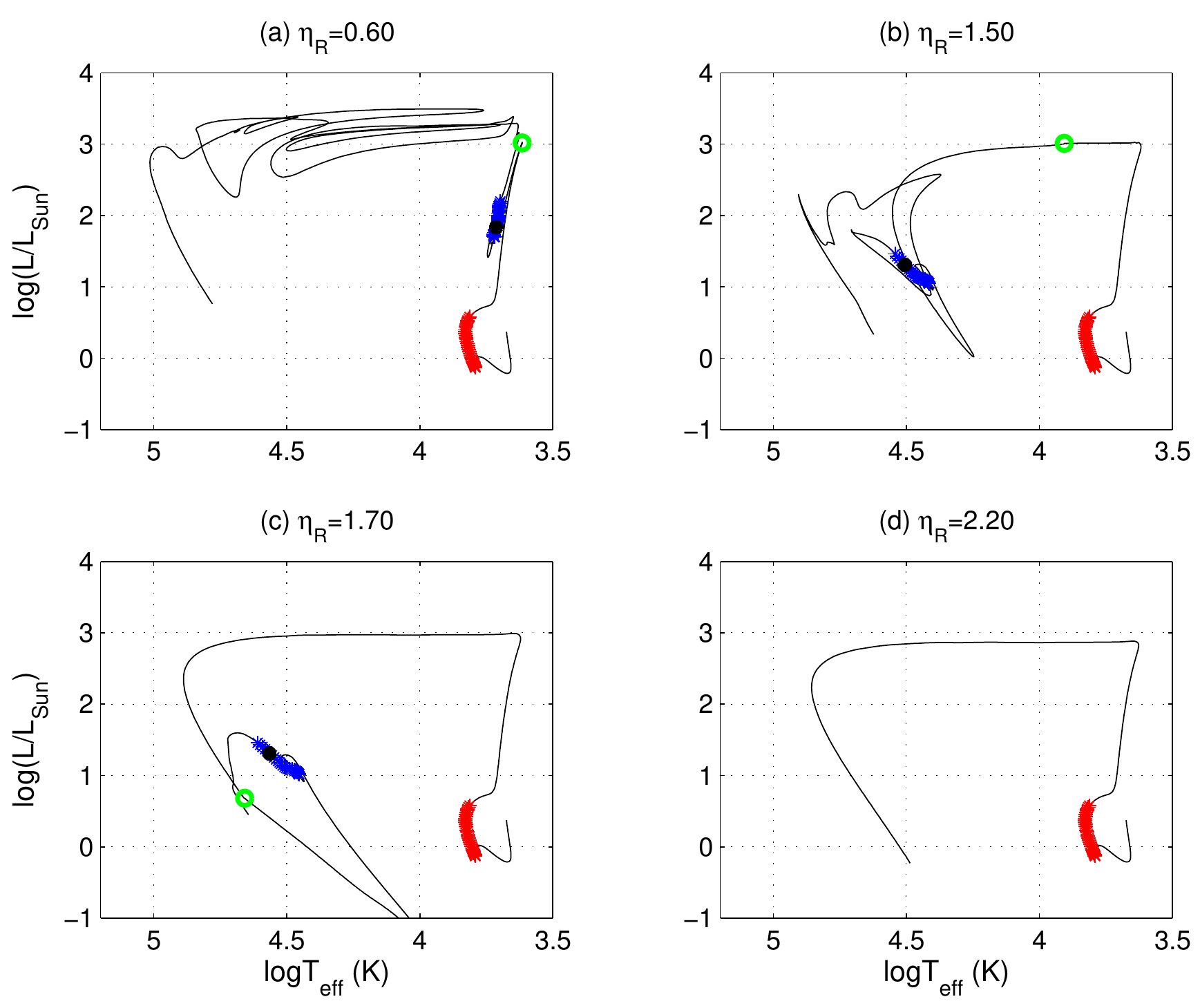}}
\fbox{ \includegraphics[scale=.45]{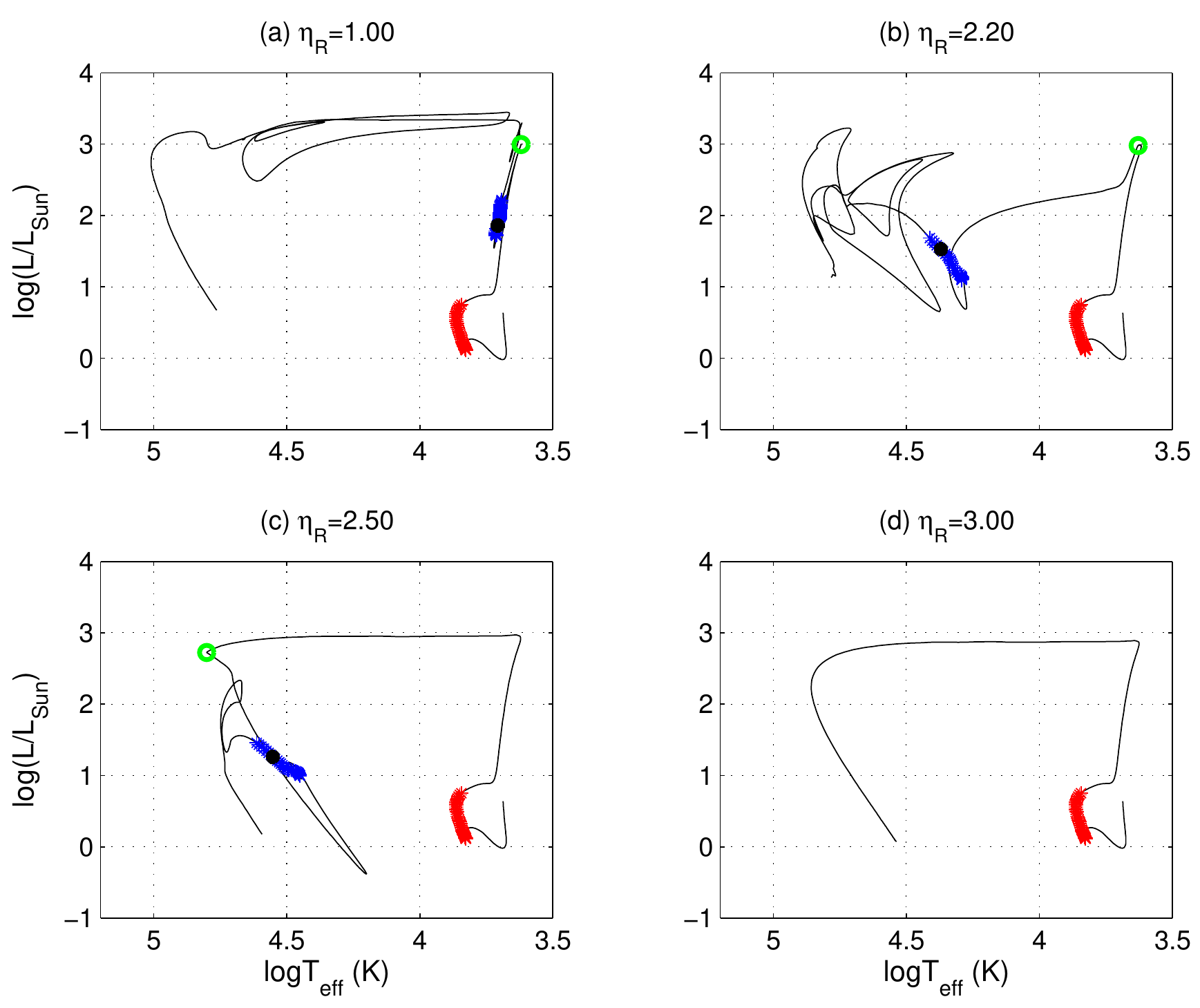}}}
{\fbox{\includegraphics[scale=.45]{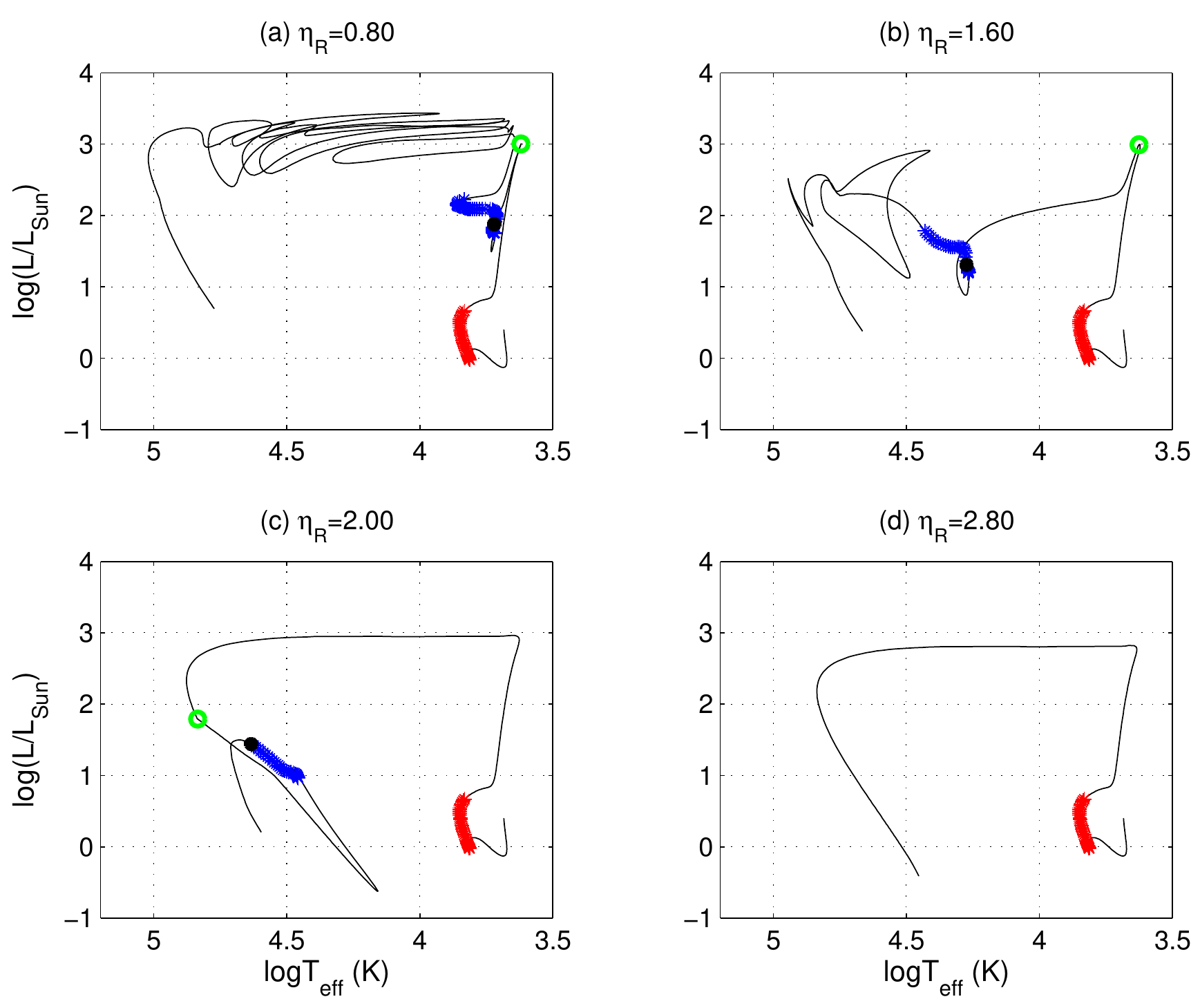}}
\fbox{ \includegraphics[scale=.45]{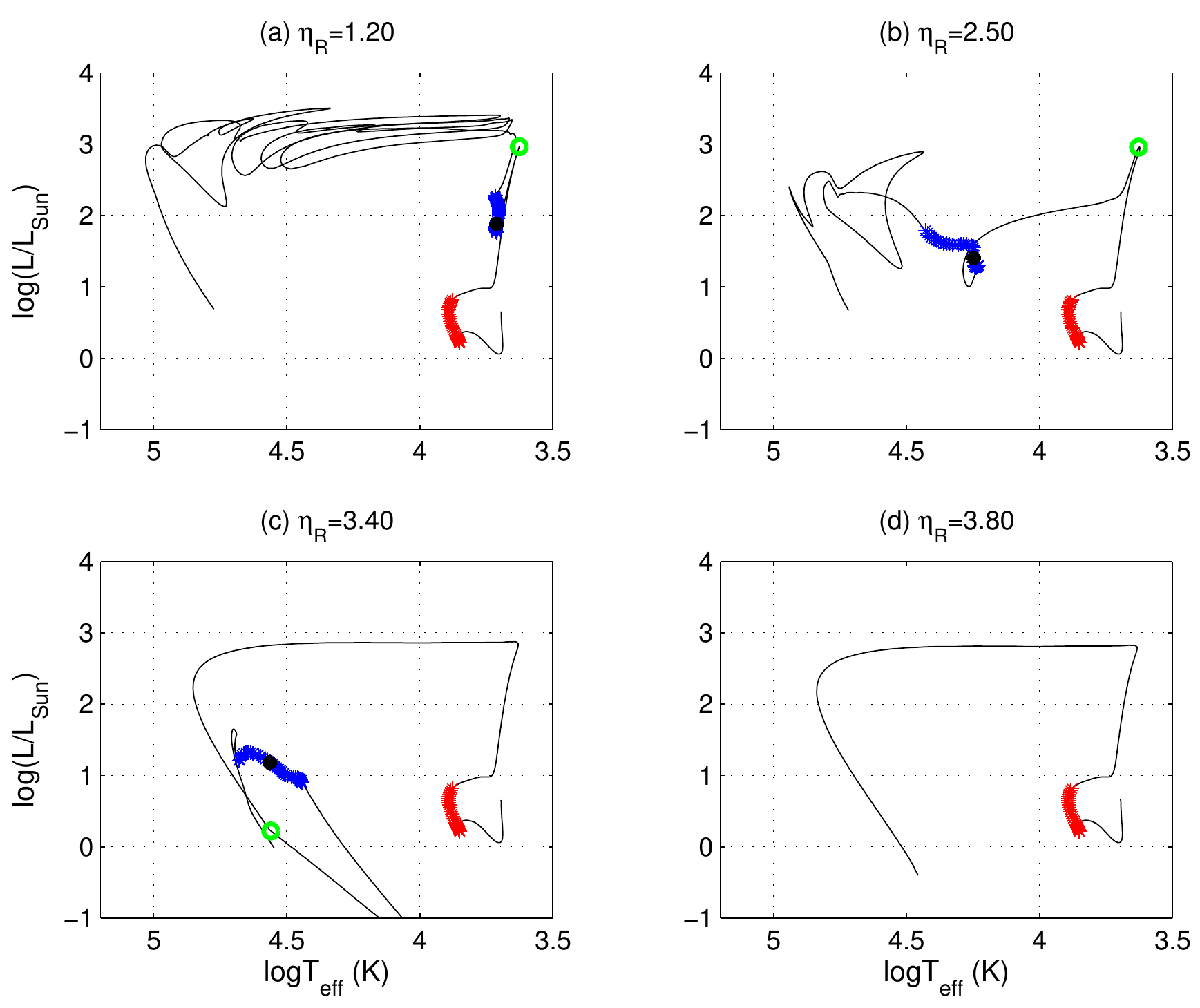}}}
{\fbox{\includegraphics[scale=.45]{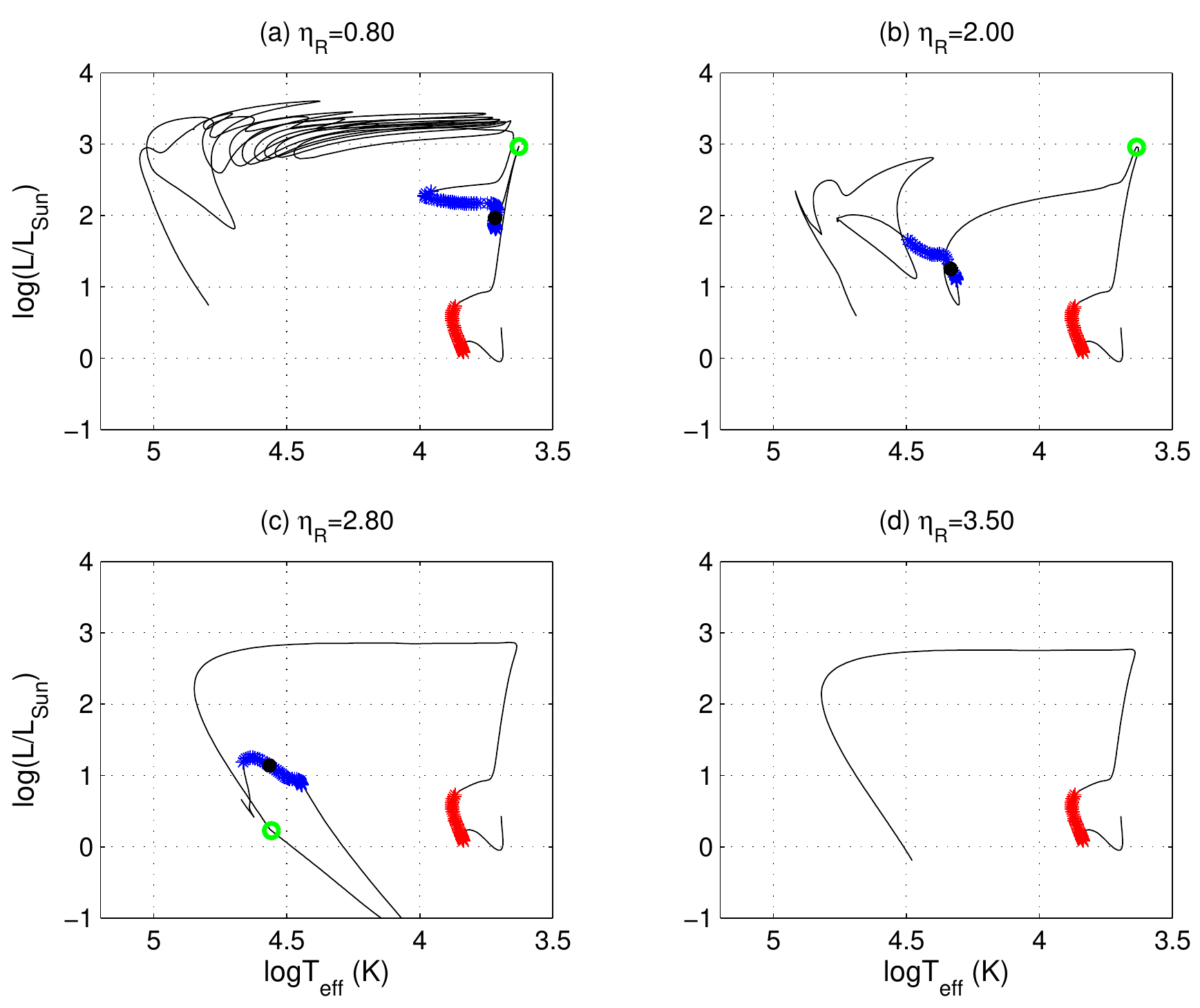}}
\fbox{ \includegraphics[scale=.45]{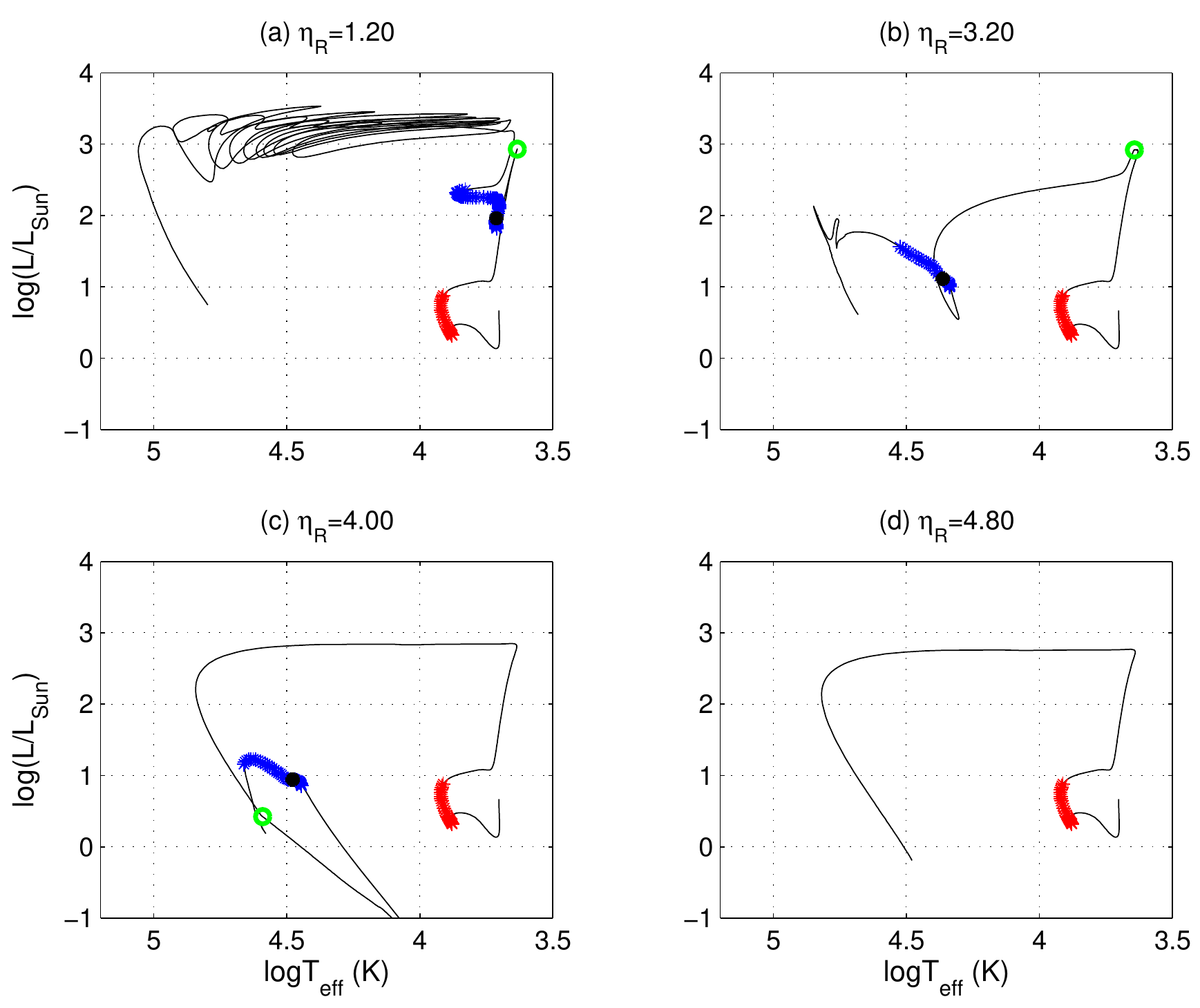}}}
\end{center}
\caption{Pop. II -- $Z=0.001$ -- He-enriched -- Complete evolutionary tracks on HRDs for increasing MLR ($\etareim$ values, shown in title of each HRD).
Left and right columns are for masses of $0.8$ and $0.9\ \Msun$, respectively;
within each column, the three panels relate to increasing initial He abundance---$Y$---of $0.32,0.36$ and $0.40$, top down.
(See figure~\ref{fig:hrds-001} for explanation of the plotted symbols.)
For the $0.8\ \Msun$ models, ages at ZAHB are $9.2, 7.0$ and $5.2$ Gyr for $Y=0.32, 0.36$ and $0.40$, respectively.
For the $0.9\ \Msun$ models, ages at ZAHB are $5.9, 4.5$ and $3.4$ Gyr for the three increasing $Y$'s.
See table~3 for more details related to the displayed sequences.
}
\label{fig:hrds-001He}
\end{figure}

Presented in Table~3 are the resulting properties for four $(Z,M)$ combinations:
$(Z,M)$=(0.001,0.80), (0.001,0.90), (0.02,0.90) and (0.02,1.00),
for which we test enhanced He-abundances of $0.32, 0.36$ and $0.40$.
For each of the twelve $(Z,Y,M)$ combinations we calculate evolutionary tracks for a sequence of increasing MLR.
As before, we show results for representative $\etareim$ values that exhibit all the different evolutionary patterns.
For $(Z,Y,M)=(0.001, 0.36, 0.90)$ we calculate an extended set of $\etareim$ values,
from $1.2$ to $3.8$ at intervals of $0.2$.
Displayed in Figures~\ref{fig:hrds-001He} and \ref{fig:hrds-02He} are the complete evolutionary tracks corresponding
to the parameter combinations presented in Table~3.

Our results show that the highest $\Teff$ obtained for each ($Z,M$) combination increases with increasing $Y$ and is obtained for the highest value of $\etareim$ adopted for that combination, in agreement with the trend exhibited by GC 
observations. We may thus conclude that a higher mass loss rate, as implied by a higher  $\etareim$  value is more consistent with reality. As an example we mention the hottest EHB stars observed in $\omega$ Cen (\cite{2011A&A...526A.136M}), where $\Teff\sim43,000$~K and $Y\sim0.38$. This is excellent agreement with our $(Z,Y,M)=(0.001, 0.36, 0.80)$ model, with $\etareim=2.0$, the highest for that sequence, which yields $\Teff=43,000$~K (at TAHB) for $Y=0.36$.

\begin{figure}   
\begin{center}
{\includegraphics[scale=.42]{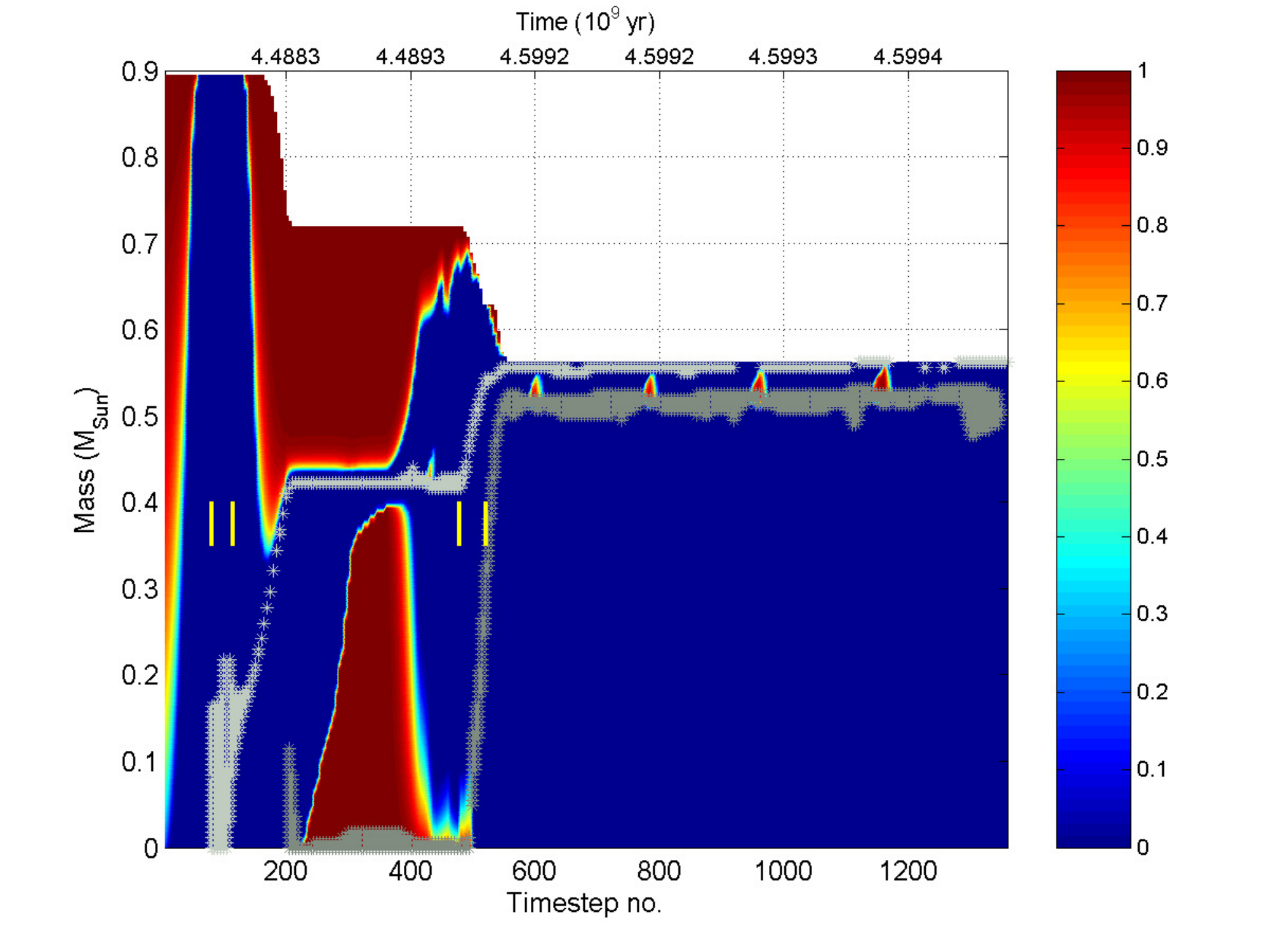}
 \includegraphics[scale=.42]{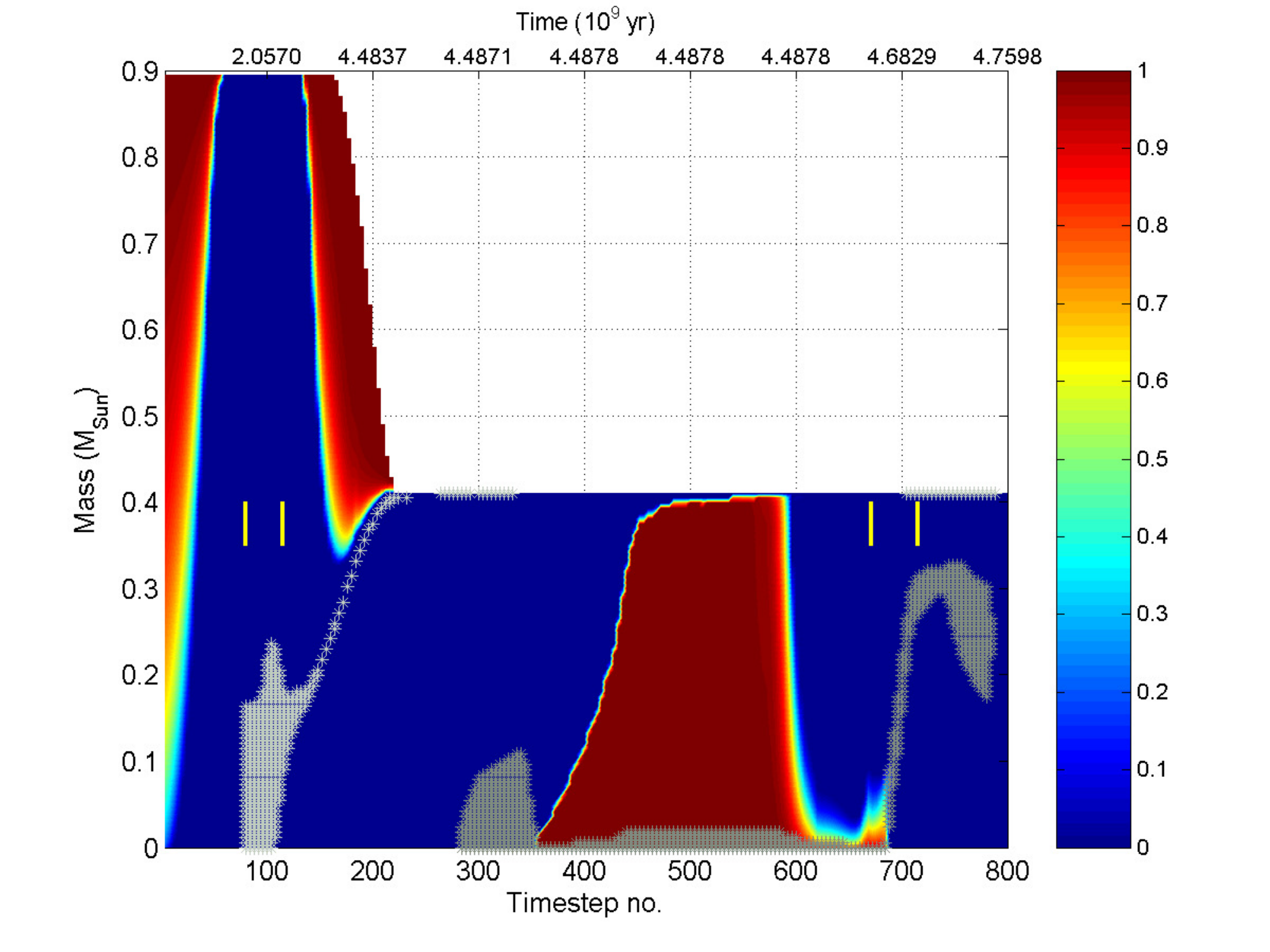}}
{\includegraphics[scale=.42]{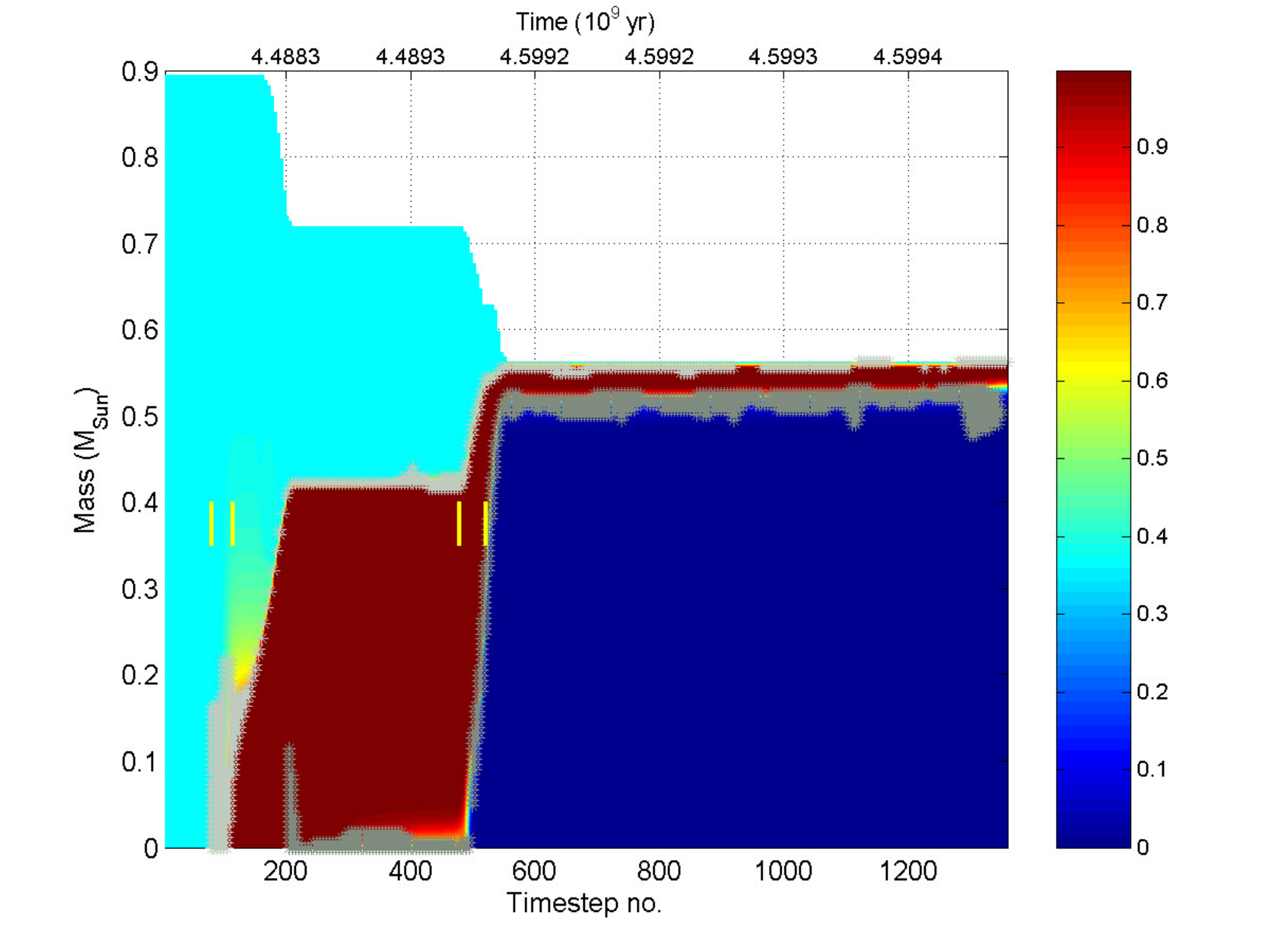}
 \includegraphics[scale=.42]{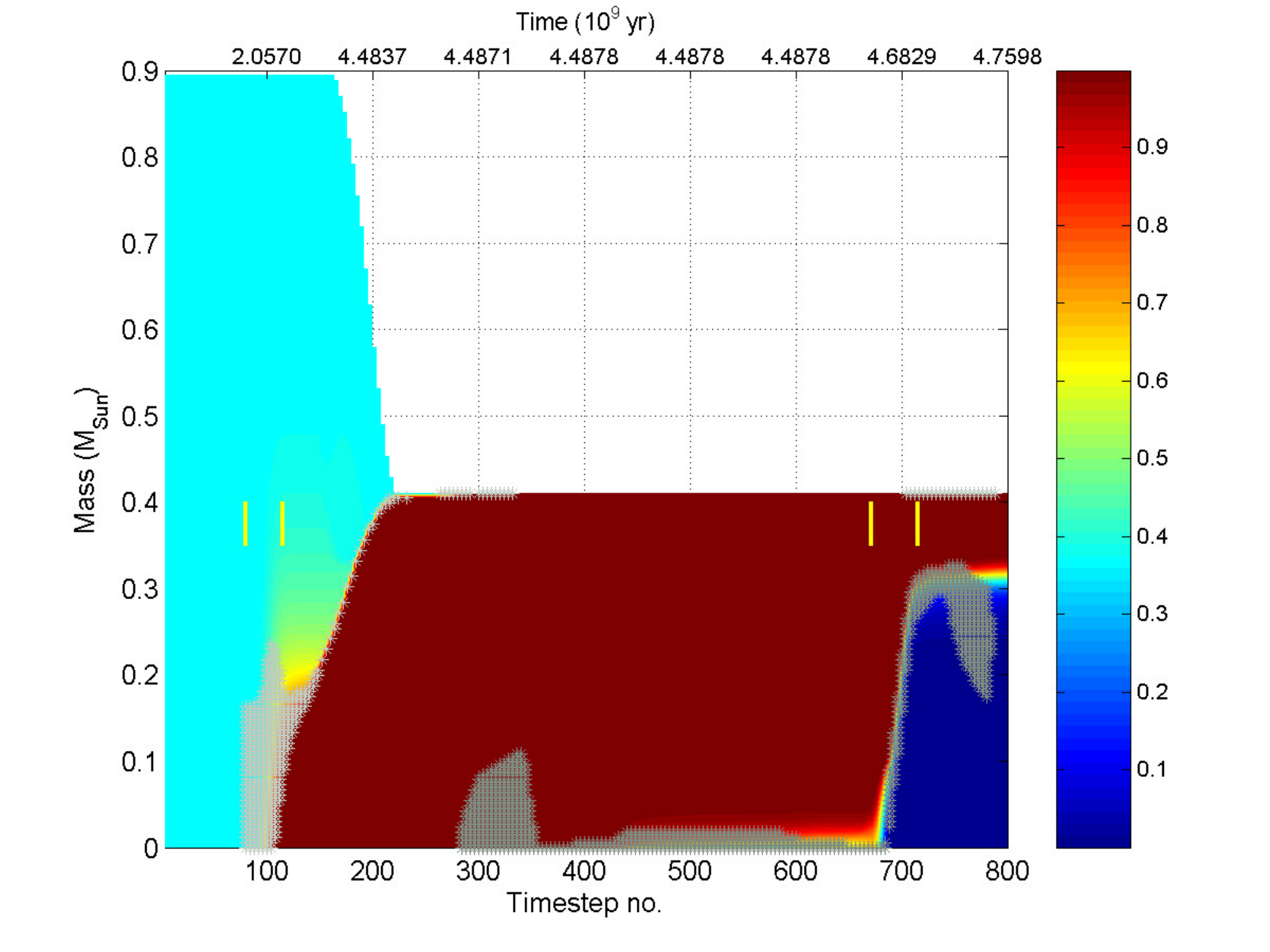}}
\end{center}
\caption{Kippenhahn diagrams (evolution in time of stellar configuration) displaying - 
top: convective regions (the ratio of convective to total flux: $F_{conv}/F_{tot}$),
bottom: the He mass fraction; for the He-enriched Pop. II model $(Z,Y,M)=(0.001,0.36,0.90)$ 
with $\etareim=1.20$ (left column) -- producing a `normal' Tip Flasher, 
and a high $\etareim$ value of $3.40$ (right column) -- producing a WD Flasher and consequently a very blue HB position.
The two couples of vertical yellow lines in each plot delimit MS and HB phases.
Light and dark gray asterisks, plotted on top of the colour maps, denote the hydrogen and the helium burning shells, respectively.
}
\label{fig:kipp_090_2}
\end{figure}

In Figure~\ref{fig:kipp_090_2} Kippenhahn diagrams show the evolution of 
convective regions (top panels) and Y (bottom panels) throughout the star---center to surface---for the
$(Z,Y,M)=(0.001,0.36,0.90)$ model, for $\etareim=1.20$ (left column) -- yielding a normal `Tip-Flasher' evolution, and
$\etareim=3.40$ (right column) -- leading to a `WD Flasher' (see corresponding tracks in Figure~\ref{fig:hrds-001He}).
We note that the abscissa is time step number, which is not equivalent to time, since time step lengths vary considerably. The
purpose is to amplify phases where changes take place (and time steps are correspondingly shorter). 
Also marked are the hydrogen and helium peak of the burning regions.

Comparing the results obtained for low and high MLR, we note that in both cases, an extended convective region develops with the violent ignition of He in the core and disappears by the time the star settles on HB.
For the low MLR, multiple last shell flashes are clearly apparent during post-HB phase close to the surface, where 
small convective regions develop during the unstable double-shell burning.
For the high MLR (`WD Flasher'), however, the envelope is too thin to sustain any double-shell burning and post-HB evolution is that of
`AGB-$\rm{manqu\grave{e}}$' -- continuing to WD cooling. In this case, a more massive He envelope remains on top of the CO core.

\subsection{HB Durations} \label{dur}

As stressed, e.g., by \cite{2007arXiv0708.2445C}, for a given chemical composition, the bolometric luminosity is roughly constant along the ZAHB. Therefore, they argue, the hotter a star becomes when it reaches the ZAHB, the higher its potential contribution to the population's total UV output.
In order to test the significance of hot HB stars and their progeny to the total UV flux emanating from a given stellar population,
it is also of importance to examine the durations of the HB phase for the various models.
\cite{1993ApJ...419..596D} have produced in their study a grid of calculations of stellar models---from metal-poor to solar metallicity and above---from ZAHB until some point in the late post-HB or WD cooling track.
Based on HB and post-HB lifetimes for EHB models, they conclude that
it is the EHB objects (if present in an old stellar population) and their progeny, that will provide largest contribution 
to the UV flux, due to their longevity at high temperatures, largely exceeding the contribution from normal post-AGB stars.
They derive HB lifetimes in the range $\sim 1.20-1.50\ \times10^8$~yr at luminosities of $\sim 10-100\ \Lsun$.

We find that within each $(Z,M)$ sequence, the HB lifetime increases with increasing MLR.
Minimal, maximal and average HB lifetimes (out of all combinations, in units of $10^8$~yr) are 
$\tau_{8,min}\simeq 0.9,\ \tau_{8,max}\simeq 1.7,\ \tau_{8,avg}\simeq 1.3$, respectively.
For the enriched He abundances, too, we find that
within each $(Z,Y,M)$ sequence, the HB lifetime increases with increasing MLR,
but lifetimes are longer for the higher $Y$.
Minimal, maximal and average HB lifetimes (in units of $10^8$~yr) are now 
$\tau_{8,min}\simeq 0.9,\ \tau_{8,max}\simeq 2.7,\ \tau_{8,avg}\simeq 1.6$, respectively.

To summarize, the average HB lifetimes of $1.3$ and $1.6 \times10^8$~yr (which fall below the corresponding MS lifetimes by 1 to 2 orders of magnitude)
are in good agreement with the range of HB durations estimated by \cite{1993ApJ...419..596D}.
We can further conclude that the bluer the position of the HB (higher MLR),
the longer the duration of the HB phase, to a significant extent.
This trend strengthens the significance of hot EHB stars,
members of old stellar populations, as major contributors to the integrated lifetime UV flux. Finally, the post-HB lifetimes obtained in our calculations --- from TAHB to the base of the WD cooling track ---
fall in the range $\lesssim10\,-\,\lesssim40$~Myr, 
again, following the trend that the bluer the HB position, the longer the post-AGB (or either P-EAGB or AGB-$\rm{manqu\grave{e}}$) duration.

\begin{figure}
\begin{center}
\includegraphics[scale=.5]{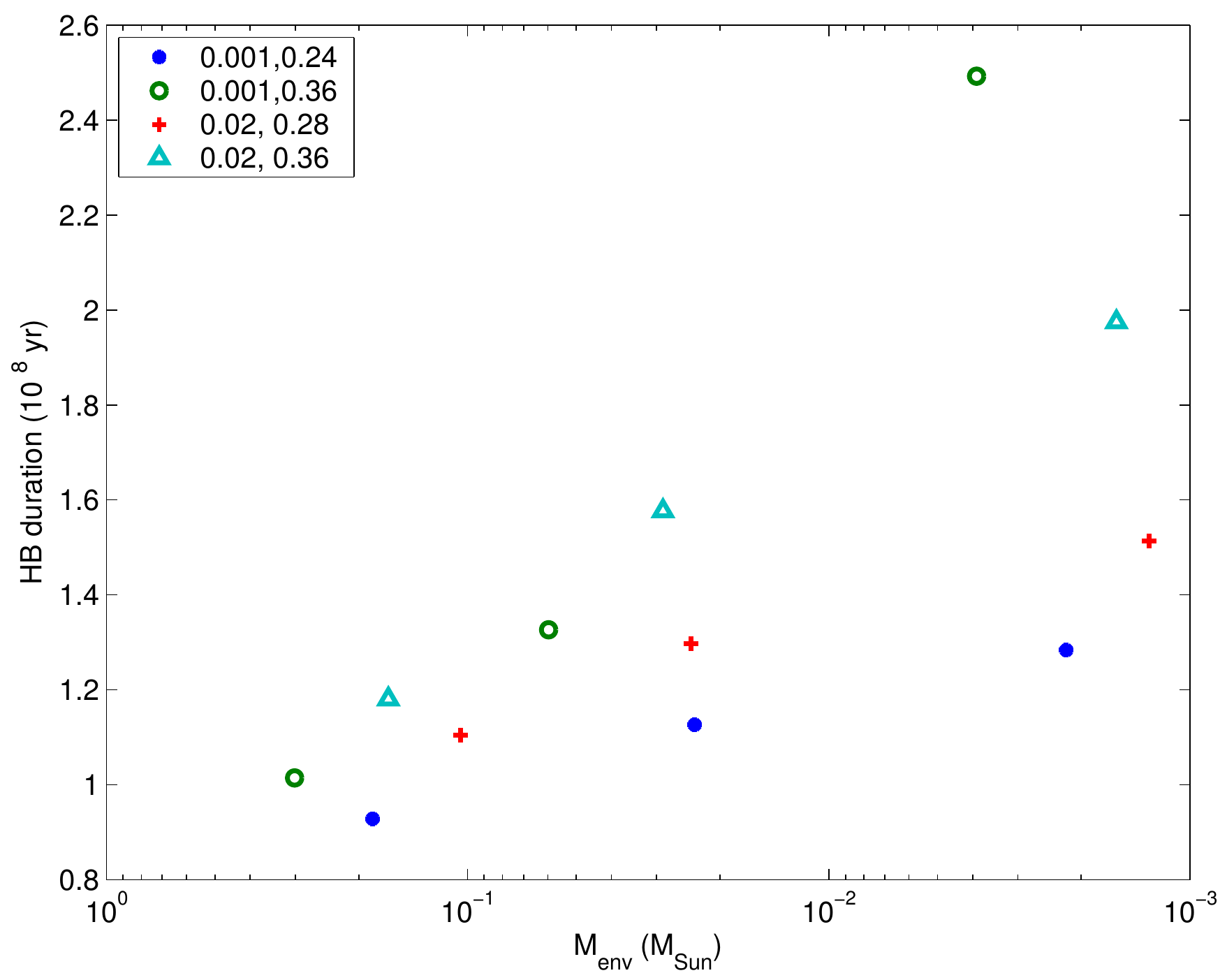}
\end{center}
\caption{HB durations as function of $\Menv$ (decreasing left to right) for $M=0.9\ \Msun$ and four $(Z,Y)$ combinations (legend).
HB lifetime increases significantly for decreasing envelope masses.
}
\label{fig:HBdur_Menv}
\end{figure}

In Figure~\ref{fig:HBdur_Menv} we plot HB duration versus envelope mass for four illustrative $(Z,Y)$ combinations ($M=0.90\ \Msun$).
We see that the scatter in HB durations is strongly affected by all parameters, including the metallicity and He-abundance, especially at low $\Menv$.
However, it seems that, in agreement with the conclusions of e.g. \cite{1994astro.ph..2046D},
it is the envelope mass (and hence the MLR) that affects the HB lifetime most significantly (especially for the He-enriched models),
and hence, has the greatest influence on the integrated lifetime UV output.


\section{Extended `WD Flasher' Results} \label{extend}

So far we have shown results for sequences of increasing MLR, for various $(Z,Y,M)$ combinations,
adopting the more common metallicity values representing populations I \& II: from $Z=0.001$ at the lower end,
to the slightly super-solar value of $Z=0.02$ at the high end.
A question remains -- what type of EHB stars would be produced for metallicities outside this range?
Especially, would it be possible to obtain even hotter HB stars for still lower metallicities?
Will the trend continue, so that with decreasing $Z$, bluer EHB stars be produced (for sequences of increasing MLRs)?
\cite{2009Ap&SS.tmp...18C} note that the most metal-poor GCs are \emph{not} necessarily the ones with the bluest observed HBs,
and even mention the possibility of a \emph{reversal} in trend at some critical metallicity around [Fe/H]$\approx -1.8$. 

This has led us to examine additional models for: {\it (1)} two additional low-$Z$ values: $Z=10^{-4}$, and 
$Z=1.4\times10^{-4}$ -- according to the estimated metallicity of the massive, EHB-rich cluster NGC 2419
(\cite{2008AJ....136.2259S} and references therein);
{\it (2)} an additional super-solar value of $Z=0.03$; and {\it (3)} an intermediate $Z$ of $0.005$.
Results for the extended set of metallicities 
are shown in Tables 4 \& 5 together with the results for $Z$ values  of $0.001,0.01$ and $0.02$ from Tables 1 \& 3, 
but only for the highest MLR values---those leading to a `WD Flasher'---to serve as a complete dataset of the bluest HB positions
obtained by our numerical calculations. 

Table~4 shows models of normal $Y(Z)$ ($Y=0.24$ for the Pop. II metallicities; $0.28$ for all Pop. I models; 
and $0.26$ for the intermediate $Z=0.005$), and for masses within the range $0.80$ and $1.20\ \Msun$ 
(omitting the higher masses for the low-Z models and adding a $M=1.30\ \Msun$ model for the highest Pop. I Z value). 
An additional model, for $M=0.85\ \Msun$, is calculated for $Z=1.4\times10^{-4}$, 
so as to collect more results around the MSTO mass of NGC 2419, which --- according to
our calculations --- is around $0.80-0.81\ \Msun$ for the estimated age of the cluster -- $\approx 13$~Gyr.

Table~5 lists the He-enriched models, adopting two representative initial masses for each metallicity (similarly to Table~3):
$0.80$ and $0.90\ \Msun$ for $0.0001\leq Z \leq 0.005$; 
$0.90$ and $1.00\ \Msun$ for the Pop. I $Z$'s of $0.01$ and $0.02$;
and masses of $1.00$ and $1.10\ \Msun$ for the highest $Z$ of $0.03$.
These choices are guided by the requirement of acceptable ZAHB ages.

\begin{figure}
\begin{center}
\includegraphics[scale=.6]{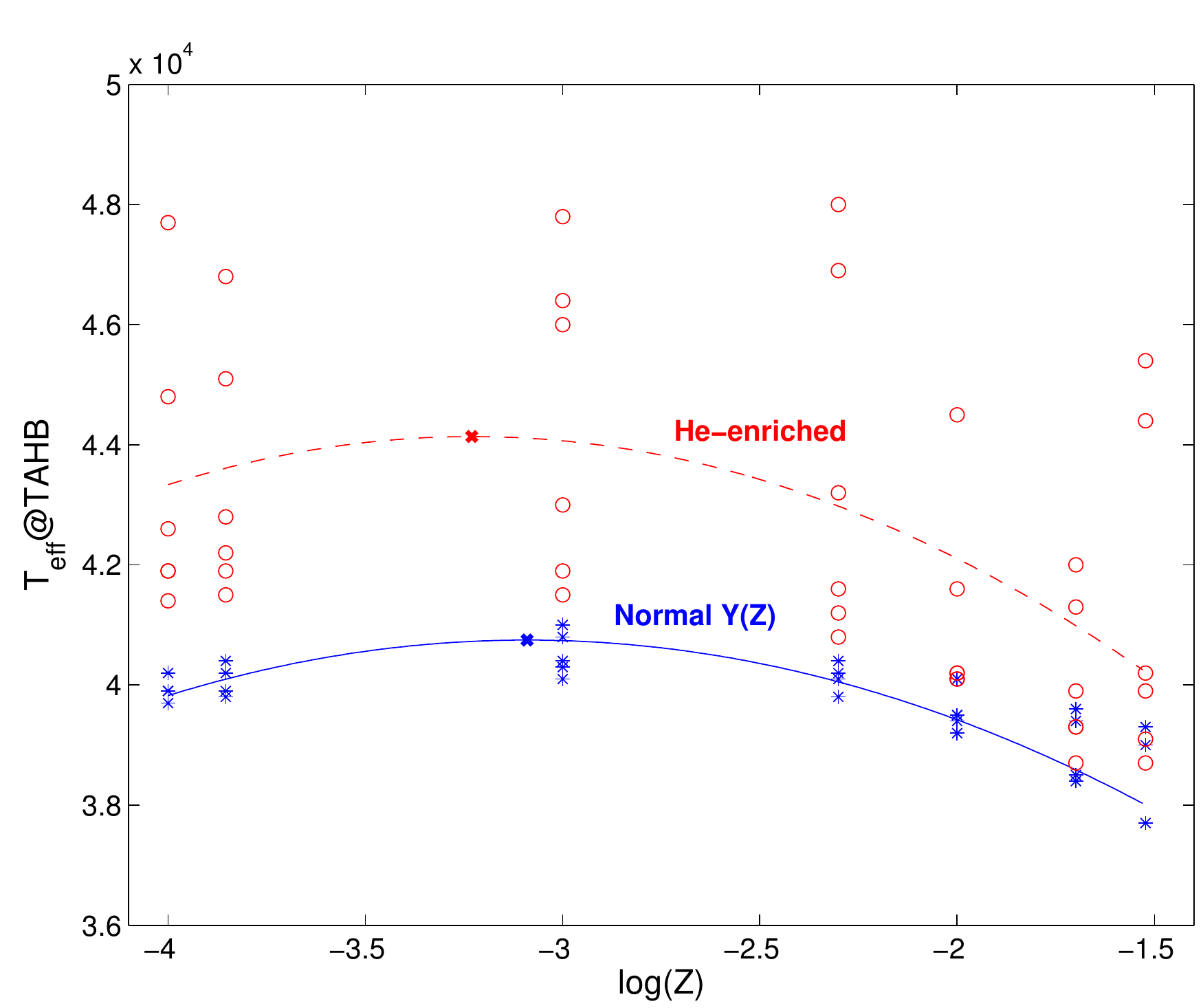}
\end{center}
\caption{$T_{eff}$ at TAHB for the WD Flasher calculations (bluest obtained HB positions) as function of metallicity (based on the data of tables 4 \& 5).
Blue asterisks: normal $Y(Z)$ (table~4); red open circles: He-enriched models (table~5).
Solid and dashed lines are quadratic fits to the normal-Y and He-enriched data points, respectively; 
displaying a peak at $Z\simeq8.1\times10^{-4}$ for the normal-Y,
and at $Z\simeq5.9\times10^{-4}$ for the He-enriched models. 
}
\label{fig:wdflash}
\end{figure}

We obtain high effective temperatures at TAHB in excess of $40,000$~K for the additional low-$Z$ models 
and even up to $\simeq48,000$~K for some He-enriched combinations, such as, $(Z,Y,M)=(0.0001,0.40,0.80),\ (0.00014,0.40,0.80)$. 
However, we note that the hottest HB effective temperatures for the lowest-$Z$ models are not surpassing those obtained for the $Z=0.001$ models.
Based on all the results presented in Tables 4 \& 5, we show in Figure~\ref{fig:wdflash} the bluest 
HB positions---$\Teff$ at TAHB as function of $Z$---over the complete parameter space, together with 
quadratic fits to the normal and enhanced He data points.
Two trends emerge: {\it (1)} the He-enriched models indeed yield significantly hotter HB effective temperatures,
with a large spread in the He-enriched $\Teff$ values; 
{\it (2)} there is a peak in $\Teff$, for both the normal and He-enhanced models, 
at a $Z$ just below $0.001$, indicating a halt or even a possible reversal in trend.
We note that although Figure~\ref{fig:wdflash} is based on effective temperatures corresponding to the TAHB, 
a similar trend emerges for the mid-HB temperatures, as listed in Tables 4 and 5.

Thus, it appears that the trend of HB stars becoming bluer with decreasing metallicity ceases (and even reverses) 
at some critical low $Z$, probably around a few $10^{-4}$. This phenomenon may have a simple explanation. 
The nuclear luminosity generated within the star is radiated at the surface, determining the surface boundary conditions $R$ and $T_{eff}$. 
Decreasing the metallicity $Z$ results in a decreased opacity and hence a more contracted radius and higher effective temperature. 
However, as the temperature increases and approaches 40,000~K, the opacity increases again, as a result of helium ionization. 
This will tend to increase the radius and lower the effective temperature (negative feedback).


\section{Comparison with Observations} \label{comp}

Figure~\ref{fig:hbpos_comb} shows HB positions in the HRD and in the $[\Teff,\log g]$ ($g=GM/R^2$) plane 
obtained for both the normal and He-enriched models.
There is no significant difference between positions of Pop. I and Pop. II models,
except that it is for the Pop. II (and the He-enriched) ones that the greatest extent blueward is achieved.
It is also apparent, especially at the red end of HB, that the He-enriched models are located at slightly higher luminosities
for both ZAHB and TAHB bands.
A conspicuous gap opens up in the distribution, which is reminiscent of gaps observed in CMDs of GCs.

The question whether there are gaps between the red HB (RHB) and the BHB within the EHB, or between EHB and `blue-hook' stars 
is still open: for example, a division into three groups of the blue HB of NGC 2808 was pointed out by \cite{2006A&A...446..569C}, 
while \cite{2007arXiv0711.2761C} cast doubts on the existence of real gaps, and suggested instead that there was evidence for `multi-modality' 
within HB structures (see also discussion in \cite{2001ApJ...562..368B}).
In order to investigate the occurrence of gaps and its source, we have run two sequences of 
models -- $(Z,M)=(0.001,0.90)$, $Y=0.24, 0.36$ -- covering the same range of $\etareim$, but with denser sampling: intervals of $0.1$ or $0.2$ throughout. 
The results presented in Figure~\ref{fig:hbpos_dense} show that the gap almost disappears.
Figure~\ref{fig:hbpos_dense} indicates that the morphology of the HB depends on the MLR (treated here as a free parameter) 
in a complicated manner; gaps may arise from constraints on the MLR.

Based on our resulting HB positions for the normal and He-enriched models, as displayed in Figure~\ref{fig:hbpos_comb} 
(and for the more common Pop. I and II metallicity values, presented in Tables 1 \& 3),
we show in Figure~\ref{fig:hbpos_obs} cubic fits to the derived ZAHB and TAHB positions in the $[\Teff,\log g]$ plane.
Superposed are points corresponding to HB stars derived from observations of four galactic 
GCs, all of which are known to have extensively populated blue HB tails: \cite{1995A&A...294...65M} for M15,
\cite{1997A&A...317L..83M, 1997A&A...319..109M} for NGC 6752, and \cite{2004A&A...415..313M} for NGC 2808 and $\omega$ Cen. 
Additional data points correspond to a sample of 35 Milky Way field Subdwarf B (sdB) stars (including also some SdOB), all with $\Teff$ in excess of $20,000$~K,
according to atmospheric parameters obtained by \cite{2005A&A...430..223L} (from the ESO Supernova Ia Progenitor Survey);
and also, a sample of 38 sdB stars from \cite{2001MNRAS.326.1391M}, based on sdB-classified stars from the Palomar-Green catalogue (of UV-excess stellar objects).

The HB region obtained from our results, extending between the ZAHB and TAHB bands, does cover most of the observed HB positions.
The majority of the outliers -- the stars lying above the TAHB curve -- are specifically mentioned as probable EHB progeny; e.g.
the three stars of NGC 6752 that lie above the TAHB at $\Teff>30,000$ are noted in the dataset as being \emph{Post}-EHB;
the field-sdB outlier at $(\Teff,\log g)=(3.28\times10^4,5.12)$, for instance, may well be considered a SdOB-class member: 
a hot subdwarf no longer burning helium in the core.

SdBs (and SdOBs) have long been identified with the bluest of the EHB stars and are regarded 
as the field analogues of the hot HB stars residing in clusters 
(see e.g. the review by \cite{2008arXiv0804.0507H}).
The hotter and more luminous sisters of SdBs are the (pre-WD/AGB-$\rm{manqu\grave{e}}$) SdO stars, consisting of He-dominated atmospheres; 
these may also be related to a more rare (`peculiar') class of blue HB stars, the `blue hook' stars (BHk), 
which are detected at the very hot end of the HB in CMDs of mainly far-UV
(see \cite{2007A&A...474..105B} regarding NGC 6441 and NGC 6388, and also
\cite{2009arXiv0901.1309D} who discuss possible mechanisms for the production of BHk stars in GCs, 
by examining whether or not clear correlations exist between BHk populations and properties of their host clusters).

When we compare the results of evolutionary calculations with observations, with the question in mind of how blue may EHB stars get,
the occurrence of a high $\Teff$ by itself is not sufficient. The high $\Teff$ must be maintained for a sufficient length of time for stars
to be detected in this phase. Thus the relevant question is somewhat different: How blue can observable EHBs get?
Since $\Teff$ does not remain constant during EHB evolution, but rather changes continually, although at 
very different rates at various stages, the answer to this question is not straightforward.  

In Table~6 we present effective temperatures at TAHB alongside the maximal post-HB $\Teff$ for sequences of increasing MLRs
for four representative $(M,Z,Y)$ combinations (shown in Tables~1 \& 3).
Passing from Tip, through Post-tip, till WD Flasher (for each $(Z,Y,M)$ combination), 
we note that although the maximal post-HB $\Teff$ decreases, the duration around the maximal temperature 
(defined as the time interval for being within $1\%$ below the peak value) significantly increases.
Thus, although for the WD Flashers the maximal post-HB temperatures are lower (than for the Tip or Post-tip flashers),
the time that is spent around the maximal $\Teff$ is longer, with the consequent influence of this 
on the observed EHB frequency and contribution to the integrated UV flux.
With respect to the latter, the bluest obtained HB/post-HB positions can be considered as 
somewhere in between the TAHB $\Teff$ and maximal post-HB $\Teff$ of the WD Flasher models.

\begin{figure}
\begin{center}
 \includegraphics[scale=.6]{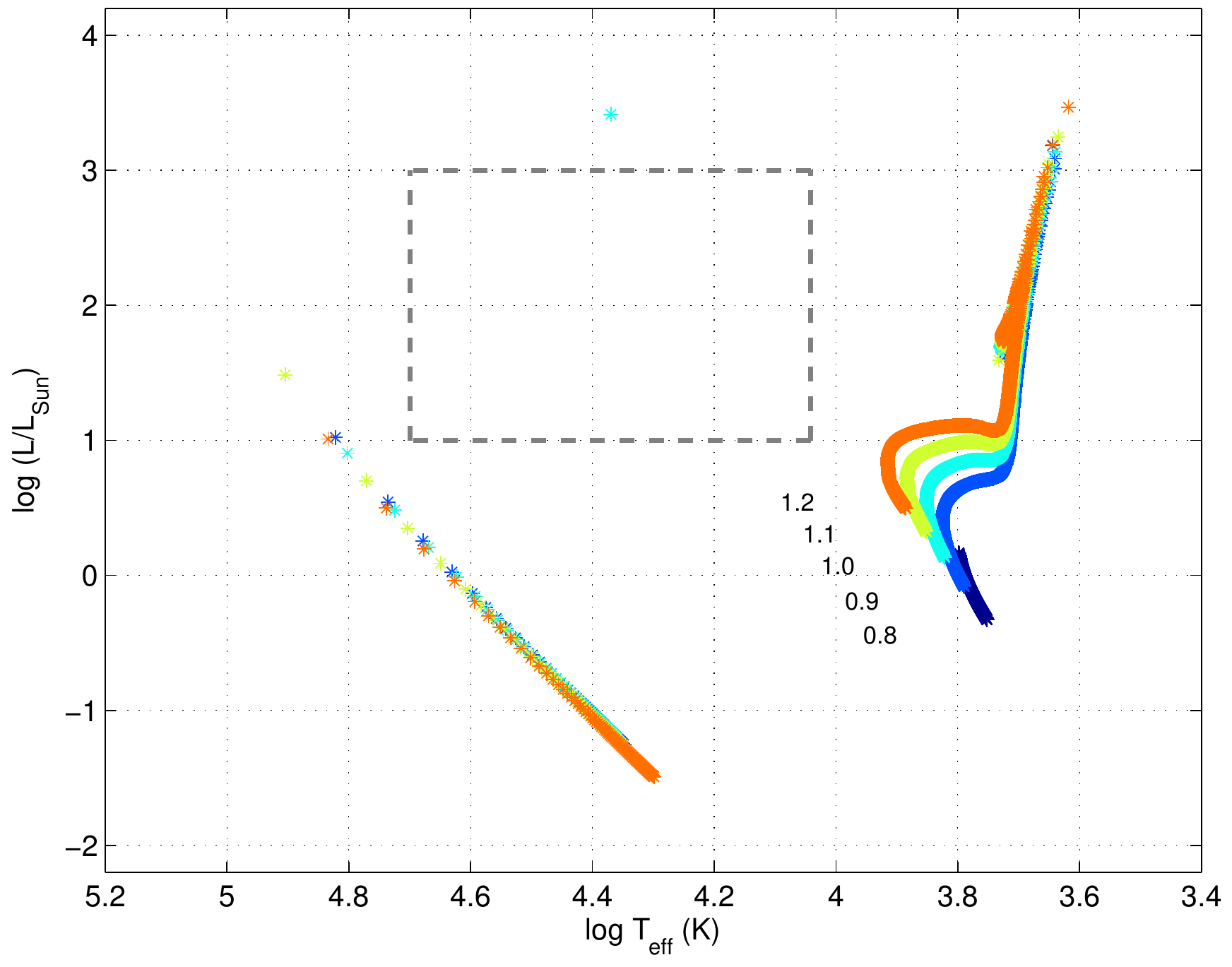}
 \includegraphics[scale=.5]{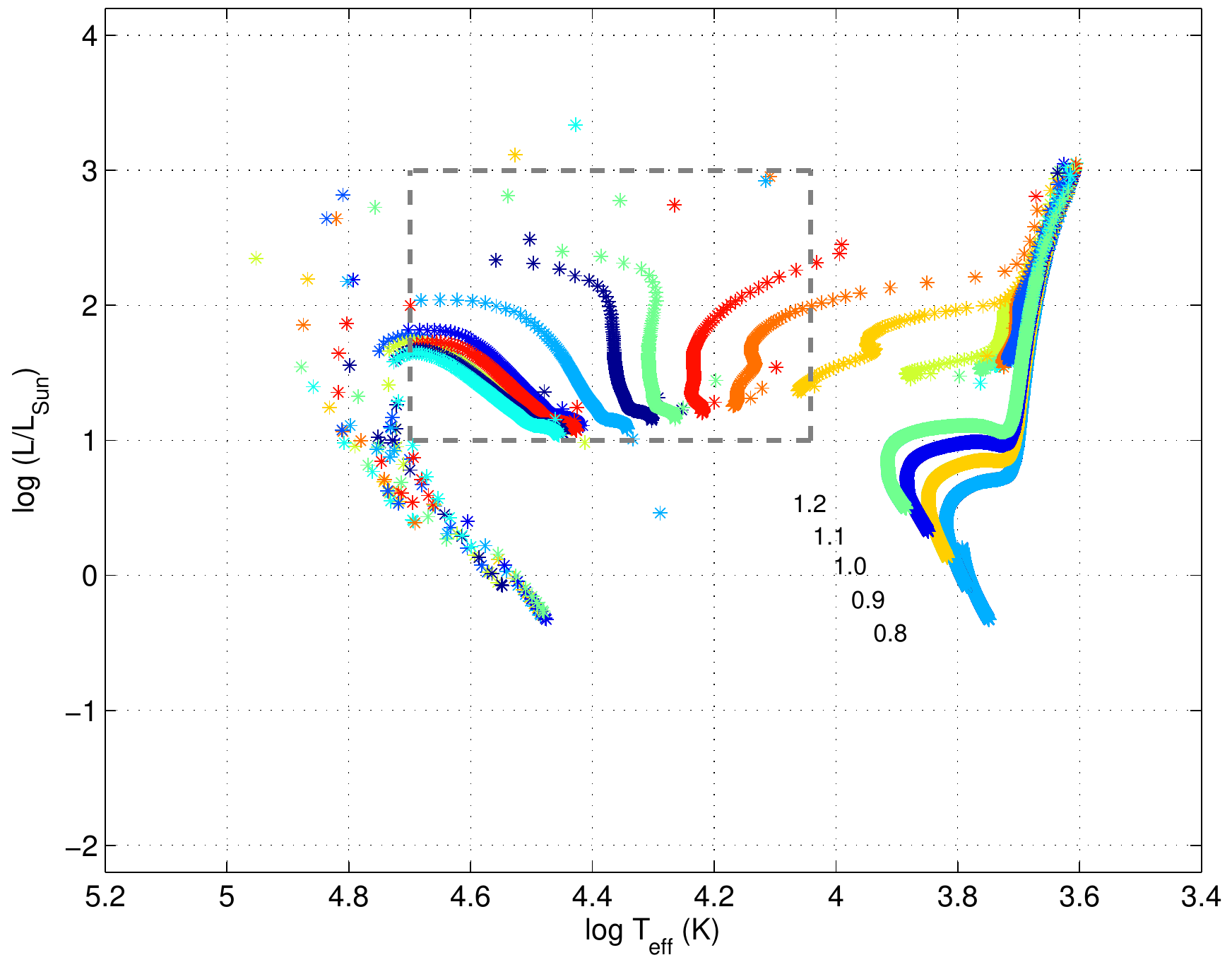}
 \includegraphics[scale=.5]{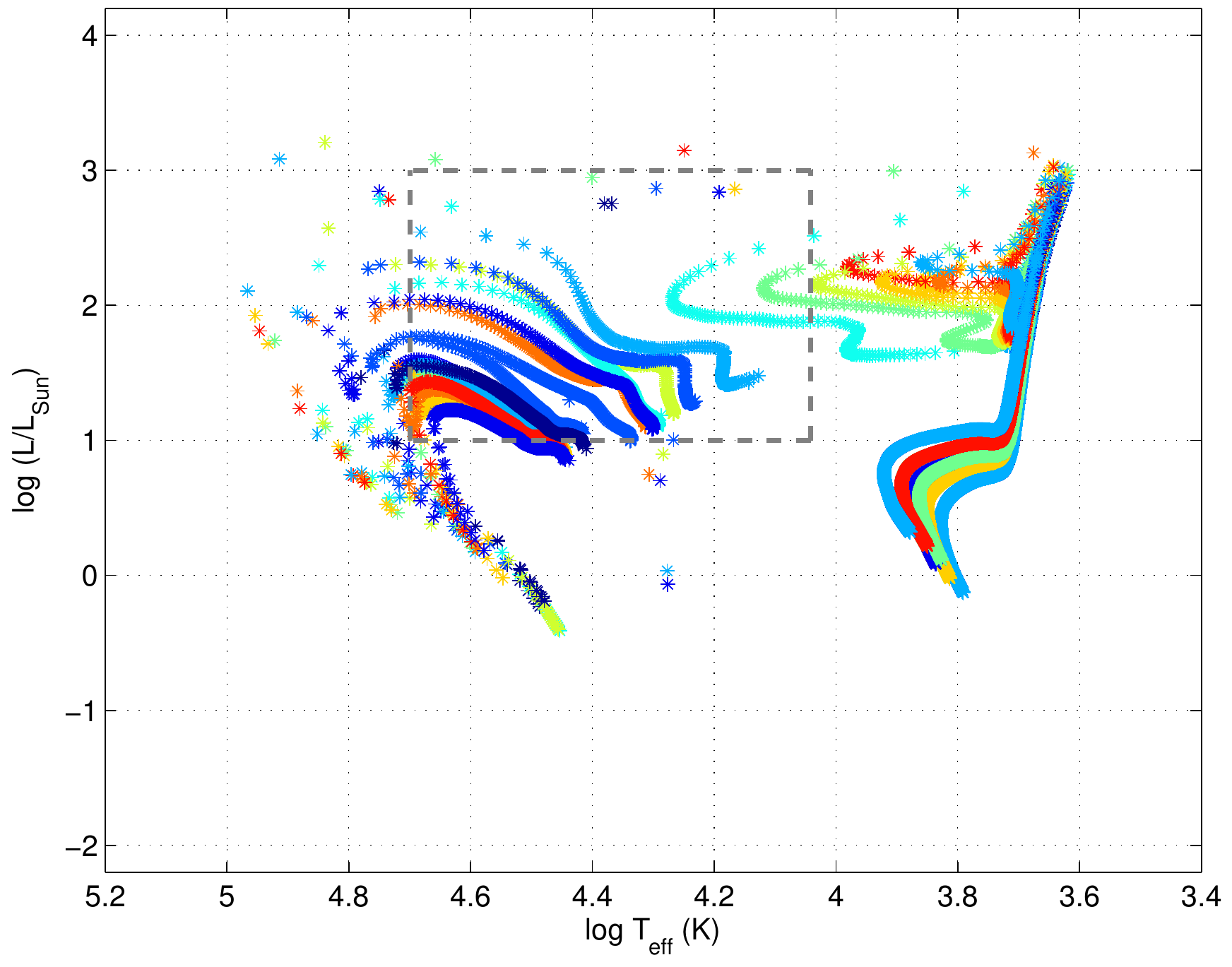}
\end{center}
\caption{Evolutionary tracks on HRD for common Pop. II ($Z=0.001$) stars 
at constant time intervals of $\Delta t=10^6$~yrs, over a $13$~Gyr timespan.
Top: "Canonical" evolution of stars in the range $M=0.80-1.20\ \Msun$ (to serve as reference);
bottom left: Normal initial He abundance ($Y=0.24$) in the same mass range,
but with enhanced mass-loss rates on the RGB such that the EHB is populated;
bottom right: He-enriched EHB stars of masses $0.80,0.90\ \Msun$, with initial He content of $0.32,0.36,0.40$.
The dashed gray rectangle in all panels highlight the area that is populated by BHB/EHB stars, obtained only
when enhanced mass-loss is applied.
(In the top panel, the WD cooling curve is more extended simply because we carried the calculations for longer times.) 
}
\label{fig:iso_ehb}
\end{figure}

In order to further answer the question of the frequency of EHB stars 
and their potential contribution to the UV output, we show in Figure~\ref{fig:iso_ehb} data points along full evolutionary ($13$~Gyr) 
tracks in the HRD at fixed time intervals. Thus long phases of evolution will result in a large number of points, 
while fleeting episodes will barely leave their marks.  
The top panel displays --- for reference --- `canonical' tracks for stars in the range $M=0.80-1.20\ \Msun$, 
corresponding to `normal' (non-enhanced) mass-loss rates.
The middle panel shows tracks corresponding to normal He abundance, covering the same mass range $M=0.80-1.20\ \Msun$. 
Here, we note that each mass or $(M,Y)$ combination consists of a sequence of models with increasing MLR ($\etareim$ values). 
The tracks within each sequence overlap for the MS and RGB phases,
but open up for the HB phase; the smaller the envelope mass at ZAHB (i.e. the higher the mass that was lost on giant branch),
the bluer the position the star is "thrown" to.
The lowest panel shows tracks of He-enriched models
for masses $0.80$ and $0.90\ \Msun$, with $Y=0.32,0.36,0.40$ for each mass, and again, a sequence of increasing MLR for each $(M,Y)$ pair.
The overcrowding of points at high $\Teff$ is conspicuous, meaning that stars spend relatively long periods of time in this region of the HRD, and thus
at any given time this region of the diagram should be more densely populated. This is particularly noteworthy in view of the complete lack of stars
in this region for the reference case. Clearly, $\etareim$ must exceed some limiting value for a HB to be produced at all, and certainly for an EHB
to appear; for the latter, He-enhancement is probably required as well.   

\section{Discussion and Conclusions} \label{conc}

The aim of the current study is to re-examine the production and subsequent evolution of HB and EHB stars,
showing the  diversity that can be obtained for given inputs.
A wide range of parameter combinations is examined in a self-consistent manner, that is, 
based on the same input physics, using a code that robustly follows full evolutionary sequences.

The HB appears mainly as a sequence in $\Teff$, but as emphasized by e.g. \cite{2001ApJ...562..368B}, it is also, and primarily, 
a sequence in $\Menv$, so that large envelope masses occupy the RHB while small ones occupy the EHB;
this, in turn, is a result of mass-loss during the preceding RGB phase. Thus, the mass loss rate is a crucial parameter.
The diversity of morphological differences among
HBs of GCs can be accounted for by changes in metallicity, another parameter, but not entirely. Thus, two leading factors 
(`second parameter' candidates) have been suggested for producing the observed features, such as the varying blue tails and EHB stars:
the He abundance and (again) the mass loss rate on the RGB.
Whereas in some clusters there is strong evidence for the existence of multiple populations among the MS stars or red giants, 
thus providing the grounds for considering the He self-enrichment scenario, 
others, such as NGC 2419, do not show such a feature, and the question remains whether the dominant mechanism for producing hot HB stars
is that involving higher mass-loss rates on first giant branch, and if so, 
how does this relate to the dynamical characteristics and history of the host cluster.

Therefore, the parameters considered here besides the mass, are $Z$, $Y$ and $\etareim$ and the parameter space spans 
combinations of $(Z, Y, M,\etareim)$ such that:
\begin{enumerate}
\item High and low metallicity populations were simulated - $0.001\le Z\le0.02$, extending the range to $0.0001\le Z\le0.03$ in some cases.
\item The initial helium abundance is either `normal' for the corresponding $Z$, or enhanced up to $Y=0.40$.
\item Initial masses are around and slightly above typical GCs MSTO masses, $0.80$ to $1.20\ \Msun$.
\item Values of $\etareim$ on the RGB are such that, depending on the $(Z,Y,M)$ combination,
a complete spectrum of evolutionary patterns is obtained. 
\end{enumerate}

\begin{figure}
\begin{center}
{\includegraphics[scale=.45]{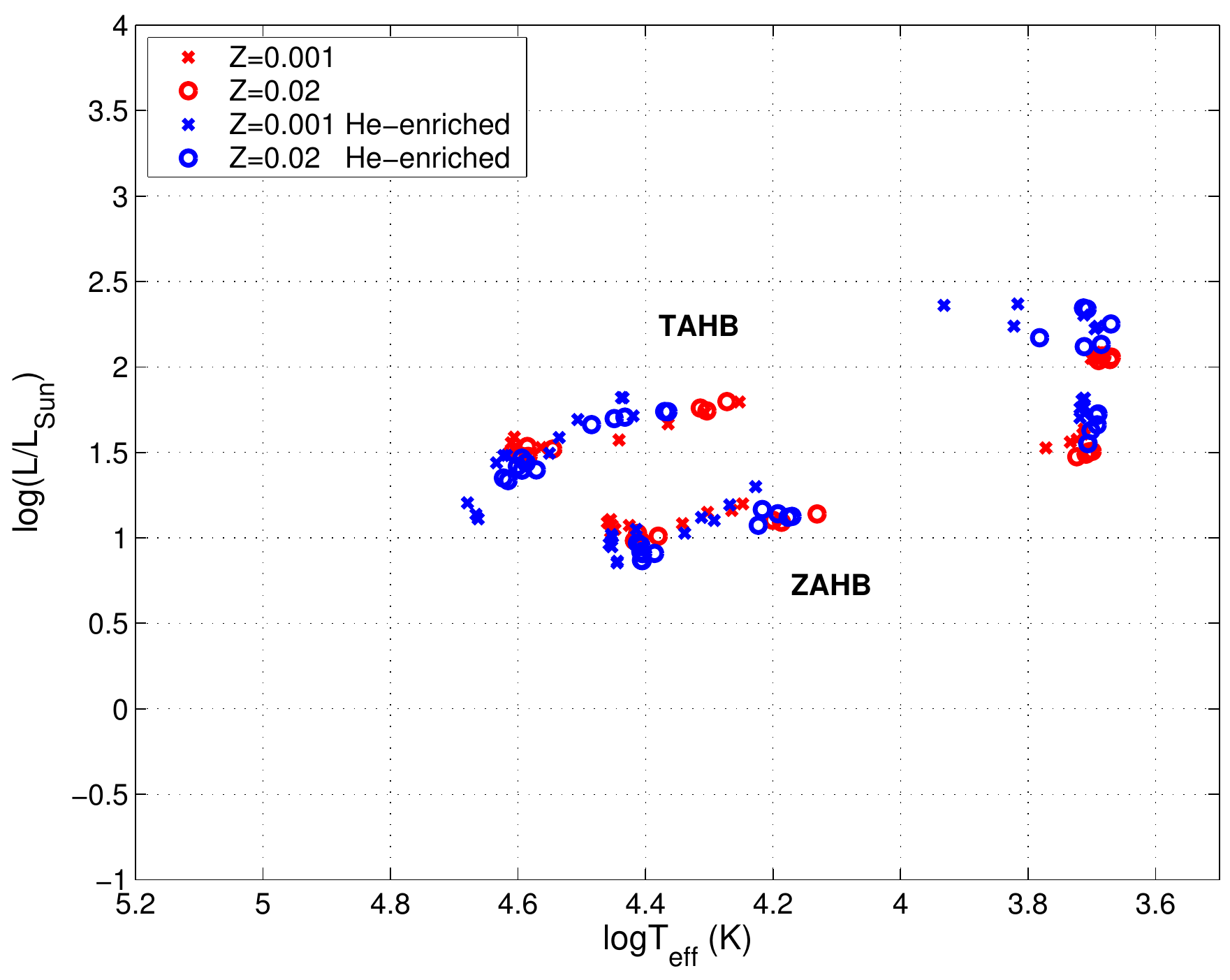}
 \includegraphics[scale=.45]{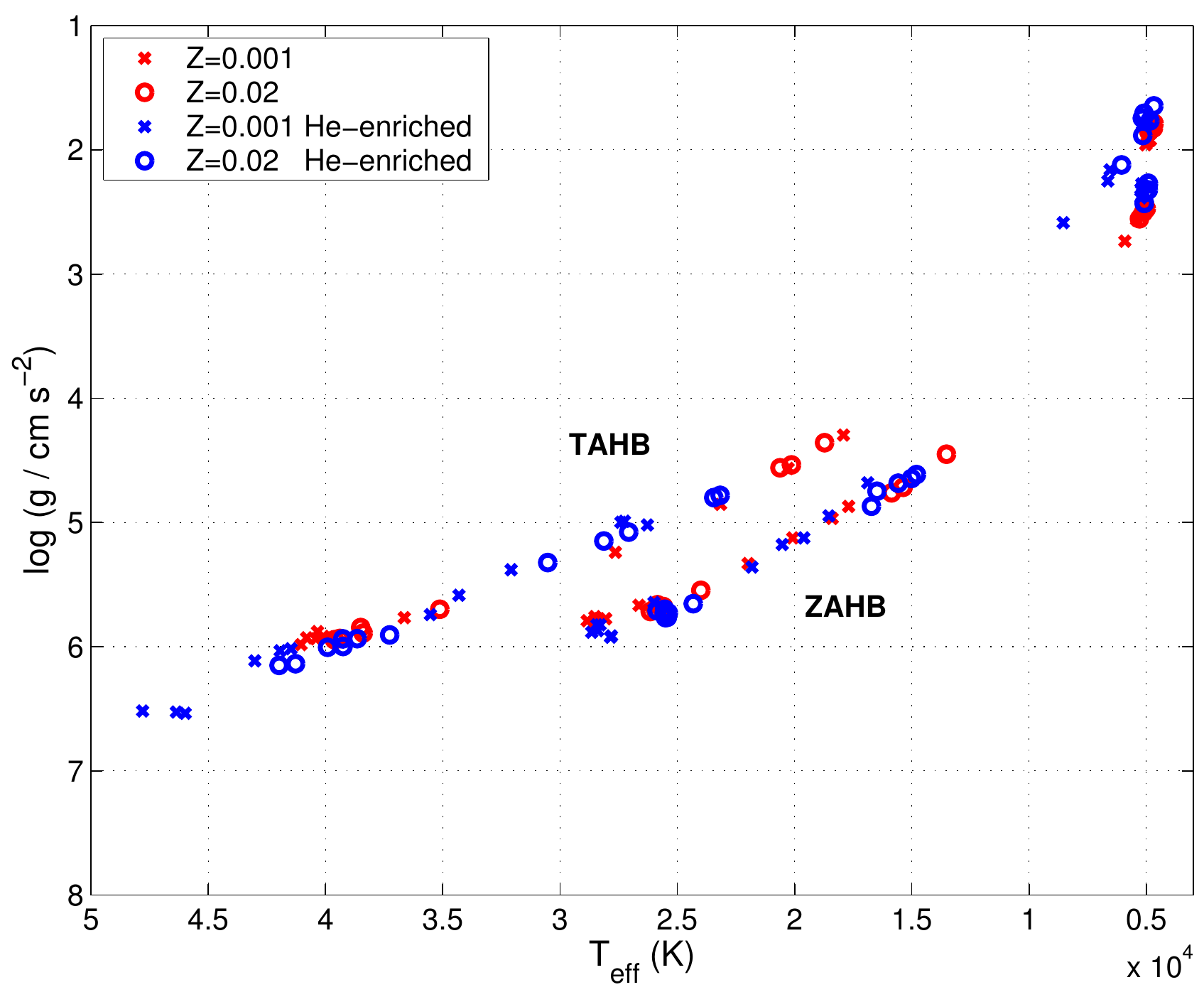}}
\end{center}
\caption{HB positions on HRD (left) and on $\log g-\Teff$ plane (right) as obtained for both normal-Y and He-enriched models,
 Pops. I \& II. The full extent of the HB can be regarded as occupying the area in between the ZAHB and TAHB bands.
 }
 \label{fig:hbpos_comb}
\end{figure}

\begin{figure}
\begin{center}
{\includegraphics[scale=.45]{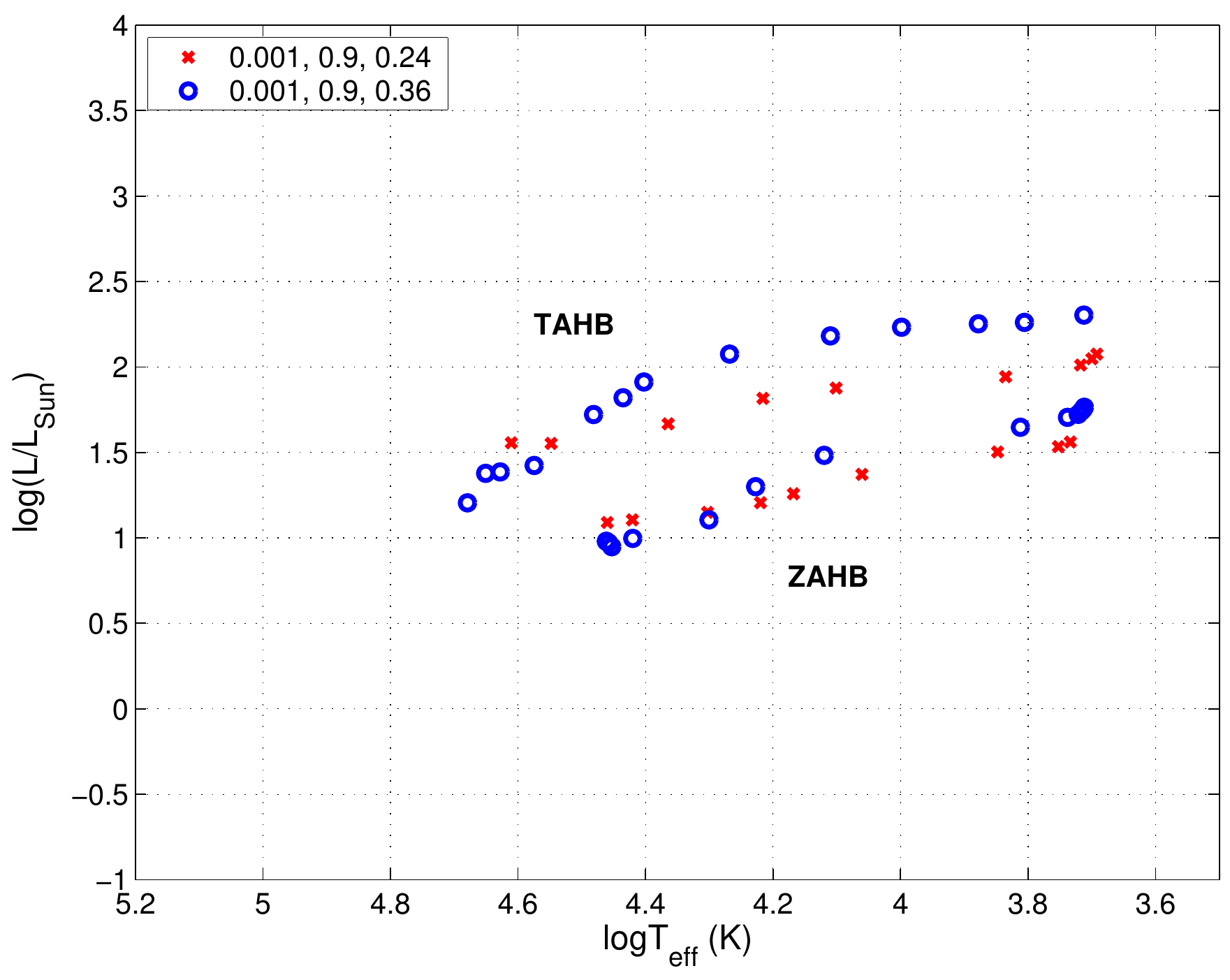}
 \includegraphics[scale=.45]{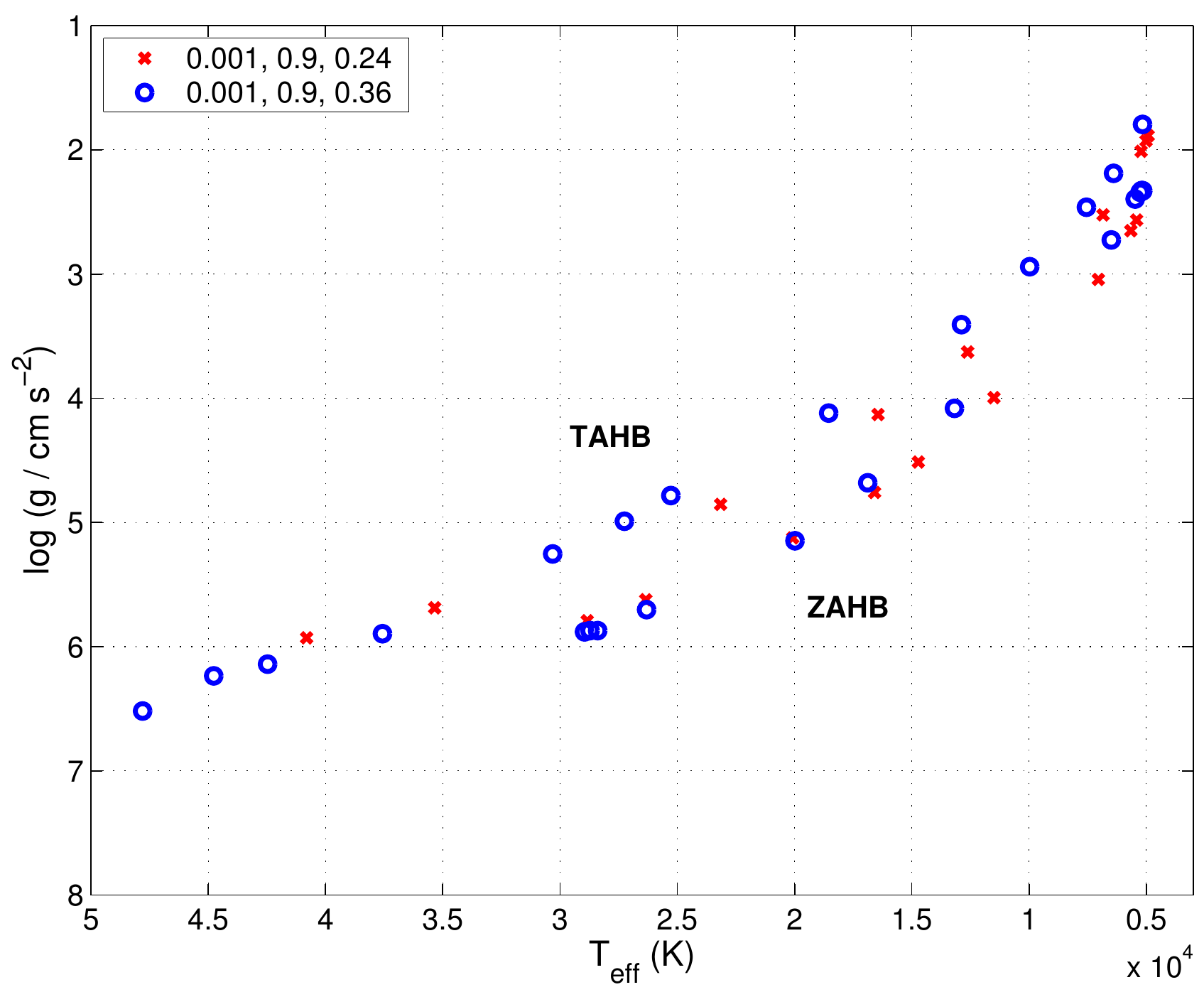}}
\end{center}
\caption{HB positions on HRD (left) and on $\log g-\Teff$ plane (right) 
as obtained for increasing MLR ($\etareim$ values) of two selected Pop. I models - $(Z,M)=(0.001, 0.9\,\Msun)$,
one of `normal' He abundance -- $Y=0.24$, the other with $Y=0.36$.
The $\etareim$ values are ranging from $1.0$ to $1.8$ in jumps of $0.1$ for $Y=0.24$,
and from $1.2$ to $3.4$ in jumps of $0.2$ for the He-enriched model of $Y=0.36$.
The gaps that appear in figure~\ref{fig:hbpos_comb} seem to disappear for a denser sampling of the MLR ($\etareim$ values).
}
\label{fig:hbpos_dense}
\end{figure}

\begin{figure}
\begin{center}
\includegraphics[scale=.7]{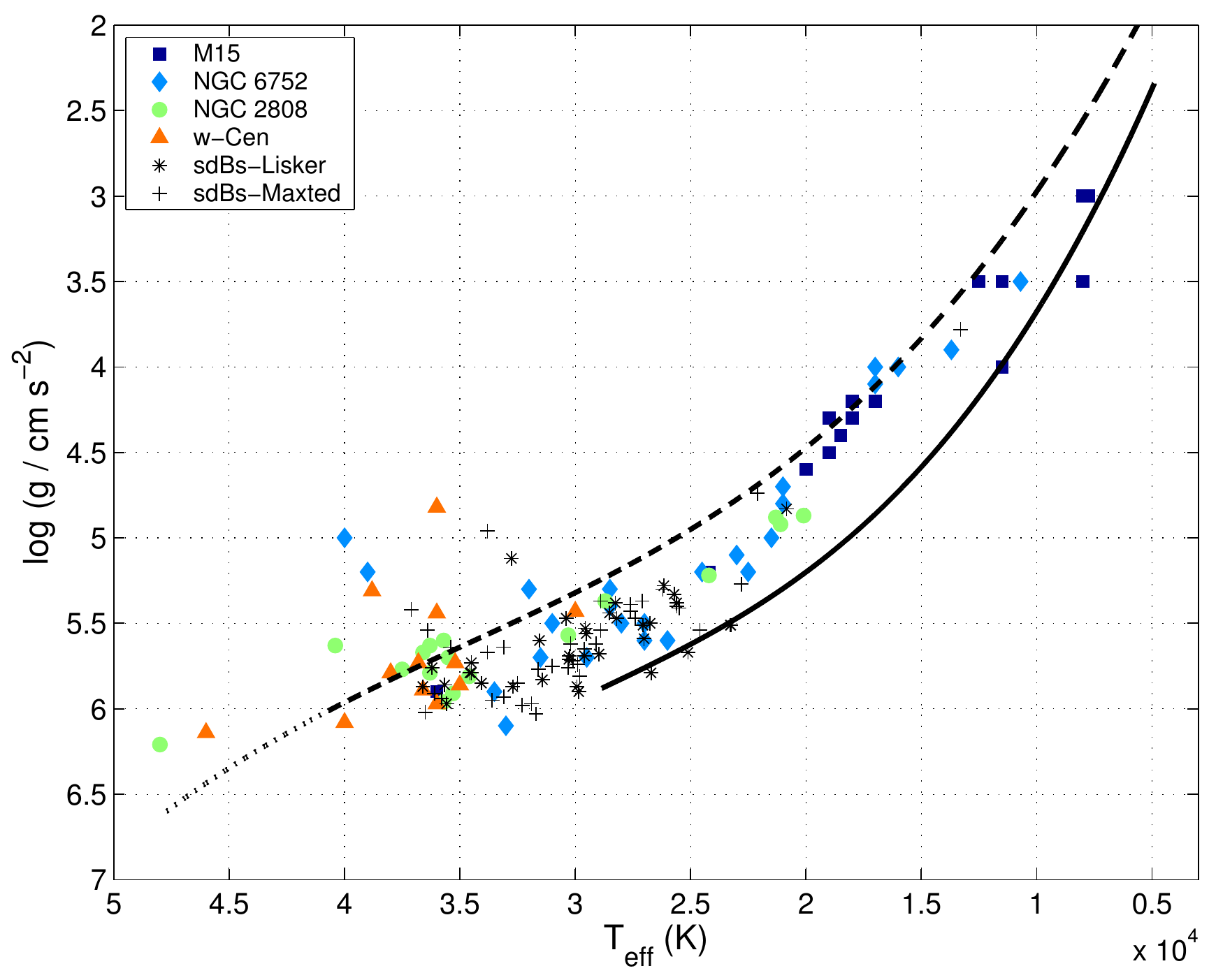}
\end{center}
\caption{HB positions on $\log g-\Teff$ plane. Data of the four GCs is from a set of papers by Moehler et al.;
black asterisks are according to a sample of field sdB stars as obtained by \cite{2005A&A...430..223L};
plus signs are a sample of sdBs obtained from \cite{2001MNRAS.326.1391M} (see text).
Solid line is a cubic fit to our derived ZAHB of both `normal' and He-enriched models, Pops. I \& II (figure~\ref{fig:hbpos_comb}).
Similarly, the dashed line is a fit to our TAHB; the dotted section at high $(\log g,\Teff)$ marks the contribution of only the He-enriched models.
}
\label{fig:hbpos_obs}
\end{figure}

Our main conclusion is that for most combinations of $(Z,Y,M)$, all the different patterns of behaviour are obtained 
within a \emph{reasonable} range of $\etareim$ values,
from core He flash at the tip of the RGB to a core flash taking place while already on the WD cooling curve. 
The HB positions of all our $(Z,Y,M,\etareim)$ sequences form a wide band in the HRD, 
overlapping the extent of the observed HB for $\Teff$ ranging from $\sim5,000$ to $\sim50,000$~K (see Figure~\ref{fig:hbpos_comb}).
The resulting HB region in the $[\Teff,\log g]$ plane covers the region obtained from observations (based on four galactic GCs and tens of field sdB stars).

We find that for a given $(Z,Y)$ combination, 
increasing $\etareim$ values are required with increasing mass $(M)$ for deriving Post-Tip and WD Flashers
(within the normal Y models -- up to a maximum value of $\etareim=4.4$ for $(Z,Y,M)=(0.001,0.24,1.20)$), 
while for a given $M$, decreasing $\etareim$ values are required within combinations of increasing $(Z,Y)$
(down to a minimum value of $\etareim=0.8$ for $(Z,Y,M)=(0.02,0.28,0.90)$).
Thus, for the metal-richer, lower-mass stars of our sample, smaller values of $\etareim$, as well as a narrower range, 
are required for obtaining the complete spectrum of behaviour. By contrast, 
for the metal-poorer, higher-mass stars, higher $\etareim$ values, spanning a wider range, are required for the same purpose.

Average lifetimes on the HB are $1.3$ and $1.6 \times 10^8$~yr for the normal $Y$ and He-enriched models, respectively;
maximal HB durations are in excess of $2\times10^8$~yr;
The HB lifetime increases within each increasing $\etareim$ sequence by factors ranging between $\sim1.5$ and approaching $3.0$;
thus, the bluer the position on the HB, the longer the HB lifetime.

A significant fraction of HB stars may be EHB stars, situated blueward of the instability strip at the extreme end of the HB in the HRD,
for example (see \cite{2008AJ....136.2259S} and references therein):
$\gtrsim38\%$ for the massive halo GC - NGC 2419; around $30\%$ or more for $\omega$ Cen (\cite{2000ApJ...530..352D});
around $15\%$ for M54 (NGC 6715); $\sim12\%$ for NGC 2808. 
It should be noted that for all these GCs, masses have been estimated to be in excess of $10^6\ \Msun$ 
(see e.g. compilation of \cite{1993ASPC...50..357P});
therefore one may argue that it is especially for massive GCs that the population of EHB stars becomes significant.

It should also be noted that in many cases there is a high probability that positions on CMDs of EHB stars 
(and even more so, of BHk stars) may overlap the domains occupied by 
another unique group of stars, that of Blue Stragglers - commonly referred to as `rejuvenated' MS stars,
lying at brighter and bluer positions than the MSTO on CMDs of star clusters
(see, for example, \cite{2002A&A...391..945P}, displaying CMDs of GCs as observed with the HST/WFPC2,
and also the optical+UV CMDs of NGC 6441 and NGC 6388 as presented by \cite{2007A&A...474..105B}). 
Nevertheless, the two populations have different origins and
for both, there is considerable ambiguity concerning the production mechanisms within the constraints imposed by observations.

\cite{2001MNRAS.326.1391M} find in a survey based on field populations
that about two thirds of all EHB stars are short-period binaries and 
conclude that this provides strong evidence for binary star evolution to play a crucial role in the formation of the majority of EHB stars.
It may well be that the enhanced mass loss rate required to produce EHB stars is a result of binary interaction, 
such as a common envelope phase (see also, e.g., \cite{2010A&A...519A..25G, 2015A&A...576A.123S}, regarding sdB's binarity).
On the other hand, examination of EHB stars in galactic GCs revealed a clear lack of close binaries
(\cite{2006A&A...451..499M, 2008A&A...480L...1M, 2015ApJ...812L..31M});
the binary fraction being significantly lower (in cases almost vanishing) compared with the fraction amongst the field hot subdwarfs counterparts.
A relation between binary fraction and cluster age may provide a possible solution to this difference,
if the binary fraction is found to decrease with increasing age of the stellar population (\cite{2008A&A...480L...1M, 2008A&A...484L..31H}).
It seems, however, that there may be differences between evolutionary paths leading to EHBs  
within field and cluster populations.

Our main conclusions regarding EHB stars are as follows:
\begin{itemize}
\item For both high and low metallicities, it is possible to produce EHB stars by increasing the amount of mass that is lost on the RGB.
\item Envelope masses (at core He flash) below $0.01\ \Msun$ all yield $\Teff$ in excess of $20,000$ K at ZAHB, and in excess of $30,000$ K at TAHB; 
the Pop. II ($Z=0.001$) models attaining higher $\Teff$ values as function of $\Menv$.
\item The highest $\Teff\simeq48,000$ K (at TAHB) was obtained in our calculations for He-enriched Pop. II models.
\item Increasing $Y$ for \emph{fixed} $(Z,M)$ combinations produces bluer HB positions for the highest value of 
the $\etareim$ range corresponding to each combination.
\item The longest HB lifetimes, in excess of $2.3\times10^8$~yr, are all obtained for the He-enriched models, 
which also attain the highest effective temperatures on the HB, $\Teff@TAHB\gtrsim40,000$~K.  
This trend strengthens the significance of EHB stars as major contributors to the total UV output emanating from their host system.
\item While $\Teff$ increases with decreasing $Z$ down to just below $10^{-3}$, a halt or even a slight reversal in trend occurs as $Z$ decreases further towards $10^{-4}$.
\end{itemize}

In summary, it seems that the \emph{combined} effect of He self-enrichment \emph{and} 
enhanced mass-loss on the RGB is required for producing the \emph{hottest} HB stars.
As shown in Table~2, none of the sequences of increasing He abundances manage to produce EHB stars when adopting mass-loss rates 
that are on the low-ish end.
This conclusion is confirmed by other studies: regarding the production of the hottest HB stars in $\omega$ Cen,
\cite{2007A&A...475L...5M} conclude that, whereas He-enhancement is not ruled out, "additional processes are required", with strong support for the 
late hot flasher scenario. Similarly, \cite{2007A&A...474..105B} conclude that both high mass-loss efficiency on RGB, and a He-spread of up to $Y\approx0.40$,
are required for explaining the extended HB and BHk populations in the metal-rich clusters NGC 6388 and 6441.
Our theoretical study shows that Pop. II stars are more apt to yield the bluest HB positions. Thus, it is
the metal-poorer clusters that may have a higher potential for producing the bluemost-extending EHB populations.
This, however, leads to a puzzle: adopting the hot He flash scenario as a major channel for production of EHB stars,
a scenario that relies on enhanced mass-loss on the RGB, how do we reconcile the requirements of low-$Z$ and enhanced mass loss (given that
low-$Z$ is usually associated with a lower rate of mass-loss)?

Therefore, the key problem that remains to be solved is finding the mass-loss laws or mechanisms, 
as a function of the leading characteristics -- stellar mass and composition, 
be it the result of the wind of a single star, binary evolution (\cite{2006A&A...446..569C}), 
interactions with planets or envelope stripping due to collisions in a dense stellar environment.
Only then shall we be able to identify among the grid of models presented in this study those that are most compatible with physical reality.

\acknowledgments

This work was supported in part by the Israeli Science Foundation grant 388/07.

\bibliographystyle{apj}
\bibliography{ehbbibfile}

\clearpage


\begin{deluxetable}{ll|cc|ccc|cc|ccl}
\label{tab:tab1}
\tablecolumns{12}
\tablewidth{0pc}
\tabletypesize{\scriptsize}
\tablecaption{Characteristics at core He flash and Horizontal Branch for $[Z,Y,M]$ combinations with increasing mass-loss rates---$\etareim$}
\tablehead{
\multicolumn{2}{c}{} & \multicolumn{2}{c}{Core-Flash} & \multicolumn{3}{c}{ZAHB}                       & \multicolumn{2}{c}{TAHB}     & `mid'-HB  & \multicolumn{2}{c}{} \\
$M$  & $\etareim$  & $M_{core}$ & $\Menv$     & Time      & $\Teff$   & $L$             & $\Teff$  & $L$             & $\Teff$  & $M_{f}$  & Flash
}
\startdata
\sidehead{[Z,Y] = [0.001, 0.24]}
\hline
 0.80 & 0.6  &  0.454  & 1.55E-01 &  15.7      & 5.92E+03 & 3.36E+01 &  4.97E+03 & 1.11E+02 &  5.08E+03 & 0.53    & Tip      \\ 
      & 1.0  &  0.453  & 3.77E-02 &  15.7      & 1.77E+04 & 1.58E+01 &  1.79E+04 & 6.24E+01 &  1.88E+04 & 0.49    & Post     \\ 
      & 1.2  &  0.451  & 2.50E-03 &  15.7      & 2.85E+04 & 1.28E+01 &  4.03E+04 & 3.88E+01 &  3.54E+04 & 0.45    & WD       \\ 
      & 1.5  &  ---    & ---      &  ---       & ---      & ---      &  ---      & ---      &  ---      & 0.44    & He-WD    \\ 
 0.90 & 1.0* &  0.451  & 1.83E-01 &  10.1      & 5.42E+03 & 3.64E+01 &  4.92E+03 & 1.19E+02 &  5.06E+03 & 0.53    & Tip      \\ 
      & 1.1  &  0.450  & 1.61E-01 &  10.1      & 5.66E+03 & 3.42E+01 &  5.01E+03 & 1.12E+02 &  5.12E+03 & 0.53    & Tip      \\ 
      & 1.2  &  0.450  & 1.32E-01 &  10.1      & 7.05E+03 & 3.19E+01 &  5.22E+03 & 1.03E+02 &  5.30E+03 & 0.52    & Tip      \\ 
      & 1.3  &  0.451  & 9.38E-02 &  10.1      & 1.15E+04 & 2.35E+01 &  6.84E+03 & 8.76E+01 &  1.08E+04 & 0.51    & Tip/Post \\ 
      & 1.4  &  0.452  & 6.34E-02 &  10.1      & 1.47E+04 & 1.81E+01 &  1.26E+04 & 7.54E+01 &  1.45E+04 & 0.50    & Post     \\ 
      & 1.5  &  0.450  & 4.58E-02 &  10.1      & 1.66E+04 & 1.61E+01 &  1.64E+04 & 6.56E+01 &  1.71E+04 & 0.49    & Post     \\ 
      & 1.6* &  0.449  & 2.35E-02 &  10.1      & 2.01E+04 & 1.41E+01 &  2.32E+04 & 4.65E+01 &  2.28E+04 & 0.47    & Post     \\ 
      & 1.7  &  0.449  & 4.90E-03 &  10.1      & 2.63E+04 & 1.28E+01 &  3.53E+04 & 3.57E+01 &  3.21E+04 & 0.45    & Post     \\ 
      & 1.8* &  0.448  & 2.20E-03 &  10.1      & 2.88E+04 & 1.23E+01 &  4.08E+04 & 3.60E+01 &  3.69E+04 & 0.45    & WD       \\ 
      & 2.2* &  ---    & ---      &  ---       & ---      & ---      &  ---      & ---      &  ---      & 0.44    & He-WD    \\ 
 1.00 & 1.0  &  0.448  & 3.21E-01 &  6.8       & 5.15E+03 & 4.33E+01 &  4.83E+03 & 1.22E+02 &  5.07E+03 & 0.54    & Tip      \\ 
      & 2.2  &  0.447  & 3.31E-02 &  6.8       & 1.84E+04 & 1.45E+01 &  2.03E+04 & 5.41E+01 &  2.02E+04 & 0.48    & Post     \\ 
      & 2.5  &  0.444  & 3.10E-03 &  6.8       & 2.84E+04 & 1.19E+01 &  4.01E+04 & 3.42E+01 &  3.63E+04 & 0.45    & WD       \\ 
      & 3.1  &  ---    & ---      &  ---       & ---      & ---      &  ---      & ---      &  ---      & 0.44    & He-WD    \\ 
 1.10 & 2.3  &  0.443  & 2.12E-01 &  4.8       & 5.29E+03 & 3.77E+01 &  5.02E+03 & 1.12E+02 &  5.20E+03 & 0.53    & Tip      \\ 
      & 3.2  &  0.443  & 4.30E-03 &  4.8       & 2.66E+04 & 1.18E+01 &  3.66E+04 & 3.38E+01 &  3.35E+04 & 0.45    & Post     \\ 
      & 3.4  &  0.439  & 2.70E-03 &  4.8       & 2.81E+04 & 1.12E+01 &  4.04E+04 & 3.33E+01 &  3.59E+04 & 0.44    & WD       \\ 
      & 4.0  &  ---    & ---      &  ---       & ---      & ---      &  ---      & ---      &  ---      & 0.43    & He-WD    \\ 
 1.20 & 3.0  &  0.440  & 2.62E-01 &  3.5       & 5.18E+03 & 4.05E+01 &  4.96E+03 & 1.19E+02 &  5.11E+03 & 0.53    & Tip      \\ 
      & 4.0  &  0.439  & 1.56E-02 &  3.5       & 2.20E+04 & 1.21E+01 &  2.76E+04 & 3.73E+01 &  2.58E+04 & 0.45    & Post     \\ 
      & 4.4  &  0.437  & 2.50E-03 &  3.5       & 2.85E+04 & 1.09E+01 &  4.10E+04 & 3.17E+01 &  3.57E+04 & 0.44    & WD       \\ 
      & 5.5  &  ---    & ---      &  ---       & ---      & ---      &  ---      & ---      &  ---      & 0.42    & He-WD    \\ 
%
%
\sidehead{[Z,Y] = [0.01, 0.28]}
\hline
 0.80 & 0.4  &  0.448  & 1.05E-01 &  21.9      & 5.49E+03 & 3.13E+01 &  5.16E+03 & 1.12E+02 &  5.59E+03 & 0.53    & Tip      \\ 
      & 0.6  &  0.447  & 4.45E-02 &  21.9      & 1.32E+04 & 1.72E+01 &  1.56E+04 & 7.51E+01 &  1.26E+04 & 0.49    & Post     \\ 
      & 0.9  &  0.441  & 1.60E-03 &  21.9      & 2.69E+04 & 1.07E+01 &  3.94E+04 & 3.19E+01 &  2.78E+04 & 0.44    & WD       \\ 
      & 1.3  &  ---    & ---      &  ---       & ---      & ---      &  ---      & ---      &  ---      & 0.42    & He-WD    \\ 
 0.90 & 0.8  &  0.445  & 1.04E-01 &  13.9      & 5.48E+03 & 3.15E+01 &  5.60E+03 & 1.12E+02 &  5.61E+03 & 0.52    & Tip      \\ 
      & 1.1  &  0.444  & 2.29E-02 &  13.9      & 1.76E+04 & 1.27E+01 &  2.20E+04 & 5.11E+01 &  1.77E+04 & 0.47    & Post     \\ 
      & 1.3  &  0.442  & 1.90E-03 &  13.9      & 2.67E+04 & 1.08E+01 &  3.95E+04 & 3.36E+01 &  2.83E+04 & 0.44    & WD       \\ 
      & 1.8  &  ---    & ---      &  ---       & ---      & ---      &  ---      & ---      &  ---      & 0.43    & He-WD    \\ 
 1.00 & 1.0  &  0.444  & 1.88E-01 &  9.3       & 5.17E+03 & 3.71E+01 &  4.86E+03 & 1.22E+02 &  5.22E+03 & 0.53    & Tip      \\ 
      & 1.7  &  0.442  & 4.10E-03 &  9.3       & 2.41E+04 & 1.10E+01 &  3.38E+04 & 3.28E+01 &  2.53E+04 & 0.45    & Post     \\ 
      & 1.9  &  0.437  & 2.10E-03 &  9.3       & 2.65E+04 & 1.03E+01 &  3.92E+04 & 3.06E+01 &  2.86E+04 & 0.44    & WD       \\ 
      & 2.4  &  ---    & ---      &  ---       & ---      & ---      &  ---      & ---      &  ---      & 0.42    & He-WD    \\ 
 1.10 & 2.0  &  0.441  & 8.03E-02 &  6.4       & 5.85E+03 & 2.89E+01 &  7.39E+03 & 8.38E+01 &  7.39E+03 & 0.51    & Tip      \\ 
      & 2.2  &  0.441  & 1.76E-02 &  6.4       & 1.89E+04 & 1.17E+01 &  2.42E+04 & 4.25E+01 &  1.88E+04 & 0.46    & Post     \\ 
      & 2.5  &  0.436  & 2.10E-03 &  6.4       & 2.65E+04 & 1.01E+01 &  3.95E+04 & 3.09E+01 &  2.74E+04 & 0.44    & WD       \\ 
      & 3.2  &  ---    & ---      &  ---       & ---      & ---      &  ---      & ---      &  ---      & 0.42    & He-WD    \\ 
 1.20 & 2.5  &  0.440  & 1.06E-01 &  4.7       & 5.45E+03 & 3.11E+01 &  6.59E+03 & 9.75E+01 &  5.53E+03 & 0.51    & Tip      \\ 
      & 2.7  &  0.440  & 3.45E-02 &  4.7       & 1.54E+04 & 1.36E+01 &  2.00E+04 & 6.21E+01 &  1.50E+04 & 0.47    & Post     \\ 
      & 3.0  &  0.438  & 2.00E-03 &  4.7       & 2.68E+04 & 1.04E+01 &  4.01E+04 & 3.26E+01 &  2.89E+04 & 0.44    & WD       \\ 
      & 4.0  &  ---    & ---      &  ---       & ---      & ---      &  ---      & ---      &  ---      & 0.42    & He-WD    \\ 
%
%
\sidehead{[Z,Y] = [0.02, 0.28]}
\hline
 0.90 & 0.3  &  0.452  & 1.05E-01 &  19.8      & 5.29E+03 & 2.98E+01 &  4.89E+03 & 1.09E+02 &  5.37E+03 & 0.53    & Tip      \\ 
      & 0.5  &  0.449  & 2.41E-02 &  19.8      & 1.59E+04 & 1.27E+01 &  2.01E+04 & 5.54E+01 &  1.54E+04 & 0.47    & Post     \\ 
      & 0.8  &  0.446  & 1.30E-03 &  19.8      & 2.58E+04 & 1.06E+01 &  3.85E+04 & 3.43E+01 &  2.77E+04 & 0.45    & WD       \\ 
      & 1.4  &  ---    & ---      &  ---       & ---      & ---      &  ---      & ---      &  ---      & 0.43    & He-WD    \\ 
 1.00 & 0.6  &  0.446  & 1.54E-01 &  13.1      & 5.07E+03 & 3.18E+01 &  4.67E+03 & 1.14E+02 &  5.12E+03 & 0.53    & Tip      \\ 
      & 1.1  &  0.444  & 2.61E-02 &  13.1      & 1.54E+04 & 1.23E+01 &  2.06E+04 & 5.75E+01 &  1.50E+04 & 0.47    & Post     \\ 
      & 1.5  &  0.437  & 1.60E-03 &  13.2      & 2.62E+04 & 9.64E+00 &  3.96E+04 & 3.02E+01 &  2.77E+04 & 0.44    & WD       \\ 
      & 2.0  &  ---    & ---      &  ---       & ---      & ---      &  ---      & ---      &  ---      & 0.42    & He-WD    \\ 
 1.10 & 1.0  &  0.446  & 1.37E-01 &  9.0       & 5.12E+03 & 3.09E+01 &  4.83E+03 & 1.16E+02 &  5.13E+03 & 0.53    & Tip      \\ 
      & 1.6  &  0.443  & 3.35E-02 &  9.0       & 1.35E+04 & 1.38E+01 &  1.87E+04 & 6.27E+01 &  1.29E+04 & 0.48    & Post     \\ 
      & 2.0  &  0.436  & 1.50E-03 &  9.0       & 2.59E+04 & 9.55E+00 &  3.94E+04 & 3.01E+01 &  2.65E+04 & 0.44    & WD       \\ 
      & 2.7  &  ---    & ---      &  ---       & ---      & ---      &  ---      & ---      &  ---      & 0.42    & He-WD    \\ 
 1.20 & 1.5  &  0.443  & 1.78E-01 &  6.5       & 5.01E+03 & 3.21E+01 &  4.69E+03 & 1.11E+02 &  5.05E+03 & 0.53    & Tip      \\ 
      & 2.2  &  0.442  & 2.80E-02 &  6.5       & 2.40E+04 & 1.02E+01 &  3.51E+04 & 3.31E+01 &  2.55E+04 & 0.44    & Post     \\ 
      & 2.5  &  0.437  & 1.60E-03 &  6.5       & 2.56E+04 & 9.67E+00 &  3.84E+04 & 3.00E+01 &  2.76E+04 & 0.44    & WD       \\ 
      & 3.4  &  ---    & ---      &  ---       & ---      & ---      &  ---      & ---      &  ---      & 0.42    & He-WD    \\ 
\enddata
\tablecomments{Masses and luminosities are in solar units; time (=age at ZAHB) in Gyr; $M_{f}$ is the final WD mass.} 
\end{deluxetable}

\begin{deluxetable}{ll|cc|ccc|cc|cc}
\label{tab:tab2}
\tablecolumns{11}
\tablewidth{0pc}
\tabletypesize{\scriptsize}
\tablecaption{Characteristics at core He flash and Horizontal Branch for $[Z,M,\etareim]$ combinations with increasing initial He abundance}
\tablehead{
\multicolumn{2}{c}{} & \multicolumn{2}{c}{Core-Flash} & \multicolumn{3}{c}{ZAHB}                       & \multicolumn{2}{c}{TAHB}     & \multicolumn{2}{l}{`mid'-HB} \\
$(M,\etareim)$ & $Y$ & $M_{core}$ & $\Menv$     & Time      & $\Teff$   & $L$             & $\Teff$  & $L$             & $\Teff$  & $M_{f}$
}
\startdata
\sidehead{Z=0.001}
\hline
%
 (0.90, 0.8)  & 0.20 &   0.460  & 1.90E-01 &   12.9      & 5.44E+03 & 3.28E+01 &   4.87E+03 & 1.01E+02 &   5.30E+03 & 0.53    \\ 
             & 0.24 &   0.451  & 2.32E-01 &   10.0      & 5.26E+03 & 3.95E+01 &   4.85E+03 & 1.26E+02 &   5.16E+03 & 0.54    \\ 
             & 0.28 &   0.441  & 2.84E-01 &   7.8       & 5.18E+03 & 4.70E+01 &   4.85E+03 & 1.48E+02 &   5.12E+03 & 0.55    \\ 
             & 0.32 &   0.432  & 3.25E-01 &   5.9       & 5.13E+03 & 5.46E+01 &   4.84E+03 & 1.84E+02 &   5.08E+03 & 0.56    \\ 
             & 0.36 &   0.421  & 3.58E-01 &   4.5       & 4.56E+03 & 3.59E+02 &   4.88E+03 & 2.19E+02 &   5.10E+03 & 0.57    \\ 
             & 0.40 &   0.410  & 3.94E-01 &   3.4       & 4.59E+03 & 3.83E+02 &   4.92E+03 & 2.82E+02 &   5.10E+03 & 0.59    \\ 
 (1.00, 1.0) & 0.20 &   0.457  & 2.72E-01 &   8.7       & 5.22E+03 & 3.68E+01 &   4.81E+03 & 1.09E+02 &   5.12E+03 & 0.53    \\ 
             & 0.24 &   0.448  & 3.21E-01 &   6.8       & 5.15E+03 & 4.33E+01 &   4.83E+03 & 1.22E+02 &   5.07E+03 & 0.54    \\ 
             & 0.28 &   0.437  & 3.84E-01 &   5.2       & 5.12E+03 & 5.02E+01 &   4.80E+03 & 1.59E+02 &   5.06E+03 & 0.55    \\ 
             & 0.32 &   0.427  & 4.27E-01 &   4.0       & 5.09E+03 & 5.88E+01 &   4.81E+03 & 1.91E+02 &   5.07E+03 & 0.56    \\ 
             & 0.36 &   0.416  & 4.65E-01 &   3.1       & 4.59E+03 & 3.64E+02 &   4.82E+03 & 2.36E+02 &   5.07E+03 & 0.58    \\ 
             & 0.40 &   0.404  & 4.97E-01 &   2.3       & 4.61E+03 & 3.80E+02 &   4.84E+03 & 3.02E+02 &   5.08E+03 & 0.60    \\ 
 (1.10, 2.0) & 0.20 &   0.453  & 1.89E-01 &   6.1       & 5.42E+03 & 3.24E+01 &   4.93E+03 & 9.89E+01 &   5.34E+03 & 0.52    \\ 
             & 0.24 &   0.443  & 2.77E-01 &   4.8       & 5.18E+03 & 4.11E+01 &   4.87E+03 & 1.19E+02 &   5.14E+03 & 0.53    \\ 
             & 0.28 &   0.435  & 3.42E-01 &   3.7       & 5.12E+03 & 4.92E+01 &   4.84E+03 & 1.53E+02 &   5.08E+03 & 0.54    \\ 
             & 0.32 &   0.422  & 4.26E-01 &   2.8       & 4.57E+03 & 3.33E+02 &   4.83E+03 & 1.90E+02 &   5.06E+03 & 0.56    \\ 
             & 0.36 &   0.410  & 4.87E-01 &   2.2       & 4.60E+03 & 3.51E+02 &   4.86E+03 & 2.24E+02 &   5.08E+03 & 0.57    \\ 
             & 0.40 &   0.399  & 5.38E-01 &   1.7       & 4.63E+03 & 3.66E+02 &   4.91E+03 & 2.95E+02 &   5.09E+03 & 0.59    \\ 
\sidehead{Z=0.018}
\hline
 (1.00, 0.6) & 0.24 &   0.460  & 1.96E-02 &   16.5      & 1.64E+04 & 1.38E+01 &   1.84E+04 & 5.02E+01 &   1.63E+04 & 0.48    \\ 
             & 0.26 &   0.454  & 1.09E-01 &   14.3      & 5.29E+03 & 2.89E+01 &   4.78E+03 & 1.06E+02 &   5.32E+03 & 0.53    \\ 
             & 0.28 &   0.447  & 2.36E-01 &   12.4      & 4.96E+03 & 3.56E+01 &   4.53E+03 & 1.27E+02 &   4.97E+03 & 0.54    \\ 
             & 0.32 &   0.438  & 3.66E-01 &   9.3       & 4.89E+03 & 4.45E+01 &   4.51E+03 & 1.55E+02 &   4.89E+03 & 0.56    \\ 
             & 0.36 &   0.430  & 4.38E-01 &   6.8       & 4.91E+03 & 5.20E+01 &   4.50E+03 & 2.00E+02 &   4.89E+03 & 0.57    \\ 
             & 0.40 &   0.422  & 4.89E-01 &   5.1       & 4.93E+03 & 6.04E+01 &   4.51E+03 & 2.52E+02 &   4.89E+03 & 0.59    \\ 
\sidehead{Z=0.02}
\hline
 (0.90, 0.3) & 0.28 &   0.452  & 1.05E-01 &   19.8      & 5.29E+03 & 2.98E+01 &   4.89E+03 & 1.09E+02 &   5.37E+03 & 0.53    \\ 
             & 0.32 &   0.441  & 2.32E-01 &   14.6      & 4.92E+03 & 4.03E+01 &   4.53E+03 & 1.47E+02 &   4.97E+03 & 0.55    \\ 
             & 0.36 &   0.431  & 3.39E-01 &   10.8      & 4.87E+03 & 4.90E+01 &   4.50E+03 & 1.82E+02 &   4.91E+03 & 0.57    \\ 
             & 0.40 &   0.423  & 3.82E-01 &   7.9       & 4.88E+03 & 5.68E+01 &   4.50E+03 & 2.35E+02 &   4.90E+03 & 0.59    \\ 
 (1.00, 0.6) & 0.28 &   0.446  & 1.54E-01 &   13.1      & 5.07E+03 & 3.18E+01 &   4.67E+03 & 1.14E+02 &   5.12E+03 & 0.53    \\ 
             & 0.32 &   0.437  & 2.88E-01 &   9.7       & 4.88E+03 & 4.13E+01 &   4.50E+03 & 1.50E+02 &   4.95E+03 & 0.55    \\ 
             & 0.36 &   0.430  & 3.60E-01 &   7.2       & 4.86E+03 & 4.96E+01 &   4.47E+03 & 1.95E+02 &   4.86E+03 & 0.57    \\ 
             & 0.40 &   0.423  & 4.26E-01 &   5.3       & 4.89E+03 & 5.78E+01 &   4.49E+03 & 2.41E+02 &   4.92E+03 & 0.59    \\ 
 (1.10, 1.0) & 0.24 &   0.454  & 1.11E-02 &   12.0      & 1.88E+04 & 1.23E+01 &   2.37E+04 & 4.28E+01 &   1.91E+04 & 0.47    \\ 
             & 0.28 &   0.446  & 1.37E-01 &   9.0       & 5.12E+03 & 3.09E+01 &   4.83E+03 & 1.16E+02 &   5.13E+03 & 0.53    \\ 
             & 0.32 &   0.436  & 3.04E-01 &   6.8       & 4.87E+03 & 4.17E+01 &   4.50E+03 & 1.48E+02 &   4.88E+03 & 0.55    \\ 
             & 0.36 &   0.428  & 4.05E-01 &   5.2       & 4.87E+03 & 4.99E+01 &   4.48E+03 & 1.86E+02 &   4.92E+03 & 0.57    \\ 
             & 0.40 &   0.421  & 4.60E-01 &   3.8       & 4.89E+03 & 5.78E+01 &   4.48E+03 & 2.42E+02 &   4.93E+03 & 0.59    \\ 
\enddata
\end{deluxetable}

\begin{deluxetable}{lll|cc|ccc|cc|ccl}
\label{tab:tab3}
\tablecolumns{13}
\tablewidth{0pc}
\tabletypesize{\scriptsize}
\tablecaption{He-enriched models - Characteristics at core He flash and Horizontal Branch for $[Z,M,Y]$ combinations with increasing mass-loss rates---$\etareim$}
\tablehead{
\multicolumn{3}{c}{} & \multicolumn{2}{c}{Core-Flash} & \multicolumn{3}{c}{ZAHB}                       & \multicolumn{2}{c}{TAHB}     & `mid'-HB  & \multicolumn{2}{c}{} \\
$M$ & $Y$  & $\etareim$  & $M_{core}$ & $\Menv$     & Time      & $\Teff$   & $L$             & $\Teff$  & $L$             & $\Teff$  & $M_{f}$  & Flash
}
\startdata
\sidehead{Z = 0.001}
\hline
0.80 & 0.32 & 0.6  &   0.436  & 2.34E-01 &   9.2       & 5.23E+03 & 5.04E+01 &   4.95E+03 & 1.67E+02 &   5.16E+03 & 0.55    & Tip      \\ 
     &      & 1.5  &   0.436  & 6.80E-03 &   9.2       & 2.60E+04 & 1.12E+01 &   3.55E+04 & 3.12E+01 &   3.20E+04 & 0.44    & Post     \\ 
     &      & 1.7  &   0.433  & 2.80E-03 &   9.2       & 2.84E+04 & 1.04E+01 &   4.15E+04 & 3.04E+01 &   3.67E+04 & 0.43    & WD       \\ 
     &      & 2.2  &   ---    & ---      &   ---       & ---      & ---      &   ---      & ---      &   ---      & 0.42    & He-WD    \\ 
     & 0.36 & 0.8  &   0.428  & 2.18E-01 &   7.0       & 5.23E+03 & 5.74E+01 &   6.64E+03 & 1.73E+02 &   5.25E+03 & 0.56    & Tip      \\ 
     &      & 1.6  &   0.426  & 5.21E-02 &   7.0       & 1.85E+04 & 1.56E+01 &   2.74E+04 & 6.62E+01 &   1.88E+04 & 0.48    & Post     \\ 
     &      & 2.0  &   0.425  & 3.00E-03 &   7.0       & 2.87E+04 & 9.21E+00 &   4.30E+04 & 2.75E+01 &   3.66E+04 & 0.43    & WD       \\ 
     &      & 2.8  &   ---    & ---      &   ---       & ---      & ---      &   ---      & ---      &   ---      & 0.41    & He-WD    \\ 
     & 0.40 & 0.8  &   0.416  & 2.60E-01 &   5.2       & 5.19E+03 & 6.39E+01 &   8.55E+03 & 2.29E+02 &   5.21E+03 & 0.57    & Tip      \\ 
     &      & 2.0  &   0.416  & 4.14E-02 &   5.2       & 2.05E+04 & 1.32E+01 &   3.21E+04 & 4.92E+01 &   2.16E+04 & 0.46    & Post     \\ 
     &      & 2.8  &   0.405  & 3.90E-03 &   5.2       & 2.78E+04 & 7.30E+00 &   4.64E+04 & 1.37E+01 &   3.67E+04 & 0.41    & WD       \\ 
     &      & 3.5  &   ---    & ---      &   ---       & ---      & ---      &   ---      & ---      &   ---      & 0.40    & He-WD    \\ 
0.90 & 0.32 & 1.0  &   0.432  & 2.88E-01 &   5.9       & 5.15E+03 & 5.35E+01 &   4.90E+03 & 1.73E+02 &   5.10E+03 & 0.55    & Tip      \\ 
     &      & 2.2  &   0.430  & 3.55E-02 &   5.9       & 1.96E+04 & 1.26E+01 &   2.63E+04 & 5.17E+01 &   2.35E+04 & 0.47    & Post     \\ 
     &      & 2.5  &   0.429  & 3.10E-03 &   5.9       & 2.83E+04 & 1.02E+01 &   4.19E+04 & 3.04E+01 &   3.56E+04 & 0.43    & WD       \\ 
     &      & 3.0  &   ---    & ---      &   ---       & ---      & ---      &   ---      & ---      &   ---      & 0.42    & He-WD    \\ 
     & 0.36 & 1.2 *&   0.421  & 3.01E-01 &   4.5       & 5.15E+03 & 5.82E+01 &   5.16E+03 & 2.01E+02 &   5.15E+03 & 0.56    & Tip      \\ 
     &      & 1.4  &   0.421  & 2.68E-01 &   4.5       & 5.17E+03 & 5.67E+01 &   6.39E+03 & 1.82E+02 &   5.19E+03 & 0.56    & Tip      \\ 
     &      & 1.6  &   0.421  & 2.31E-01 &   4.5       & 5.21E+03 & 5.51E+01 &   7.55E+03 & 1.79E+02 &   5.21E+03 & 0.55    & Tip      \\ 
     &      & 1.8  &   0.421  & 1.96E-01 &   4.5       & 5.27E+03 & 5.29E+01 &   9.97E+03 & 1.71E+02 &   5.32E+03 & 0.55    & Tip      \\ 
     &      & 2.0  &   0.421  & 1.55E-01 &   4.5       & 5.47E+03 & 5.08E+01 &   1.29E+04 & 1.52E+02 &   6.32E+03 & 0.54    & Tip      \\ 
     &      & 2.2  &   0.420  & 1.23E-01 &   4.5       & 6.49E+03 & 4.44E+01 &   1.85E+04 & 1.19E+02 &   9.27E+03 & 0.53    & Tip      \\ 
     &      & 2.4  &   0.420  & 7.68E-02 &   4.5       & 1.32E+04 & 3.04E+01 &   2.53E+04 & 8.18E+01 &   1.53E+04 & 0.50    & Post     \\ 
     &      & 2.5 *&   0.420  & 5.97E-02 &   4.5       & 1.69E+04 & 1.99E+01 &   2.73E+04 & 6.61E+01 &   1.77E+04 & 0.48    & Post     \\ 
     &      & 2.6  &   0.420  & 3.85E-02 &   4.5       & 2.00E+04 & 1.28E+01 &   3.03E+04 & 5.28E+01 &   2.12E+04 & 0.46    & Post     \\ 
     &      & 2.8  &   0.420  & 7.40E-03 &   4.5       & 2.63E+04 & 9.94E+00 &   3.76E+04 & 2.65E+01 &   3.33E+04 & 0.43    & Post     \\ 
     &      & 3.0  &   0.419  & 3.30E-03 &   4.5       & 2.90E+04 & 9.55E+00 &   4.25E+04 & 2.43E+01 &   3.03E+04 & 0.42    & WD       \\ 
     &      & 3.2  &   0.415  & 3.30E-03 &   4.5       & 2.87E+04 & 9.36E+00 &   4.48E+04 & 2.39E+01 &   3.04E+04 & 0.42    & WD       \\ 
     &      & 3.4 *&   0.410  & 3.90E-03 &   4.5       & 2.84E+04 & 8.87E+00 &   4.78E+04 & 1.60E+01 &   3.66E+04 & 0.41    & WD       \\ 
     &      & 3.8 *&   ---    & ---      &   ---       & ---      & ---      &   ---      & ---      &   ---      & 0.41    & He-WD    \\ 
     & 0.40 & 1.2  &   0.411  & 3.41E-01 &   3.4       & 5.15E+03 & 6.58E+01 &   6.55E+03 & 2.34E+02 &   5.15E+03 & 0.58    & Tip      \\ 
     &      & 3.2  &   0.410  & 3.03E-02 &   3.4       & 2.18E+04 & 1.07E+01 &   3.43E+04 & 3.87E+01 &   2.31E+04 & 0.44    & Post     \\ 
     &      & 4.0  &   0.403  & 4.30E-03 &   3.4       & 2.78E+04 & 7.14E+00 &   4.60E+04 & 1.29E+01 &   3.00E+04 & 0.41    & WD       \\ 
     &      & 4.8  &   ---    & ---      &   ---       & ---      & ---      &   ---      & ---      &   ---      & 0.40    & He-WD    \\ 
\sidehead{Z = 0.02}
\hline
0.90 & 0.32 & 0.6  &   0.438  & 1.51E-01 &   14.6      & 5.08E+03 & 3.57E+01 &   4.84E+03 & 1.36E+02 &   5.18E+03 & 0.54    & Tip      \\ 
     &      & 1.0  &   0.437  & 3.22E-02 &   14.6      & 1.50E+04 & 1.32E+01 &   2.32E+04 & 5.48E+01 &   1.45E+04 & 0.47    & Post     \\ 
     &      & 1.3  &   0.432  & 1.60E-03 &   14.6      & 2.55E+04 & 9.03E+00 &   3.87E+04 & 2.75E+01 &   2.66E+04 & 0.43    & WD       \\ 
     &      & 1.8  &   ---    & ---      &   ---       & ---      & ---      &   ---      & ---      &   ---      & 0.42    & He-WD    \\ 
     & 0.36 & 0.8  &   0.431  & 1.66E-01 &   10.8      & 5.03E+03 & 4.25E+01 &   6.05E+03 & 1.48E+02 &   5.12E+03 & 0.54    & Tip      \\ 
     &      & 1.3  &   0.430  & 2.88E-02 &   10.8      & 1.67E+04 & 1.19E+01 &   2.81E+04 & 4.99E+01 &   1.62E+04 & 0.46    & Post     \\ 
     &      & 1.6  &   0.425  & 1.60E-03 &   10.8      & 2.54E+04 & 8.10E+00 &   3.93E+04 & 2.49E+01 &   2.65E+04 & 0.43    & WD       \\ 
     &      & 2.2  &   ---    & ---      &   ---       & ---      & ---      &   ---      & ---      &   ---      & 0.41    & He-WD    \\ 
     & 0.40 & 0.8  &   0.423  & 2.46E-01 &   7.9       & 4.93E+03 & 5.20E+01 &   5.10E+03 & 2.18E+02 &   5.03E+03 & 0.57    & Tip      \\ 
     &      & 1.7  &   0.423  & 3.70E-03 &   7.9       & 2.43E+04 & 8.11E+00 &   3.73E+04 & 2.50E+01 &   2.52E+04 & 0.43    & Post     \\ 
     &      & 2.0  &   0.415  & 2.00E-03 &   7.9       & 2.54E+04 & 7.35E+00 &   4.13E+04 & 2.16E+01 &   2.61E+04 & 0.42    & WD       \\ 
     &      & 2.6  &   ---    & ---      &   ---       & ---      & ---      &   ---      & ---      &   ---      & 0.40    & He-WD    \\ 
1.00 & 0.32 & 1.0  &   0.437  & 1.48E-01 &   9.7       & 5.09E+03 & 3.56E+01 &   5.15E+03 & 1.32E+02 &   5.16E+03 & 0.53    & Tip      \\ 
     &      & 1.5  &   0.436  & 3.28E-02 &   9.7       & 1.48E+04 & 1.34E+01 &   2.34E+04 & 5.50E+01 &   1.45E+04 & 0.47    & Post     \\ 
     &      & 1.7  &   0.435  & 1.50E-03 &   9.7       & 2.59E+04 & 9.31E+00 &   3.93E+04 & 2.93E+01 &   2.72E+04 & 0.43    & WD       \\ 
     &      & 2.4  &   ---    & ---      &   ---       & ---      & ---      &   ---      & ---      &   ---      & 0.41    & He-WD    \\ 
     & 0.36 & 1.0  &   0.430  & 2.45E-01 &   7.2       & 4.92E+03 & 4.59E+01 &   4.68E+03 & 1.79E+02 &   4.99E+03 & 0.55    & Tip      \\ 
     &      & 1.8  &   0.429  & 3.41E-02 &   7.2       & 1.56E+04 & 1.38E+01 &   2.71E+04 & 5.09E+01 &   1.59E+04 & 0.46    & Post     \\ 
     &      & 2.1  &   0.426  & 1.70E-03 &   7.2       & 2.56E+04 & 8.29E+00 &   3.99E+04 & 2.62E+01 &   2.68E+04 & 0.43    & WD       \\ 
     &      & 2.8  &   ---    & ---      &   ---       & ---      & ---      &   ---      & ---      &   ---      & 0.41    & He-WD    \\ 
     & 0.40 & 1.2  &   0.422  & 2.83E-01 &   5.3       & 4.90E+03 & 5.31E+01 &   5.17E+03 & 2.21E+02 &   5.00E+03 & 0.57    & Tip      \\ 
     &      & 2.2  &   0.422  & 3.37E-02 &   5.3       & 1.65E+04 & 1.47E+01 &   3.05E+04 & 4.59E+01 &   1.68E+04 & 0.45    & Post     \\ 
     &      & 2.6  &   0.416  & 2.10E-03 &   5.3       & 2.55E+04 & 7.41E+00 &   4.20E+04 & 2.24E+01 &   2.72E+04 & 0.42    & WD       \\ 
     &      & 3.3  &   ---    & ---      &   ---       & ---      & ---      &   ---      & ---      &   ---      & 0.40    & He-WD    \\ 
\enddata
\end{deluxetable}


\begin{deluxetable}{ll|cc|ccc|cc|ccl}
\label{tab:tab4}
\tablecolumns{12}
\tablewidth{0pc}
\tabletypesize{\scriptsize}
\tablecaption{Extended `WD Flasher' results -- Characteristics at core He flash and Horizontal Branch for only the WD Flashers; $Z$ from $0.0001$ to $0.03$}
\tablehead{
\multicolumn{2}{c}{} & \multicolumn{2}{c}{Core-Flash} & \multicolumn{3}{c}{ZAHB}                       & \multicolumn{2}{c}{TAHB}     & `mid'-HB  & \multicolumn{2}{c}{} \\
$M$  & $\etareim$  & $M_{core}$ & $\Menv$     & Time      & $\Teff$   & $L$             & $\Teff$  & $L$             & $\Teff$  & $M_{f}$  & Flash
}
\startdata
\sidehead{[Z,Y] = [0.0001, 0.24]}
\hline
 0.80 & 1.5  &   0.461  & 3.10E-03 &   14.1      & 2.91E+04 & 1.42E+01 &   4.02E+04 & 4.36E+01 &   3.62E+04 & 0.46    & WD       \\ 
 0.85 & 2.0  &   0.452  & 3.50E-03 &   11.2      & 2.88E+04 & 1.31E+01 &   3.99E+04 & 3.74E+01 &   3.52E+04 & 0.45    & WD       \\ 
 0.90 & 2.3  &   0.454  & 3.80E-03 &   9.1       & 2.89E+04 & 1.33E+01 &   3.97E+04 & 3.81E+01 &   3.56E+04 & 0.46    & WD       \\ 
 1.00 & 3.4  &   0.447  & 3.50E-03 &   6.1       & 2.85E+04 & 1.23E+01 &   3.99E+04 & 3.65E+01 &   3.50E+04 & 0.45    & WD       \\ 
\sidehead{[Z,Y] = [0.00014, 0.24]}
\hline
 0.80 & 1.5  &   0.459  & 3.20E-03 &   14.2      & 2.90E+04 & 1.41E+01 &   4.02E+04 & 4.21E+01 &   3.61E+04 & 0.46    & WD       \\ 
 0.85 & 1.9  &   0.453  & 4.00E-03 &   11.3      & 2.87E+04 & 1.33E+01 &   3.99E+04 & 3.90E+01 &   3.51E+04 & 0.46    & WD       \\ 
 0.90 & 2.3  &   0.453  & 3.20E-03 &   9.1       & 2.90E+04 & 1.31E+01 &   3.98E+04 & 3.77E+01 &   3.57E+04 & 0.46    & WD       \\ 
 1.00 & 3.3  &   0.446  & 3.10E-03 &   6.2       & 2.88E+04 & 1.22E+01 &   4.04E+04 & 3.56E+01 &   3.48E+04 & 0.45    & WD       \\ 
\sidehead{[Z,Y] = [0.001, 0.24]}
\hline
 0.80 & 1.2  &   0.451  & 2.50E-03 &   15.7      & 2.85E+04 & 1.28E+01 &   4.03E+04 & 3.88E+01 &   3.54E+04 & 0.45    & WD       \\ 
 0.90 & 1.8  &   0.448  & 2.20E-03 &   10.1      & 2.88E+04 & 1.23E+01 &   4.08E+04 & 3.60E+01 &   3.69E+04 & 0.45    & WD       \\ 
 1.00 & 2.5  &   0.444  & 3.10E-03 &   6.8       & 2.84E+04 & 1.19E+01 &   4.01E+04 & 3.42E+01 &   3.63E+04 & 0.45    & WD       \\ 
 1.10 & 3.4  &   0.439  & 2.70E-03 &   4.8       & 2.81E+04 & 1.12E+01 &   4.04E+04 & 3.33E+01 &   3.59E+04 & 0.44    & WD       \\ 
 1.20 & 4.4  &   0.437  & 2.50E-03 &   3.5       & 2.85E+04 & 1.09E+01 &   4.10E+04 & 3.17E+01 &   3.57E+04 & 0.44    & WD       \\ 
\sidehead{[Z,Y] = [0.005, 0.26]}
\hline
 0.80 & 1.0  &   0.445  & 2.20E-03 &   19.2      & 2.74E+04 & 1.16E+01 &   3.98E+04 & 3.48E+01 &   2.92E+04 & 0.45    & WD       \\ 
 0.90 & 1.4  &   0.443  & 2.20E-03 &   12.2      & 2.73E+04 & 1.14E+01 &   4.01E+04 & 3.52E+01 &   2.89E+04 & 0.45    & WD       \\ 
 1.00 & 1.9  &   0.442  & 2.40E-03 &   8.2       & 2.73E+04 & 1.12E+01 &   4.01E+04 & 3.51E+01 &   2.91E+04 & 0.44    & WD       \\ 
 1.10 & 2.5  &   0.440  & 2.20E-03 &   5.7       & 2.72E+04 & 1.09E+01 &   4.02E+04 & 3.38E+01 &   2.91E+04 & 0.44    & WD       \\ 
 1.20 & 3.2  &   0.439  & 2.20E-03 &   4.2       & 2.72E+04 & 1.08E+01 &   4.04E+04 & 3.40E+01 &   2.86E+04 & 0.44    & WD       \\ 
\sidehead{[Z,Y] = [0.01, 0.28]}
\hline
 0.80 & 0.9  &   0.441  & 1.60E-03 &   21.9      & 2.69E+04 & 1.07E+01 &   3.94E+04 & 3.19E+01 &   2.78E+04 & 0.44    & WD       \\ 
 0.90 & 1.3  &   0.442  & 1.90E-03 &   13.9      & 2.67E+04 & 1.08E+01 &   3.95E+04 & 3.36E+01 &   2.83E+04 & 0.44    & WD       \\ 
 1.00 & 1.9  &   0.437  & 2.10E-03 &   9.3       & 2.65E+04 & 1.03E+01 &   3.92E+04 & 3.06E+01 &   2.86E+04 & 0.44    & WD       \\ 
 1.10 & 2.5  &   0.436  & 2.10E-03 &   6.4       & 2.65E+04 & 1.01E+01 &   3.95E+04 & 3.09E+01 &   2.74E+04 & 0.44    & WD       \\ 
 1.20 & 3.0  &   0.438  & 2.00E-03 &   4.7       & 2.68E+04 & 1.04E+01 &   4.01E+04 & 3.26E+01 &   2.89E+04 & 0.44    & WD       \\ 
\sidehead{[Z,Y] = [0.02, 0.28]}
\hline
 0.90 & 0.8  &   0.446  & 1.30E-03 &   19.8      & 2.58E+04 & 1.06E+01 &   3.85E+04 & 3.43E+01 &   2.77E+04 & 0.45    & WD       \\ 
 1.00 & 1.5  &   0.437  & 1.60E-03 &   13.2      & 2.62E+04 & 9.64E+00 &   3.96E+04 & 3.02E+01 &   2.77E+04 & 0.44    & WD       \\ 
 1.10 & 2.0  &   0.436  & 1.50E-03 &   9.0       & 2.59E+04 & 9.55E+00 &   3.94E+04 & 3.01E+01 &   2.65E+04 & 0.44    & WD       \\ 
 1.20 & 2.5  &   0.437  & 1.60E-03 &   6.5       & 2.56E+04 & 9.67E+00 &   3.84E+04 & 3.00E+01 &   2.76E+04 & 0.44    & WD       \\ 
\sidehead{[Z,Y] = [0.03, 0.28]}
\hline
 1.00 & 0.9  &   0.445  & 2.50E-03 &   16.4      & 2.36E+04 & 1.01E+01 &   3.43E+04 & 3.21E+01 &   2.41E+04 & 0.45    & WD       \\ 
 1.10 & 1.6  &   0.440  & 1.30E-03 &   11.2      & 2.50E+04 & 9.46E+00 &   3.77E+04 & 3.04E+01 &   2.54E+04 & 0.44    & WD       \\ 
 1.20 & 2.2  &   0.436  & 1.60E-03 &   8.0       & 2.54E+04 & 9.04E+00 &   3.93E+04 & 2.96E+01 &   2.56E+04 & 0.44    & WD       \\ 
 1.30 & 2.8  &   0.434  & 1.50E-03 &   6.1       & 2.57E+04 & 8.84E+00 &   3.90E+04 & 2.72E+01 &   2.59E+04 & 0.44    & WD       \\ 
\enddata
\end{deluxetable}

\begin{deluxetable}{lll|cc|ccc|cc|ccl}
\label{tab:tab5}
\tablecolumns{13}
\tablewidth{0pc}
\tabletypesize{\scriptsize}
\tablecaption{Extended `WD Flasher' results, He-enriched models -- Characteristics at core He flash and Horizontal Branch for only the WD Flashers; $Z$ from $0.0001$ to $0.03$}
\tablehead{
\multicolumn{3}{c}{} & \multicolumn{2}{c}{Core-Flash} & \multicolumn{3}{c}{ZAHB}                       & \multicolumn{2}{c}{TAHB}     & `mid'-HB  & \multicolumn{2}{c}{} \\
$M$ & $Y$  & $\etareim$  & $M_{core}$ & $\Menv$     & Time      & $\Teff$   & $L$             & $\Teff$  & $L$             & $\Teff$  & $M_{f}$  & Flash
}
\startdata
\sidehead{Z = 0.0001}
\hline
0.80 & 0.32 & 2.2  &   0.437  & 4.90E-03 &   8.4       & 2.87E+04 & 1.11E+01 &   4.14E+04 & 3.40E+01 &   3.55E+04 & 0.44    & WD       \\ 
     & 0.36 & 2.8  &   0.423  & 5.50E-03 &   6.3       & 2.84E+04 & 9.44E+00 &   4.19E+04 & 2.75E+01 &   3.51E+04 & 0.43    & WD       \\ 
     & 0.40 & 3.6  &   0.410  & 5.90E-03 &   4.7       & 2.83E+04 & 8.07E+00 &   4.77E+04 & 1.56E+01 &   3.52E+04 & 0.42    & WD       \\ 
0.90 & 0.32 & 3.4  &   0.431  & 4.30E-03 &   5.4       & 2.87E+04 & 1.02E+01 &   4.19E+04 & 3.10E+01 &   3.57E+04 & 0.43    & WD       \\ 
     & 0.36 & 4.7  &   0.412  & 5.50E-03 &   4.1       & 2.80E+04 & 8.17E+00 &   4.26E+04 & 2.26E+01 &   3.53E+04 & 0.42    & WD       \\ 
     & 0.40 & 6.0  &   0.401  & 5.50E-03 &   3.0       & 2.77E+04 & 7.09E+00 &   4.48E+04 & 1.19E+01 &   2.98E+04 & 0.41    & WD       \\ 
\sidehead{Z = 0.00014}
\hline
0.80 & 0.32 & 2.2  &   0.434  & 4.60E-03 &   8.4       & 2.85E+04 & 1.08E+01 &   4.15E+04 & 3.28E+01 &   3.49E+04 & 0.44    & WD       \\ 
     & 0.36 & 2.9  &   0.418  & 5.30E-03 &   6.4       & 2.82E+04 & 8.70E+00 &   4.19E+04 & 2.50E+01 &   3.56E+04 & 0.42    & WD       \\ 
     & 0.40 & 3.6  &   0.408  & 5.30E-03 &   4.8       & 2.80E+04 & 7.69E+00 &   4.68E+04 & 1.38E+01 &   3.61E+04 & 0.41    & WD       \\ 
0.90 & 0.32 & 3.4  &   0.428  & 4.60E-03 &   5.4       & 2.85E+04 & 9.67E+00 &   4.22E+04 & 3.04E+01 &   3.04E+04 & 0.43    & WD       \\ 
     & 0.36 & 4.4  &   0.413  & 5.50E-03 &   4.1       & 2.81E+04 & 8.24E+00 &   4.28E+04 & 2.30E+01 &   3.55E+04 & 0.42    & WD       \\ 
     & 0.40 & 5.6  &   0.401  & 5.30E-03 &   3.1       & 2.78E+04 & 7.14E+00 &   4.51E+04 & 1.22E+01 &   2.96E+04 & 0.41    & WD       \\ 
\sidehead{Z = 0.001}
\hline
0.80 & 0.32 & 1.7  &   0.433  & 2.80E-03 &   9.2       & 2.84E+04 & 1.04E+01 &   4.15E+04 & 3.04E+01 &   3.67E+04 & 0.43    & WD       \\ 
     & 0.36 & 2.0  &   0.425  & 3.00E-03 &   7.0       & 2.87E+04 & 9.21E+00 &   4.30E+04 & 2.75E+01 &   3.66E+04 & 0.43    & WD       \\ 
     & 0.40 & 2.8  &   0.405  & 3.90E-03 &   5.2       & 2.78E+04 & 7.30E+00 &   4.64E+04 & 1.37E+01 &   3.67E+04 & 0.41    & WD       \\ 
0.90 & 0.32 & 2.5  &   0.429  & 3.10E-03 &   5.9       & 2.83E+04 & 1.02E+01 &   4.19E+04 & 3.04E+01 &   3.56E+04 & 0.43    & WD       \\ 
     & 0.36 & 3.4  &   0.410  & 3.90E-03 &   4.5       & 2.84E+04 & 8.87E+00 &   4.78E+04 & 1.60E+01 &   3.66E+04 & 0.41    & WD       \\ 
     & 0.40 & 4.0  &   0.403  & 4.30E-03 &   3.4       & 2.78E+04 & 7.14E+00 &   4.60E+04 & 1.29E+01 &   3.00E+04 & 0.41    & WD       \\ 
\sidehead{Z = 0.005}
\hline
0.80 & 0.32 & 1.3  &   0.431  & 2.00E-03 &   12.6      & 2.72E+04 & 9.83E+00 &   4.12E+04 & 2.98E+01 &   2.91E+04 & 0.43    & WD       \\ 
     & 0.36 & 1.5  &   0.424  & 2.70E-03 &   9.4       & 2.73E+04 & 9.01E+00 &   4.16E+04 & 2.66E+01 &   2.93E+04 & 0.43    & WD       \\ 
     & 0.40 & 1.9  &   0.412  & 2.50E-03 &   7.0       & 2.72E+04 & 7.88E+00 &   4.80E+04 & 1.52E+01 &   2.91E+04 & 0.42    & WD       \\ 
0.90 & 0.32 & 1.9  &   0.428  & 2.00E-03 &   8.0       & 2.72E+04 & 9.48E+00 &   4.08E+04 & 2.77E+01 &   2.90E+04 & 0.43    & WD       \\ 
     & 0.36 & 2.2  &   0.421  & 2.60E-03 &   6.0       & 2.71E+04 & 8.63E+00 &   4.32E+04 & 2.64E+01 &   2.93E+04 & 0.42    & WD       \\ 
     & 0.40 & 2.7  &   0.410  & 3.20E-03 &   4.5       & 2.71E+04 & 7.65E+00 &   4.69E+04 & 1.41E+01 &   2.80E+04 & 0.41    & WD       \\ 
\sidehead{Z = 0.01}
\hline
0.90 & 0.32 & 1.6  &   0.433  & 2.30E-03 &   10.4      & 2.67E+04 & 9.75E+00 &   4.01E+04 & 2.97E+01 &   2.68E+04 & 0.44    & WD       \\ 
     & 0.36 & 2.0  &   0.422  & 2.00E-03 &   7.7       & 2.64E+04 & 8.44E+00 &   4.02E+04 & 2.42E+01 &   2.67E+04 & 0.42    & WD       \\ 
     & 0.40 & 2.4  &   0.413  & 2.60E-03 &   5.7       & 2.66E+04 & 7.84E+00 &   4.16E+04 & 2.09E+01 &   2.66E+04 & 0.42    & WD       \\ 
1.00 & 0.32 & 2.2  &   0.431  & 2.20E-03 &   7.0       & 2.65E+04 & 9.47E+00 &   4.01E+04 & 2.89E+01 &   2.67E+04 & 0.43    & WD       \\ 
     & 0.36 & 2.7  &   0.421  & 2.00E-03 &   5.2       & 2.64E+04 & 8.27E+00 &   4.02E+04 & 2.36E+01 &   2.69E+04 & 0.42    & WD       \\ 
     & 0.40 & 3.3  &   0.411  & 2.50E-03 &   4.0       & 2.64E+04 & 7.49E+00 &   4.45E+04 & 1.98E+01 &   2.65E+04 & 0.41    & WD       \\ 
\sidehead{Z = 0.02}
\hline
0.90 & 0.32 & 1.3  &   0.432  & 1.60E-03 &   14.6      & 2.55E+04 & 9.03E+00 &   3.87E+04 & 2.75E+01 &   2.66E+04 & 0.43    & WD       \\ 
     & 0.36 & 1.6  &   0.425  & 1.60E-03 &   10.8      & 2.54E+04 & 8.10E+00 &   3.93E+04 & 2.49E+01 &   2.65E+04 & 0.43    & WD       \\ 
     & 0.40 & 2.0  &   0.415  & 2.00E-03 &   7.9       & 2.54E+04 & 7.35E+00 &   4.13E+04 & 2.16E+01 &   2.61E+04 & 0.42    & WD       \\ 
1.00 & 0.32 & 1.7  &   0.435  & 1.50E-03 &   9.7       & 2.59E+04 & 9.31E+00 &   3.93E+04 & 2.93E+01 &   2.72E+04 & 0.43    & WD       \\ 
     & 0.36 & 2.1  &   0.426  & 1.70E-03 &   7.2       & 2.56E+04 & 8.29E+00 &   3.99E+04 & 2.62E+01 &   2.68E+04 & 0.43    & WD       \\ 
     & 0.40 & 2.6  &   0.416  & 2.10E-03 &   5.3       & 2.55E+04 & 7.41E+00 &   4.20E+04 & 2.24E+01 &   2.72E+04 & 0.42    & WD       \\ 
\sidehead{Z = 0.03}
\hline
1.00 & 0.32 & 1.6  &   0.430  & 1.50E-03 &   12.1      & 2.53E+04 & 8.38E+00 &   3.87E+04 & 2.54E+01 &   2.55E+04 & 0.43    & WD       \\ 
     & 0.36 & 2.1  &   0.420  & 1.50E-03 &   8.9       & 2.46E+04 & 7.46E+00 &   3.99E+04 & 2.35E+01 &   2.47E+04 & 0.42    & WD       \\ 
     & 0.40 & 2.6  &   0.411  & 1.90E-03 &   6.5       & 2.46E+04 & 6.67E+00 &   4.54E+04 & 1.93E+01 &   2.46E+04 & 0.41    & WD       \\ 
1.10 & 0.32 & 2.2  &   0.428  & 1.50E-03 &   8.4       & 2.48E+04 & 8.13E+00 &   3.91E+04 & 2.66E+01 &   2.53E+04 & 0.43    & WD       \\ 
     & 0.36 & 2.7  &   0.420  & 1.50E-03 &   6.2       & 2.48E+04 & 7.40E+00 &   4.02E+04 & 2.31E+01 &   2.49E+04 & 0.42    & WD       \\ 
     & 0.40 & 3.4  &   0.409  & 1.80E-03 &   4.7       & 2.45E+04 & 6.66E+00 &   4.44E+04 & 1.30E+01 &   2.44E+04 & 0.41    & WD       \\ 
\enddata
\end{deluxetable}

\begin{deluxetable}{lllll|c|cc}
\label{tab:tab6}
\tablecolumns{8}
\tablewidth{0pc}
\tabletypesize{\scriptsize}
\tablecaption{$\Teff$@TAHB and maximal attained $\Teff$ during \emph{post-HB} for representative Pop. I and Pop. II, normal Y and He-enriched, models}
\tablehead{
$M$  &     $Z$   &    $Y$   &    $\etareim$  &   Flash &  $\Teff$@TAHB  &     Max. $\Teff$ ($10^4$~K) &  Duration around Max. $\Teff$ (yr)
}
\startdata
0.90 &  0.001 & 0.24 &  1.0  &  Tip  &  4.92e3    &     9.2      &      3.15e4  \\
     &        &      &  1.6  &  Post &  2.32e4    &     7.0      &      8.37e4  \\
     &        &      &  1.8  &  WD   &  4.08e4    &     5.6      &      2.06e6  \\
     &        & 0.36 &  1.2  &  Tip  &  5.16e3    &     11.0     &      1.98e4  \\
     &        &      &  2.5  &  Post &  2.73e4    &     6.5      &      3.62e4  \\
     &        &      &  3.4  &  WD   &  4.78e4    &     5.1      &      1.76e5  \\
\hline
1.00 &  0.02  & 0.28 &  0.6  &  Tip  &  4.67e3    &     9.1      &      1.76e4  \\
     &        &      &  1.1  &  Post &  2.06e4    &     6.9      &      1.83e5  \\
     &        &      &  1.5  &  WD   &  3.96e4    &     5.8      &      2.93e5  \\
     &        & 0.36 &  1.0  &  Tip  &  4.68e3    &     10.0     &      2.34e4  \\
     &        &      &  1.8  &  Post &  2.71e4    &     6.8      &      3.60e5  \\
     &        &      &  2.1  &  WD   &  3.99e4    &     5.0      &      5.23e5  \\
\enddata
\end{deluxetable}


\clearpage


\clearpage

\end{document}